\documentclass{nmeauth}

\usepackage[T1]{fontenc}

\usepackage{moreverb}

\usepackage{url}
\usepackage{pgfplots}

\usepackage{multirow}  
\usepackage{etoolbox}  
\usepackage{mathrsfs}
\usepackage{array}
\usepackage{arydshln}   
\usepackage{subcaption} 

\usepackage{tikz}
\usetikzlibrary{calc}
\usetikzlibrary{trees}
\usetikzlibrary{matrix}
\usepackage{pgfplots}
\usetikzlibrary{pgfplots.groupplots}
\usetikzlibrary{pgfplots.statistics}
\usetikzlibrary{positioning, shapes, shadows, arrows}
\usetikzlibrary{hobby}
\usetikzlibrary{external}
\usetikzlibrary{decorations.pathreplacing}
\usetikzlibrary{patterns}
\usetikzlibrary{fadings}

\usepackage{pgfplotstable}
\usepackage{tikz-3dplot}
\pgfplotsset{compat=newest}
\usepgfplotslibrary{patchplots}

\usepackage[sort&compress]{natbib}
\bibpunct{[}{]}{,}{n}{}{,}

\usepackage{arydshln} 

\usepackage[ruled]{algorithm2e}

\usepackage{upgreek}

\global\long\def\vct#1{\boldsymbol{#1}}

\global\long\def\discret#1{{\boldsymbol{\mathsf{#1}}}}

\global\long\def\transpose#1{#1^{\mathsf{T}}}

\global\long\def\idx#1{{\mathsf{#1}}}

\newcommand{\breakeq   }[3]{\ifthenelse{\equal{#1}{break}}{\nonumber\\&#2}{#3}}
\newcommand{\addifbreak}[2]{\ifthenelse{\equal{#1}{break}}{}{#2}}

\DeclareFontFamily{U}{mathx}{\hyphenchar\font45}
\DeclareFontShape{U}{mathx}{m}{n}{
      <5> <6> <7> <8> <9> <10>
      <10.95> <12> <14.4> <17.28> <20.74> <24.88>
      mathx10
      }{}
\DeclareSymbolFont{mathx}{U}{mathx}{m}{n}
\DeclareFontSubstitution{U}{mathx}{m}{n}
\DeclareMathAccent{\widecheck}{0}{mathx}{"71}
\DeclareMathAccent{\wideparen}{0}{mathx}{"75}

\AtBeginEnvironment{align}{\setcounter{subeqn}{0}}
\newcounter{subeqn} %

\usepackage[colorlinks,bookmarksopen,bookmarksnumbered,citecolor=red,urlcolor=black]{hyperref}

\newcommand\BibTeX{{\rmfamily B\kern-.05em \textsc{i\kern-.025em b}\kern-.08em
T\kern-.1667em\lower.7ex\hbox{E}\kern-.125emX}}

\newcommand{\secref}[1]{Section~\ref{#1}} 
\newcommand{\figref}[1]{Figure~\ref{#1}} 

\newcommand{\lat}[1]{#1} 
\newcommand{\ie}{\lat{i.e.\@}} 
\newcommand{\cf}{\lat{cf.\@}} 
\newcommand{\wrt}{\lat{w.r.t.\@}}
\newcommand{\eg}{\lat{e.g.\@}}

\providecommand{\DSCLab}{Data Science \& Computing Lab}
\providecommand{\IMCS}{Institute for Mathematics and Computer-Based Simulation}
\providecommand{\LNM}{Institute for Computational Mechanics}
\providecommand{\MHPC}{Mechanics \& High Performance Computing Group}
\providecommand{\TUM}{Technical University of Munich}
\providecommand{\UNIBW}{University of the Bundeswehr Munich}

\newcommand{\SoftwarePackage}[1]{\textsc{#1}}
\newcommand{\trilinos}{\SoftwarePackage{Trilinos}}
\newcommand{\muelu}{\SoftwarePackage{MueLu}}
\newcommand{\xpetra}{\SoftwarePackage{Xpetra}}

\providecommand{\norm}[2][]{\left\|{#2}\right\|_{#1}}

\newcommand{\aggbased}{aggregation-based } 

\newcommand{\level}{\ell}    

\newcommand{\schur}{S}

\providecommand{\diagonalmat}{\mathbf{D}}
\providecommand{\Dinv}{\diagonalmat^{-1}}

\newcommand{\meshsize}{\mathrm{h}}

\newcommand{\dualbasis}{\psi}

\newcommand{\PAAMG}{PA-AMG}
\newcommand{\SAAMG}{SA-AMG}

\newcommand{\EMIN}{Emin}
\providecommand{\paamg}{\PAAMG}
\providecommand{\saamg}{\SAAMG}

\providecommand{\cheapsimple}{CheapSIMPLE}
\providecommand{\cheapsimplec}{CheapSIMPLEC}

\newcommand{\fint}{\discret{f}_{\textnormal{int}}}
\newcommand{\fext}{\discret{f}_{\textnormal{ext}}}
\newcommand{\fco}{\discret{f}_{\textnormal{co}}}

\newcommand{\couplmat}{\discret{C}}

\providecommand{\virtualwork}{\delta\mathcal{W}}

\providecommand{\master}{\mathcal{M}}
\providecommand{\slave} {\mathcal{S}}
\providecommand{\inner}{\mathcal{N}}
\providecommand{\nione}{\inner_1}
\providecommand{\nitwo}{\inner_2}
\providecommand{\act}{\mathcal{A}}
\providecommand{\inact}{\mathcal{I}}

\providecommand{\timintscalar}{} 

\providecommand{\nonlinstep}{i}

\providecommand{\res}{\mathbf{r}}  

\providecommand{\stiffmat}{\mathsf{K}}
\providecommand{\zeromat}{\mathsf{0}}
\providecommand{\mortarMmat}{\mathsf{M}}
\providecommand{\mortarDmat}{\mathsf{D}}
\providecommand{\mortarNmat}{\mathsf{N}}  
\providecommand{\mortarFmat}{\mathsf{F}}  
\providecommand{\mortarTmat}{\mathsf{T}}  
\providecommand{\mortarCmat}{\transpose{\mathsf{C}}_1}  
\providecommand{\mortarConstrmat}{\mathsf{C}_2} 
\providecommand{\mortarZonstrmat}{{\mathsf{Z}}} 

\providecommand{\apprx}[1]{\widetilde{#1}} 

\newcommand{\slavenodes}{\slave}

\newcommand{\dofs}[1][]{\mathcal{D}_{#1}}     
\newcommand{\slavedofs}{\dofs[\slavenodes]}

\providecommand{\dispaggTOlagagg}[1][]{{\mathsf{d}^{#1}}} 

\providecommand{\det}{\det}

\providecommand{\domain}{\Omega}
\providecommand{\refdomain}{\domain_0}

\providecommand{\bdrydomain}{\partial \domain}
\providecommand{\refbdrydomain}{\bdrydomain_0}

\providecommand{\refbdry}{\Gamma}
\providecommand{\refdirichbdry}{\refbdry_{\idxdirichlet}}
\providecommand{\refneumannbdry}{\refbdry_{\idxneumann}}
\providecommand{\refcontactbdry}{\refbdry_{\idxcontact}}

\providecommand{\curbdry}{\gamma}

\providecommand{\curcontactbdry}{\curbdry_{\idxcontact}}

\providecommand{\idxcontact}{\idx{c}}

\providecommand{\idxdirichlet}{\idx{D}}
\providecommand{\idxneumann}{\idx{N}}

\providecommand{\slave}{\idx{s}}
\providecommand{\master}{\idx{m}}
\providecommand{\idxslave}{{(\slave)}}
\providecommand{\idxmaster}{{(\master)}}

\providecommand{\idxForSubdomain}{i}
\providecommand{\subdomain}{\mathcal{N}_\idxForSubdomain}

\providecommand{\idxSubdomain}{{(\idxForSubdomain)}}
\providecommand{\idxOne}{{(1)}}
\providecommand{\idxTwo}{{(2)}}

\providecommand{\dimdomain}{d}

\providecommand{\primalunknown}{u}
\providecommand{\disp}{\vct{\primalunknown}}

\providecommand{\dispi}{\Delta\disp}  

\providecommand{\testunknown}{v}
\providecommand{\testvec}{\vct{\testunknown}}

\providecommand{\lagmult}{\lambda}
\providecommand{\lagvec}{\vct{\lagmult}}
\providecommand{\lagveci}{\Delta\lagvec} 

\providecommand{\gapvec}{\textbf{g}} 

\providecommand{\normalvec}{\vct{n}}

\providecommand{\curnormal}{\normalvec}

\providecommand{\normalpart}{\mathrm{n}}
\providecommand{\tangentialpart}{\mathrm{\tau}}

\providecommand{\density}{\rho}
\providecommand{\refdensity}{\density_0}

\providecommand{\traction}{\vct{t}}

\providecommand{\curtraction}{\traction}

\providecommand{\curnormalcontacttraction}{p_{\normalpart}}

\providecommand{\ttime}{t}

\providecommand{\gapfunction}{g_{\normalpart}}

\providecommand{\gapfunctiondiscretweighted}{\widetilde{g}_{\normalpart,\meshsize}}

\providecommand{\Uspace}{\mathcal{U}}
\providecommand{\Vspace}{\mathcal{V}}
\providecommand{\Mspace}{\mathcal{M}}

\providecommand{\vecUspace}{\boldsymbol{\Uspace}}
\providecommand{\vecVspace}{\boldsymbol{\Vspace}}
\providecommand{\vecMspace}{\boldsymbol{\Mspace}}

\providecommand{\muvec}{\vct{\mu}}

\newcommand{\YoungModulus}{E} 

\providecommand{\MphysicsA}{gray!70!white}
\providecommand{\MphysicsB}{gray!50!white}

\providecommand{\MphysicsCouplingA}{lightgray!50!white}

\providecommand{\A}{\mathbf{A}}
\providecommand{\AOO}{\A_{00}}
\providecommand{\AOl}{\A_{01}}
\providecommand{\AlO}{\A_{10}}
\providecommand{\All}{\A_{11}}

\providecommand{\Prol}{\textnormal{\textbf{P}}}

\providecommand{\Restr}{\textnormal{\textbf{R}}}

\providecommand{\prolongator}{\Prol}
\providecommand{\restrictor}{\Restr}
\providecommand{\ptent}{\widehat{\prolongator}}
\providecommand{\rtent}{\widehat{\restrictor}}

\providecommand{\Aop}{\A}

\providecommand{\Afine}{\Aop^{(\level)}}
\providecommand{\Acoarse}{\Aop^{(\level+1)}}
\providecommand{\ddim}{n}

\providecommand{\residual}{r}

\providecommand{\correction}{c}

\providecommand{\sol}{x}

\providecommand{\rhs}{b}

\providecommand{\smoother}{\mathscr{S}}

\providecommand{\presmoothingsweeps}{\nu_1}
\providecommand{\postsmoothingsweeps}{\nu_2}

\providecommand{\identitymatrix}{I} 

\providecommand{\relaxationinversemat}{Q}  
\providecommand{\relaxationiterationidx}{k} 

\providecommand{\indUzawa}[1]{{#1}_{\textnormal{UZ}}}
\providecommand{\indBraessSarazin}[1]{{#1}_{\textnormal{BS}}}
\providecommand{\indSIMPLE}[1]{{#1}_{\textnormal{SIMPLE}}}

\providecommand{\aggs}[1][]{\mathscr{A_{#1}}}
\providecommand{\numnodes}[1][]{{m_{#1}}}

\providecommand{\numdofs}[1][]{\ddim_{#1}}

\providecommand{\doftonodemapping}{\mathrm{n}}
\providecommand{\doftonode}[1][]{\doftonodemapping(#1)}

\providecommand{\numaggregatedneighbors}[1][]{\mathrm{ngh}} 

\providecommand{\diag}[1][]{\textnormal{diag}#1} 

\providecommand{\linstep}{k}  

\providecommand{\prolongatordisp}{\prolongator^{\disp}}

\providecommand{\ptentlagv}{\ptent^{\lagvec}}
\providecommand{\restrictordisp}{\restrictor^{\disp}}

\providecommand{\rtentlagv}{\rtent^{\lagvec}}

\providecommand{\meshfac}{\kappa} 
\providecommand{\nproc}{n^{\mathrm{proc}}} 
\providecommand{\nDof}{n_{\mathrm{DOF}}} 
\providecommand{\nDofPrimal}{\nDof^{\disp}} 
\providecommand{\nDofDual}{\nDof^{\lagvec}} 
\providecommand{\nDofTotal}{\nDof^{\mathrm{total}}} 
\providecommand{\nDofTotalCoarsest}{\nDof^{\mathrm{total (\numlevels)}}} 
\providecommand{\operatorComplexity}{\mathcal{C}_A}

\providecommand{\numlevels}{n^{\level}} 

\def\volumeyear{2021}

\begin{document}

\runningheads{T.~A.~Wiesner et al.}{Algebraic multigrid for saddle point systems in contact mechanics}

\title{Algebraic multigrid methods for saddle point systems arising from mortar contact formulations}

\author{T.~A.~Wiesner\affil{1}$^,$\footnotemark[3], M.~Mayr\affil{2,3}$^,$\corrauth, A.~Popp\affil{2}$^,$\footnotemark[3],  M.~W.~Gee\affil{4} and W.~A. Wall\affil{5}}

\address{\centering
\affilnum{1}\ Leica Geosystems AG, Heinrich-Wild-Strasse 201, 9435 Heerbrugg, Switzerland\\
\affilnum{2}\ {\IMCS}, {\UNIBW}, \\ Werner-Heisenberg-Weg 39, D-85577 Neubiberg, Germany\\
\affilnum{3}\ {\DSCLab}, {\UNIBW}, \\ Werner-Heisenberg-Weg 39, D-85577 Neubiberg, Germany\\
\affilnum{4}\ {\MHPC}, {\TUM}, \\ Parkring 35, D-85748 Garching, Germany\\
\affilnum{5}\ {\LNM}, {\TUM}, \\ Boltzmannstr. 15, D-85748 Garching, Germany}

\corraddr{M.~Mayr, {\IMCS}, {\UNIBW}, Werner-Heisenberg-Weg 39, D-85577 Neubiberg, Germany, E-mail: matthias.mayr@unibw.de}

\begin{abstract}
In this paper, a fully aggregation-based algebraic multigrid strategy is developed
for nonlinear contact problems of saddle point type using a mortar finite element approach.
While the idea of extending multigrid methods to saddle point systems can already be found,
e.g., in the context of Stokes and Oseen equations in literature,
the main contributions of this work are
(i) the development and open-source implementation of an interface aggregation strategy specifically suited for generating Lagrange multiplier aggregates
that are required for coupling structural equilibrium equations with contact constraints and
(ii) a review of saddle point smoothers in the context of constrained interface problems.
The new interface aggregation strategy perfectly fits into an aggregation-based multigrid framework and can easily be combined with segregated transfer operators,
which allow to preserve the saddle point structure on the coarse levels.
Further analysis provides insight into saddle point smoothers applied to contact problems,
while numerical experiments illustrate the robustness of the new method.
We have implemented the proposed algorithm within the {\muelu} package of the open-source {\trilinos} project.
Numerical examples demonstrate the robustness of the proposed method in complex dynamic contact problems
as well as its scalability up to $23.9$ million unknowns on 480 MPI ranks.
\end{abstract}


\maketitle

\footnotetext[3]{This work was partially performed while these authors were affiliated with the {\LNM}, {\TUM}, Boltzmannstr. 15, D-85748 Garching, Germany.}


\section{Introduction}

Many engineering applications require the simulation of large-scale contact problems.
Therefore, it is not surprising that recent years have seen significant progress in modelling and simulation of contact interaction
and its associated phenomena, such as friction~\cite{ap_Gitterle2010,ap_Seitz2015,Popov2017a},
wear~\cite{ap_Cavalieri2013,Farah2016b,Lengiewicz2012a,Milanese2019a},
adhesion~\cite{Sauer2011a,Li2019a},
or multi-scale contact phenomena~\cite{Bonari_FEMBEM_rough_surface_contact,Vakis2018a}.
This is particularly true with regard to robust finite element based discretization techniques for finite deformations
and efficient nonlinear solution algorithms.
Above all, mortar finite element methods
--- originally introduced in the context of domain decomposition~\cite{belgacem1999,bernardi1994} ---
are meanwhile well-established as a basis for state-of-the-art contact formulations
and widely accepted among researchers as being superior to more classical discretization techniques,
such as the node-to-segment (NTS) method, the Gauss-point-to-segment (GPTS) method
and other collocation based approaches~\cite{ap_Lorenzis2014,ap_Puso2004,ap_Puso2004a,ap_Wohlmuth2011,ap_Wriggers2006}.

Nowadays, constraint enforcement in the context of mortar methods is often based on a regularized Lagrange multiplier scheme
or an augmented Lagrange method instead of a simple, yet often insufficient penalty approach.
Independent from the actual details of the constraint enforcement implementation,
the discrete Lagrange multipliers constitute an additional set of degrees of freedom in the mortar finite element contact formulation.
When using a dual mortar approach~\cite{ap_Hartmann2007, ap_Hueber2008,ap_Hueber2005,popp2009,ap_Popp2010,ap_Wohlmuth2000,wohlmuth2001},
the discrete Lagrange multiplier basis is chosen based on a 
biorthogonality condition with the underlying finite element basis.
This allows for the localization of the contact constraints and, thus, from a more algebraic point of view,
for the trivial condensation of the additional Lagrange multiplier degrees of freedom from the final linearized systems of equations.
If such a static condensation is not desired or not feasible ({\eg} when choosing a standard basis rather than dual basis functions for the Lagrange multipliers, see \eg~\cite{ap_Wohlmuth2000,Popp2014a}),
the linear system remains in its generalized saddle point format arising from the contact constraint equations.
Both the standard and the dual mortar approach have become increasingly popular in recent years,
with new contributions focusing for example on higher-order finite element interpolation~\cite{ap_Popp2012a,ap_Wohlmuth2012},
isogeometric mortar methods~\cite{ap_Lorenzis2011,ap_Temizer2011,ap_Temizer2012,Wunderlich2019a,Seitz2016a},
or improved robustness of the solution algorithms~\cite{ap_Franke2013,Hiermeier2018a,popp2013},
to name only a few particularly active research directions.

It is striking, however, that almost all current research endeavors concerned with mortar finite element methods for contact mechanics
focus exclusively on the modelling of various contact phenomena.
Yet, for large-scale and industrial applications the appropriate modelling of contact problems is not sufficient.
In fact, the demand for efficient solution strategies tailored to the specifics of contact simulations is eminent
in order to achieve optimal over-all performance.
Whereas one could use parallel direct solvers to solve the linear systems,
they are not an option for very large problems.
Iterative solvers for sparse systems ({\eg}~\cite{hackbusch1994,saad2003}) combined with good preconditioning strategies
are a far better choice with respect to computational resources. 
In particular, multigrid methods~\cite{hackbusch1985, trottenberg2001multigrid} are known to be among the most efficient solution and preconditioning strategies, at least for certain classes of problems.

From the perspective of the linear solvers and multigrid-based preconditioners,
the condensation of the Lagrange multipliers seems to be very attractive,
since it allows to circumvent the more-sophisticated saddle point formulation.
For contact problems though, we have experienced
that the resulting linear systems after condensation suffer from some challenging matrix properties
which cause severe convergence problems for standard preconditioning techniques.
In particular, the matrices tend to be non-diagonally dominant due to different (local) coordinate systems
that are typically used for the formulation of the structural equilibrium equations and the contact constraints.
In our previous work~\cite{wiesner2017},
we have developed multilevel preconditioners that address such issues
and are specifically tailored to contact problems using the dual mortar method in a condensed formulation.

On the other hand, multigrid methods already have been successfully applied to saddle point problems
as they arise from different applications 
({\eg}~Stokes flow~\cite{janka2006} or incompressible Navier-Stokes problems~\cite{griebel1998,lonsdale1991})
and even in the context of mortar finite element methods~\cite{wieners1999,Wohlmuth2000}.
The multigrid theory for this particular class of saddle point problems has evolved starting from special multigrid methods
for mortar finite element methods ({\eg}~\cite{braess1999_2,gopalakrishnan2000,krause2001,krause2000_2})
to mortar finite element methods in saddle point formulation (\eg,~\cite{braess1998,braess2000}).
Based on these ideas, specific multigrid methods for contact problems in saddle point formulation
have been developed in~\cite{wieners2003}. 
However, most of the literature available on multigrid for mortar finite element methods and contact problems in saddle point formulations
is primarily on geometric multigrid methods with abundant work on saddle point smoothers (\cf~\cite{zulehner2000}).
A first algebraic multigrid preconditioner for mortar-based contact problems has been proposed by \citet{adams2004},
performing standard aggregation on the graph of an auxiliary matrix imitating the Lagrange multipliers.
Alternatively, multigrid methods for contact problems not requiring an outer iteration loop or active set strategy have been developed in \cite{Kornhuber2001a,Wohlmuth2003a}.

In this paper, we address the case of mortar-based contact problems in saddle point formulation
and show how to tailor iterative solvers with algebraic multigrid preconditioners to such problems.
In contrast to geometric multigrid methods, algebraic multigrid methods ({\eg}~\cite{trottenberg2001}) 
do not rely on geometric user-provided mesh information, but use only purely algebraic information from the fine level matrix.
Since static condensation of Lagrange multiplier unknowns is not required, 
our approach is applicable to mortar methods using both standard or dual shape functions. 
The proposed multigrid method is based on the (smoothed) aggregation algebraic multigrid algorithms
(\cf~\cite{vanek1992,vanek1996,vanek19992,vanek2001,sala2008})
with special extensions for block matrices and some minor contact-specific adaptions.
We propose a novel aggregation strategy for the discrete Lagrange multiplier unknowns along the contact interface,
which we consider simpler to implement, computationally less expensive, and more intuitive for contact problems compared to the ideas from \cite{adams2004}.
Inspired by our prior work on fluid-structure interaction~\cite{gee2011,Mayr2020a},
where we have investigated the beneficial effect of satisfying interface constraints within the preconditioner,
we will then use segregated transfer operators suitable for block matrices
to transfer and incorporate the contact constraints in all coarse levels.
We analyze various Schur complement block smoothers and assess their suitability for satisfying the contact constraints.
Finally, we demonstrate and assess the performance of the proposed preconditioner
in several three-dimensional examples.

The remainder of this paper is organized as follows: 
Section~\ref{sec:mortarmethods} provides a brief introduction to mortar methods for finite deformation contact problems in saddle point formulation.
After the basic notation is introduced, we specifically present the resulting linear system that is arising 
if the discrete Lagrange multipliers are explicitly included into the set of unknowns to be solved for.
After a brief introduction to the general idea of multigrid methods,
Section~\ref{sec:AMGForContact} describes our strategy to tailor a multigrid preconditioner to contact problems in saddle point formulation.
It comprises the coarsening of the mortar contact constraints as detailed in Section~\ref{sec:CoarseningOfContactConstraints} 
as well as suitable block smoothers as discussed in Section~\ref{sec:BlockSmoothers}.
Finally, Section \ref{sec:NumEx} presents numerical examples that showcase the robustness, scalability, and performance of the proposed multigrid preconditioners,
before we close with some final remarks.

\section{Mortar methods for finite deformation contact}
\label{sec:mortarmethods}

As this paper is concerned with preconditioning of the system of linear equations arising from contact problems,
just a brief summary to the contact formulation and discretization is given here.
For a detailed presentation, the reader is referred to our previous work~\cite{ap_Popp2010}.

\subsection{Problem formulation and governing equations}

We consider two solid bodies, which are represented
by~$\refdomain^\idxOne$, $\refdomain^\idxTwo\subset\mathbb{R}^{\dimdomain}$
with~$\dimdomain\in\{2,3\}$ in the reference configuration.
Their surfaces~$\refbdrydomain^{\idxSubdomain}$, $\idxForSubdomain\in\left\{1,2\right\}$ are decomposed
into three disjoint subsets~$\refdirichbdry^{\idxSubdomain}$, $\refneumannbdry^{\idxSubdomain}$
and~$\refcontactbdry^{\idxSubdomain}$ denoting the Dirichlet boundary, the Neumann boundary,
and the potential contact interface with unknown contact tractions~$\curtraction_{\idxcontact}^\idxSubdomain$, respectively.
The solid bodies themselves are governed by nonlinear elasticity.
Since we are only interested in the algebraic block structure of the final system of equations after discretization and linearization,
it is sufficient to discuss a quasi-static contact problem with only two deformable bodies.

In order to describe the contact phenomenon,
we state the Hertz-Signorini-Moreau conditions
\begin{align}
\label{eq:HertzSignoriniMoreau}
\gapfunction \geq 0
\quad\wedge\quad
\curnormalcontacttraction \leq 0
\quad\wedge\quad
\gapfunction\curnormalcontacttraction = 0.
\end{align}
Therein, $\gapfunction$ defines a so-called gap function,
which measures the distance of a point 
on the slave interface~$\curcontactbdry^\idxslave$
to the projected corresponding point on the master side~$\curcontactbdry^\idxmaster$ of the contact interface in the current configuration. 
Furthermore, $\curnormalcontacttraction$ denotes the normal contact traction.
In the mathematical formulation, one introduces the negative slave side contact traction~$\curtraction_{\idxcontact}^\idxOne$ as Lagrange multiplier,
\ie,~$\lagvec=-\curtraction_{\idxcontact}^\idxOne$.
Using~$\curnormal$ to denote the outward unit normal vector,
the normal part of the contact stress can be denoted by~$\lagmult_\normalpart:=\transpose{\lagvec}\normalvec$
and the tangential part by~$\lagvec_\tangentialpart:=\lagvec-\lagmult_\normalpart\normalvec$.


We employ the usual function spaces~$\vecUspace^\idxSubdomain$ and~$\vecVspace^\idxSubdomain$
for the displacement field~$\disp$ of the the solid body and its weighting function~$\testvec$, respectively.
Furthermore, a suitable function space~$\vecMspace_{+}$ for the Lagrange multiplier field~$\lagvec$
and its weighting function~$\muvec$ is assumed.
The weak form of the governing equations then reads:
Find $\bigl(\disp,\lagvec\bigr)\in \vecUspace\times\vecMspace_{+}$ such that
\begin{subequations}
\label{eq:ch4_weakform}
\begin{align}
-\virtualwork_{\mathrm{int,ext}} + \int_{\curcontactbdry^\idxslave} \lagvec \left(\testvec^\idxOne - \testvec^\idxTwo\right) \, \mathrm{dA} &= 0,\qquad & \forall \, \testvec&\in\vecVspace,
\label{eq:ch4_weakform1}\\
\int_{\curcontactbdry^\idxslave} \left(\mu_\normalpart-\lagmult_\normalpart\right) \, \gapfunction \mathrm{dA} &\geq 0, \quad & \forall \, \muvec &\in \vecMspace_{+}.
\label{eq:ch4_weakform2}
\end{align}
\end{subequations}%
Herein, the internal and external virtual work contributions $\virtualwork_{\mathrm{int,ext}}$ are defined as usual in nonlinear solid mechanics ({\cf}~\cite{Zienkiewicz2014b} for example)
and, thus, further details are omitted.
The second term in~\eqref{eq:ch4_weakform1} can be identified as contact virtual work~$\virtualwork_{\idxcontact}$
and the expression in~\eqref{eq:ch4_weakform2} as variational inequality formulation of the contact constraints.
An extension to frictional contact based on Coulomb's law is straightforward and can be found in our previous work~\cite{ap_Gitterle2010} for example. 

\subsection{Finite element discretization}

For spatial discretization of the displacement field,
either isoparametric finite elements with first-order and second-order Lagrange interpolation 
or isogeometric analysis (IGA) with NURBS-based shape functions are employed.
After discretization, the discrete representation of displacement unkowns is given by the nodal degrees of freedom (DOFs)
\begin{equation*}
\disp=\transpose{\bigl[\disp_{\nione}, \disp_{\slave}, \disp_{\master}, \disp_{\nitwo}\bigr]}.
\end{equation*}
Therein, the $\disp_{\subdomain}$,~$\idxForSubdomain\in\{1,2\},$ contain all degrees of freedom
associated with the mesh nodes of the corresponding solid body without the nodes at the contact interface,
where we use the convention that indices~$1$ and~$2$ denote the slave and master ``body'', respectively.
The degrees of freedom associated with the contact interface on the slave and master side are represented by $\disp_{\slave}$ and $\disp_{\master}$, respectively. 

\begin{figure}
\begin{center}
\scalebox{0.8}{
\hspace{1.5cm} 
\tikzexternaldisable
\begin{tikzpicture}
\def\mycolorOne{gray!20!white}  
\def\mycolorTwo{gray!50!white}  
\def\mycolorThree{black}  
\def\mycolorFour{gray!60!white}  

\begin{scope}[rotate=90,yshift=-2cm] 

\begin{scope}[xshift=0.7cm,yshift=1.2cm]
\fill[color=\mycolorOne] (0,0) rectangle (4,4);
\draw[step=1cm, line width=0.1mm, black!90!white] (0,0) grid (4,4);
  \foreach \x in {0,...,4}{
  \foreach \y in {0,...,4}{
        \node[draw,circle,inner sep=1.5pt,fill] at (\x,\y) (LOW\x\y) {};
             Places a dot at those points
            }
      } 
      
   \foreach \x in {0,...,4}{
     \node[mark size=4pt,color=black] at (\x,4) {\pgfuseplotmark{diamond*}};
   }
\draw [rounded corners, dashed, very thick] (-0.2,-0.2) -- (4.2,-0.2) -- (4.2,3.2) -- (-0.2,3.2) -- cycle;
\draw [rounded corners, dashed, very thick] (-0.2,4.2) -- (4.2,4.2) -- (4.2,3.8) -- (-0.2,3.8) -- cycle;
\end{scope}

\begin{scope}[scale=1.25, yshift=5cm]
\fill[color=\mycolorTwo] (0,0) rectangle (4,4);
\draw[step=1cm, line width=0.1mm, black!90!white] (0,0) grid (4,4);
  \foreach \x in {0,...,4}{
  \foreach \y in {0,...,4}{
        \node[draw,circle,inner sep=1.5pt,fill] at (\x,\y) (UP\x\y) {};
             Places a dot at those points
            }
      }  
      
   \foreach \x in {0,...,4}{
     \node[mark size=3pt,color=black,xshift=0pt] at (\x,0) {\pgfuseplotmark{square*}};
   }

      \node at(0.5,-1.28) (LAGA) {};         
      \node at(3,-1.28) (LAGB) {};
\draw [rounded corners, dashed, very thick] (-0.2,0.8) -- (4.2,0.8) -- (4.2,4.2) -- (-0.2,4.2) -- cycle;
\draw [rounded corners, dashed, very thick] (-0.2,-0.2) -- (4.2,-0.2) -- (4.2,0.2) -- (-0.2,0.2) -- cycle;
\end{scope}  
      
\end{scope} 

\begin{scope}[scale=0.5,xshift=-1.4\textwidth,yshift=-3cm]  

\draw (-2.5,0.7) rectangle (25.3,-3.0);

\node[mark size=2pt,color=black] at (-2cm,0) (l1) {\pgfuseplotmark{*}};
\node[mark size=4pt,color=black,xshift=0pt] (l2) [below of = l1,node distance=0.5cm] {\pgfuseplotmark{diamond*}};
\node[mark size=3pt,color=black,xshift=0pt] (l3) [below of = l2,node distance=0.5cm] {\pgfuseplotmark{square*}};

\node[anchor=west,text width=15cm, align=left] (t1) [right of = l1,node distance=8cm] {\footnotesize{Inner nodes with displacement degrees of freedom  $\disp_{\nione}\in\refdomain^{\idxOne}\setminus\refcontactbdry^{\idxOne}$ and $\disp_{\nitwo}\in\refdomain^{\idxTwo}\setminus\refcontactbdry^{\idxTwo}$}};
\node[anchor=west,text width=15cm, align=left,xshift=0pt] (t2) [right of = l2,node distance=8cm] {\footnotesize{Nodes at slave contact interface $\refcontactbdry^{\idxslave}=\refcontactbdry^{\idxOne}$ with displacement degrees of freedom $\disp_{\act}$ and $\disp_{\inact}$}\\and Lagrange multipliers~$\lagvec$};
\node[anchor=west,text width=15cm, align=left] (t3) [right of = l3,node distance=8cm] {\footnotesize{Nodes at master contact interface $\refcontactbdry^{\idxmaster}=\refcontactbdry^{\idxTwo}$ with displacement degrees of freedom $\disp_{\master}$}};


\begin{scope}[xshift=4.0cm]
\node at(1.7,14.2) (labelA) {$\nitwo$};
\node at(7.7,14.2) (labelB) {$\master$};
\node at(9.7,13.4) (labelC) {$\slave$};
\node at(14.7,13.4) (labelA) {$\nione$};
\end{scope}

\end{scope}
\end{tikzpicture}
\tikzexternalenable
}
\caption{Schematic mesh illustrating interior, master and slave interface nodes.}
\label{fig:SchematicMesh}
\end{center}
\end{figure}
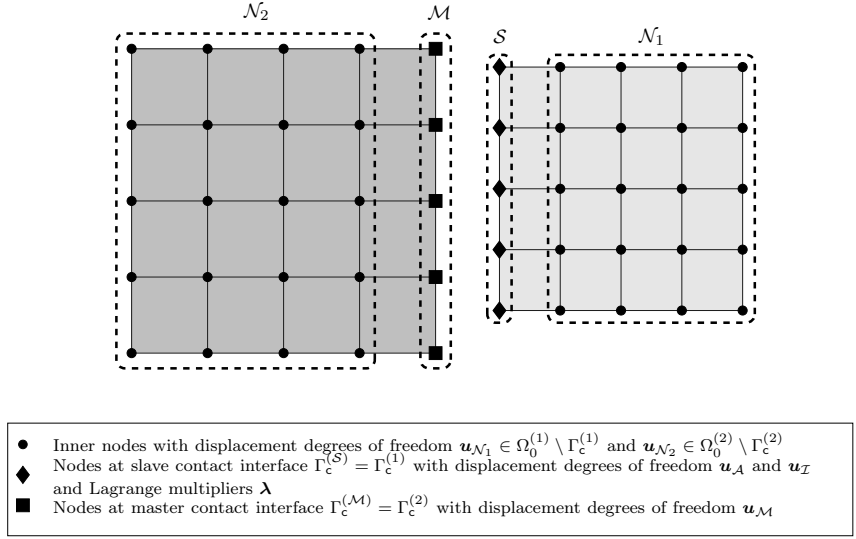

Since~\eqref{eq:ch4_weakform} represents a mixed variational form,
we also have to discretize the Lagrange multiplier field~$\lagvec$.
We choose to follow a mortar approach for its mathematical properties and its superiority
to other schemes~\cite{ap_Lorenzis2014,ap_Puso2004,ap_Puso2004a,ap_Wohlmuth2011,ap_Wriggers2006}.
As usual for mortar methods,
the Lagrange multiplier field is discretized on the slave side contact interface~$\curcontactbdry^\idxslave$ in the current configuration.
We either use standard ansatz functions, {\ie} Lagrange polynomials with a trace space relation with the underlying volume element,
or dual shape functions.
The latter satisfy a biorthogonality property and, thus,
allow for a computationally cheap condensation of the additional unknown Lagrange multipliers from the final system of equations.
For details on dual basis functions in the context of mortar-based contact discretizations,
we refer to~\cite{ap_Popp2012a,ap_Wohlmuth2000,ap_Lamichhane2005}.
Their interplay with preconditioners for iterative linear solvers has been discussed in our previous work~\cite{wiesner2017}.
The vector of discrete Lagrange multipliers is now referred to as~$\lagvec$.
A schematic mesh illustrating interior, slave interface, and master interface nodes is sketched in \figref{fig:SchematicMesh}.

The final spatially discretized formulation of the quasi-static frictionless problem \eqref{eq:ch4_weakform} using the nodal
vector representation now emerges as
\begin{subequations}
\label{eq:ch4_discreteform}
\begin{align}
\label{eq:ch4_discretealgebraicproblem}
\fint(\disp)+ \fco(\disp,\lagvec) & = \fext,\\
\label{eq:ch4_discretenormalcontact}
\bigl(\gapfunctiondiscretweighted\bigr)_j\geq 0, \quad 
\bigl(\lagmult_\normalpart\bigr)_j\geq 0, \quad 
\bigl(\gapfunctiondiscretweighted\bigr)_j\bigl(\lagmult_\normalpart\bigr)_j & = 0, \quad 
j=1,\ldots, n^\idxslave, \\
\label{eq:ch4_discretefrictionless}
\bigl(\lagvec_\tangentialpart\bigr)_j & = \boldsymbol{0}, \quad j=1,\ldots, n^\idxslave.
\end{align}
\end{subequations}
The internal forces~$\fint(\disp)$ and external forces~$\fext$ are common in nonlinear finite element methods and need no further explanation.
The discrete vector of contact forces~$\fco$ is computed based on two mortar matrices~$\mortarDmat$ and $\mortarMmat$,
arising from the integral over the slave interface~$\curcontactbdry^\idxslave$ in~\eqref{eq:ch4_weakform1},
and the discrete Lagrange multiplier vector~$\lagvec$.
For details regarding the computation of~$\mortarDmat$ and~$\mortarMmat$, see~\cite{ap_Popp2010,ap_Puso2004} for example.
Using~$\bigl(\gapfunctiondiscretweighted\bigr)_j$ to denote the discrete weighted gap function at node~$j$,
a closer look at the discrete contact constraints reveals
that~\eqref{eq:ch4_discretenormalcontact} basically represents a discrete version of the Karush--Kuhn--Tucker (KKT) type conditions in~\eqref{eq:HertzSignoriniMoreau}
with an additional weighting based on the Lagrange multiplier shape functions~$\dualbasis_{j}$,
while the nodal enforcement of frictionless sliding in~\eqref{eq:ch4_discretefrictionless} is straightforward anyway.

Since the discrete contact constraints summarized in~\eqref{eq:ch4_discretenormalcontact} are still formulated as inequalities,
an active set strategy usually referred to as primal-dual active set strategy (PDASS) is needed
in addition to the usual nonlinear solution procedure to identify the currently active and inactive contact regions~$\act$
and~$\inact=\slave \setminus \act$, respectively.
It has been demonstrated in~\cite{ap_Christensen1998,ap_Hintermueller2002,ap_Qi1993}
that the PDASS can equivalently be interpreted as a semi-smooth Newton method,
thus allowing for an integrated treatment of all nonlinearities (including the search for the active set)
within one single Newton--Raphson type iteration loop.
Meanwhile, many successful applications to small and large deformation contact problems
can be found in the literature~\cite{ap_Gitterle2010,ap_Hueber2008,ap_Hueber2005,ap_Popp2010}.

\subsection{Algebraic formulation of linear systems}
\label{sec:finalalgebraicsystem}

For efficient iterative solution strategies based on multigrid methods for nonlinear contact problems,
one is primarily interested in the structure of the linear systems arising in each nonlinear iteration step
of the underlying Newton--Raphson scheme.
For the sake of brevity, details on the linearization process 
and on the Newton--Raphson procedure are omitted here and the interested reader is instead referred to~\cite{popp2009,ap_Popp2010}. 

Consistent linearization of~\eqref{eq:ch4_discreteform}
and a subsequent update of the active set~$\act$ and inactive set~$\inact$ yields the system
\begin{equation}
\setlength{\dashlinegap}{2pt}
\begin{pmatrix}
\begin{array}{ccccc:cc}
\stiffmat_{\nione  \nione}    &   \stiffmat_{\nione \master}    & \zeromat                   & \zeromat                   &   \zeromat                  &  \zeromat   &  \zeromat   \\
\stiffmat_{\master \nione}    &   \stiffmat_{\master \master}   & \stiffmat_{\master \inact} & \stiffmat_{\master \act}   &   \zeromat                  & -\timintscalar\transpose{\mortarMmat_\inact} & -\timintscalar\transpose{\mortarMmat_\act} \\
\zeromat                      &   \stiffmat_{\inact \master}    &  \stiffmat_{\inact \inact} &  \stiffmat_{\inact \act} &   \stiffmat_{\inact \nitwo} &  \timintscalar\transpose{\mortarDmat_{\inact\inact}} &  \timintscalar\transpose{\mortarDmat_{\inact\act}} \\ 
\zeromat                      &   \stiffmat_{\act \master}      &  \stiffmat_{\act \inact} &  \stiffmat_{\act \act} &   \stiffmat_{\act \nitwo} &  \timintscalar\transpose{\mortarDmat_{\act\inact}} &  \timintscalar\transpose{\mortarDmat_{\act\act}} \\ 
\zeromat                      &   \zeromat                      &  \stiffmat_{\nitwo \inact} &  \stiffmat_{\nitwo \act} &   \stiffmat_{\nitwo \nitwo} &  \zeromat &  \zeromat  \\  \hdashline
\zeromat                      &   \zeromat                      &  \zeromat                  &  \zeromat                   &   \zeromat                  &  \mathsf{I} &  \zeromat  \\
\zeromat                      &   \mortarNmat_\master           &  \mortarNmat_\inact         &  \mortarNmat_\act         &   \zeromat                  &  \zeromat &  \zeromat  \\
\zeromat                      &   \zeromat                      &  \mortarFmat_\inact         &  \mortarFmat_\act         &   \zeromat                  &  \zeromat &  \mortarTmat_\act  \\
\end{array}
\end{pmatrix}
\begin{bmatrix}
\dispi_{\nione} \\
\dispi_{\master} \\
\dispi_{\inact} \\
\dispi_{\act} \\
\dispi_{\nitwo} \\\hdashline
\lagveci_{\inact} \\
\lagveci_{\act}
\end{bmatrix}
=
-
\begin{bmatrix}
\res^{\disp}_{\nione} \\
\res^{\disp}_\master \\
\res^{\disp}_\inact \\
\res^{\disp}_\act \\
\res^{\disp}_{\nitwo} \\\hdashline
\res^{\lagvec}_{\inact}\\
\res^{\lagvec,\normalpart}_{\act}\\
\res^{\lagvec,\tangentialpart}_{\act}
\end{bmatrix}
\label{eq:ch6_finalsaddlepointsystem}
\end{equation}
to be solved in every nonlinear iteration.
The~$2\times 2$ block matrix indicated by the dashed lines in~\eqref{eq:ch6_finalsaddlepointsystem}
describes a linear system with a typical generalized saddle point structure.
The upper left block consists of the entries of the tangential stiffness matrix (i.e. linearized internal forces)
as well as linearizations of contact forces {\wrt} displacement degrees of freedom~$\disp$.
The upper right block mirrors the discrete contact operator~$\couplmat(\disp)$,
{\ie} basically the two mortar matrices~$\mortarDmat$ and ~$\mortarMmat$,
representing the linearizations of the contact forces {\wrt} the Lagrange multiplier unknowns~$\lagmult$.
The kinematic constraints are incorporated in the bottom left block.
The very simple sixth block row emerges from~\eqref{eq:ch4_discretenormalcontact}
and~\eqref{eq:ch4_discretefrictionless} for inactive nodes,
while the last block row imposes frictionless sliding in the directions tangential to the contact interface.

The distinct pattern of zero entries in the upper left block reveals that the two solid bodies (indices $\nione$ and $\nitwo$)
are indeed only coupled through the slave and master sides of the contact interface (indices $\slave$ and $\master$).
Even though formulated for two solid bodies, the generalization to~$n$ solid bodies is straightforward and only a matter of notation.

In case of dual shape functions, the matrix~$\mortarDmat$ reduces to a diagonal matrix and, thus,
allows for a cheap condensation of the Lagrange multiplier unknowns.
Algebraic multigrid preconditioners for this type of condensed system have been proposed in our earlier paper~\cite{wiesner2017}.
Furthermore, matrices~$\mortarNmat_\master$, $\mortarNmat_\inact$, and~$\mortarNmat_\act$ denote the linearizations
of the weighted gap function of~\eqref{eq:ch4_discretenormalcontact} at all active contact nodes.
Finally, linearizations of the frictionless sliding condition~\eqref{eq:ch4_discretefrictionless} are referred to
by matrices~$\mortarFmat_\inact$, $\mortarFmat_\act$, and~$\mortarTmat_\act$, respectively.

Note that the given matrix has $8$ block rows but only $7$ block columns in our notation in order to emphasize
that the normal and tangential parts of the contact constraints for active nodes are considered separately,
{\ie} these two rows contain consistent linearizations of the active branch
of~\eqref{eq:ch4_discretenormalcontact} and of~\eqref{eq:ch4_discretefrictionless}.
Again, we point out that this separate notation is possible due to the fact that a local convective coordinate system is employed
for evaluating the contact constraints / Lagrange multiplier weights~$\muvec$,
while the standard Cartesian frame is still applied for the discrete Lagrange multiplier values~$\lagvec$
as well as the displacement unknowns.
Yet, of course, the system matrix remains a square matrix with the total numbers of rows and columns being identical.
The discrete vector~$\gapvec_\act$ contains all weighted gap values~$\bigl(\gapfunctiondiscretweighted\bigr)_j$ associated with the active nodes at the contact interface.

For ease of notation, the following short block notation is used in the remainder of the manuscript:
\begin{equation}
\begin{pmatrix}
\begin{array}{cc}
\stiffmat & \mortarCmat \\ 
\mortarConstrmat & -\mortarZonstrmat
\end{array}
\end{pmatrix}
\begin{bmatrix}
\dispi\\ \lagveci
\end{bmatrix}
=
-
\begin{bmatrix}
\res^{\disp} \\
\res^{\lagvec}
\end{bmatrix}.
\label{eq:ch6_finalsaddlepointsystem_comprehensive}
\end{equation}

\section{Multigrid scheme for contact problems in saddle point formulation}
\label{sec:AMGForContact}

Although multigrid methods can be used as standalone solvers for linear systems,
they are usually incorporated into an iterative linear solver as a preconditioning method.
Throughout this paper, we use a preconditioned generalized minimal residual (GMRES) solver~\cite{saad1986}
with one multigrid V-cycle sweep for preconditioning.
A general introduction into the idea of preconditioning is beyond the scope of this paper. The reader is referred to the literature, {\eg}~\cite{benzi2002}.


\subsection{Algebraic multigrid methods in a nutshell}
\label{sec:multigridbasics}


Multigrid methods are based on the finding that many well-known and computationally cheap iterative methods
({\eg} relaxation based iterative methods such as Jacobi or Gauss-Seidel methods) to solve linear systems~$\Aop\sol=\rhs$
effectively damp the high frequency part of an error vector but are less effective in damping out the low frequency error modes.
Multigrid methods heavily make use of this smoothing property by applying such cheap smoothing methods
on different coarsened representations of the original fine level problem.

\subsubsection{Basic multigrid cycle and algorithm}

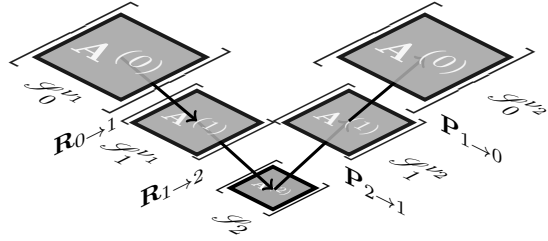
\begin{figure}
\begin{subfigure}[b]{0.48\textwidth}
\begin{minipage}{0.5\textwidth}
  \begin{tabbing}
    Links \= Mitte \= Rechts \kill
    $\textnormal{MGV}(\Afine,\sol,\rhs,\level)$:\\
    \> if $\level \neq \level_{max}$\\
    \> \> $\sol \leftarrow {\smoother^{\presmoothingsweeps}_\level} (\Afine,\sol,\rhs)$ \\
    \> \> $\residual\leftarrow \rhs-\Afine\sol$ \\
    \> \> $\correction\leftarrow 0$ \\
    \> \> $\correction\leftarrow \textnormal{MGV}(\Acoarse,\sol_{\level+1},\restrictor_{\level+1}\residual,\level+1)$ \\
    \> \> $\sol \leftarrow \sol + \prolongator_{\level+1}\correction$ \\
    \> \> $\sol \leftarrow \smoother^{\postsmoothingsweeps}_\level (\Afine, \sol, \rhs)$ \\
    \> else $\sol \leftarrow (\Afine)^{-1}\rhs$
  \end{tabbing}
\end{minipage}
\caption{Recursive multigrid algorithm.}
\label{fig:mgalgorithm}
\end{subfigure}
\begin{subfigure}[b]{0.48\textwidth}
\tikzexternaldisable
\begin{tikzpicture}
    %
	%

	\begin{scope}[shift={(8,1.3)},scale=0.2] 
     	\tikzstyle{restrict}=[line width=0.4mm,->]	
     	\tikzstyle{prolongate}=[line width=0.4mm,->]	     	
      \begin{scope}[shift={(0,0)},scale=0.4]
        \begin{scope}[every node/.append style={yslant=-0.5,xslant=1},yslant=-0.5,xslant=1]
        	\draw[line width=0.5mm,color=black,fill=\MphysicsA] (0,0) rectangle (7,-7);
        	\node[color=white] at (3.5,-3.5) {\scalebox{0.6}{$\A^{(2)}$}};        	        	
		\node (left-paren) at (-0.6,-3.5) {$\left[\vphantom{\rule{0.6cm}{0.46cm}}\right.$};
		\node (right-paren) at (7.6,-3.5) {$\left.\vphantom{\rule{0.6cm}{0.46cm}}\right]$};
		\end{scope}
      \end{scope}
      
      \draw[restrict,->] (-5,3) -- (0,-1.5); 
      \draw[prolongate,->] (0.1,-1.5) -- (5,3); 

      \begin{scope}[shift={(-5,5)},scale=0.6,opacity=0.9]
        \begin{scope}[every node/.append style={yslant=-0.5,xslant=1},yslant=-0.5,xslant=1]        
        	\draw[line width=0.5mm,color=black,fill=\MphysicsA] (0,0) rectangle (7,-7);
        	\node[color=white] at (3.5,-3.5) {$\A^{(1)}$};        	
		\node (left-paren) at (-0.6,-3.3) {$\left[\vphantom{\rule{0.7cm}{0.6cm}}\right.$};
		\node (right-paren) at (7.6,-3.3) {$\left.\vphantom{\rule{0.7cm}{0.6cm}}\right]$};
		\end{scope}
      \end{scope}
      \begin{scope}[shift={(5,5)},scale=0.6,opacity=0.9]
        \begin{scope}[every node/.append style={yslant=-0.5,xslant=1},yslant=-0.5,xslant=1]
        	\draw[line width=0.5mm,color=black,fill=\MphysicsA] (0,0) rectangle (7,-7);
        	\node[color=white] at (3.5,-3.5) {$\A^{(1)}$};        	
		\node (left-paren) at (-0.6,-3.3) {$\left[\vphantom{\rule{1cm}{0.6cm}}\right.$};
		\node (right-paren) at (7.6,-3.3) {$\left.\vphantom{\rule{1cm}{0.6cm}}\right]$};		
		\end{scope}
      \end{scope}

      \draw[restrict,->] (-10,7.2) -- (-5,3); 
      \draw[prolongate,->] (5,3) -- (10,7.2); 
                  
     \begin{scope}[shift={(-10,10)},scale=0.8,opacity=0.9]  
        \begin{scope}[every node/.append style={yslant=-0.5,xslant=1},yslant=-0.5,xslant=1]
        	\draw[line width=0.5mm,color=black,fill=\MphysicsA] (0,0) rectangle (7,-7);
        	\node[color=white] at (3.5,-3.5) {\scalebox{1.4}{$\A^{(0)}$}};
		\node (left-paren) at (-0.5,-3.5) {$\left[\vphantom{\rule{1cm}{0.9cm}}\right.$};
		\node (right-paren) at (7.5,-3.5) {$\left.\vphantom{\rule{1cm}{0.9cm}}\right]$};
		\end{scope}
      \end{scope}
      \begin{scope}[shift={(10,10)},scale=0.8,opacity=0.9]
        \begin{scope}[every node/.append style={yslant=-0.5,xslant=1},yslant=-0.5,xslant=1]
        	\draw[line width=0.5mm,color=black,fill=\MphysicsA] (0,0) rectangle (7,-7);
        	\node[color=white] at (3.5,-3.5) {\scalebox{1.4}{$\A^{(0)}$}};
		\node (left-paren) at (-0.5,-3.5) {$\left[\vphantom{\rule{1cm}{0.9cm}}\right.$};
		\node (right-paren) at (7.5,-3.5) {$\left.\vphantom{\rule{1cm}{0.9cm}}\right]$};
		\end{scope}
	  \end{scope}
	  
      \begin{scope}[every node/.append style={yslant=-0.5,xslant=1},yslant=-0.5,xslant=1]
      \node at (-12.0,-2.5) {$\smoother^{\presmoothingsweeps}_0$};
      \node at (-5.5,-4.0) {$\smoother^{\presmoothingsweeps}_1$};
      \node at (2.4,-5) {$\smoother_{2}$};
      \node at (+5.0,4.5) {$\smoother^{\postsmoothingsweeps}_1$};      
      \node at (+4.0,12) {$\smoother^{\postsmoothingsweeps}_0$};      
      \end{scope}  
      
      \begin{scope}[every node/.append style={yslant=0.5,xslant=0},yslant=0,xslant=0]
            \node at (-12.3,2.2) {$\Restr_{0\rightarrow 1}$};
            \node at (-6.6,-1.2) {$\Restr_{1\rightarrow 2}$};
	  \end{scope}  
	  
	  \begin{scope}[every node/.append style={yslant=-0.5,xslant=0.0},yslant=0,xslant=0]
            \node at (6.8,-1.6) {$\Prol_{2\rightarrow 1}$};
            \node at (13.1,1.8) {$\Prol_{1\rightarrow 0}$};
	  \end{scope}  
      
	\end{scope}
	\end{tikzpicture}%

\caption{Multigrid V-cycle in a $3$ level setting.}
\label{fig:vcycle}
\end{subfigure}
\caption{Multigrid algorithm and V-cycle.}
\end{figure}

The multigrid algorithm given in Figure~\ref{fig:mgalgorithm} is briefly described as follows: 
on each multigrid level~$\level$,
a level smoothing algorithm~$\smoother_\level$ performs~$\presmoothingsweeps$ pre-smoothing sweeps
before the residual vector~$\residual$ is transferred to the next coarser level~$\level+1$ using the restriction operator~$\restrictor$. 
After the coarse level problem has been solved on the coarsest level,
the correction~$\correction$ is then prolongated using the prolongation operator~$\prolongator$
and the solution vector is smoothed using~$\postsmoothingsweeps$ post-smoothing sweeps.
Figure~\ref{fig:vcycle} illustrates this basic multigrid V-cycle, exemplifying a three-level setting.
As one can see from Figure~\ref{fig:vcycle}, applying a multigrid method basically
means applying level smoothers on coarse representations~$\Aop_\level$ of the fine level problem~$\Aop_0$.

\subsubsection{Algebraic multigrid methods}
\label{sec:amg}
There are different strategies for defining the transfer operators~$\prolongator$ and~$\restrictor$
which are necessary to generate coarse level matrices~$\Aop_\level$~($\level>0$). 
For algebraic multigrid (AMG), the fine level operator~$\Aop_0$ is sufficient to generate coarse level matrices~$\Aop_\level$.
An important class of AMG is given with the \textit{smoothed aggregation AMG}
which is based on so-called aggregates (see {\eg}~\cite{vanek1994,vanek1996}).
The fine-level nodes are agglomerated and put into aggregates, which then represent ``supernodes" on the next coarser level.
The aggregation information together with the null space information of~$\Aop_0$
is used to construct the corresponding \textit{tentative transfer operators}~$\rtent$ (for restriction) and~$\ptent$ (for prolongation).
Transfer operators are used to restrict the fine level residual to the next coarser level
and to interpolate the coarse level correction to the next finer level via prolongation.
For an efficient multigrid method, the interaction of fine and coarse levels tackles those error modes,
which appear as low-frequency modes on the fine level and cannot effectively be reduced by iterative smoothing methods on the fine level,
but resemble high-frequency modes on a coarser level, such that iterative smoothing methods are effective again.
On the coarsest level, a direct solver can take care of the remaining error modes.
For smoothed aggregation multigrid ({\cf}~\cite{gee2008,vanek1996}),
the prolongation operator is found by applying one smoothing sweep with a damped Jacobi iteration using
\begin{equation}
\label{eq:ProlongatorSmoothing}
\prolongator = \ptent - \omega \Dinv\Aop \ptent
\end{equation}
with~$\diagonalmat$ being the diagonal part of~$\Aop$ and a damping parameter~$\omega>0$.
Depending on the symmetry of the system matrix $\Aop$,
the restriction operator is either chosen as~$\restrictor=\prolongator^T$ or smoothed independently using,
{\eg} a Petrov-Galerkin approach ({\cf}~\cite{sala2008}).

\subsection{Algebraic multigrid methods for block matrices}
\label{sec:MultigridForMultiphysicsBlock}

Block matrices usually arise if multiple types of equations are coupled together. 
In the present context of contact problems in saddle point formulation, 
two types of equations, namely the balances of linear momentum of the solid bodies and the contact constraints,
are connected via the off-diagonal blocks in~\eqref{eq:ch6_finalsaddlepointsystem_comprehensive}. 
Similarly, multiphysics problems also yield block matrices where the coupling between different physical fields
manifests itself in the off-diagonal blocks of the monolithic system matrix.

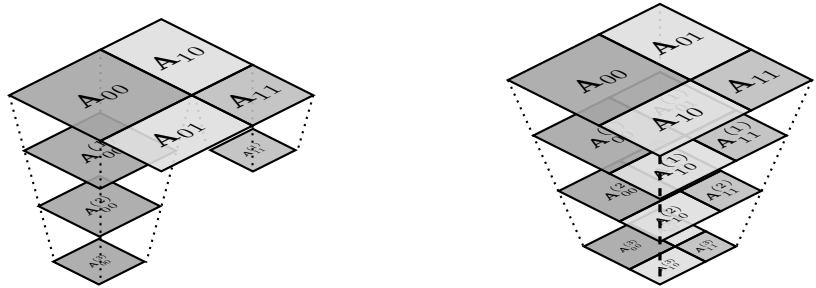
\begin{figure}[htbp]
\centering
\begin{subfigure}[t]{0.45\textwidth}\centering
\tikzexternaldisable
\begin{tikzpicture}[scale=0.4,every node/.style={minimum size=1cm}] 
\begin{scope}[xshift=0]

\draw[black,thick,dotted] (1,2.5) -- (-0.5,-3);    
\draw[black,thick,dotted] (-5,2.5) -- (-3.5,-3);   
\draw[black,thick,dotted] (-2,4) -- (-2,-2.2);     

  \draw[black,thick,dotted] (3,3.5) -- (3,1.5);    
  \draw[black,thick,dotted] (1.1,2.5) -- (1.5,0.75);    
  \draw[black,thick,dotted] (5,2.5) -- (4.5,0.75);    

    \begin{scope}[xshift=-30,yshift=-120,scale=0.5,
    	    every node/.append style={
    	    yslant=0.5,xslant=-1},yslant=0.5,xslant=-1
    	             ]
		\fill[\MphysicsA,fill opacity=.9] (0,2) rectangle (3,5);
        \draw[black,thick] (0,2) rectangle (3,5);
        \draw (1.5,3.5) node[scale=0.5]{\tiny$\AOO^{(3)}$};
    \end{scope}  
    
    \begin{scope}[xshift=-20,yshift=-80,scale=0.666,
    	    every node/.append style={
    	    yslant=0.5,xslant=-1},yslant=0.5,xslant=-1]
		\fill[\MphysicsA,fill opacity=.9] (0,2) rectangle (3,5);
        \draw[black,thick] (0,2) rectangle (3,5);
        \draw (1.5,3.5) node[scale=0.75]{\tiny$\AOO^{(2)}$};
    \end{scope}  
    
    \begin{scope}[xshift=-10,yshift=-40,scale=0.833,
    	    every node/.append style={
    	    yslant=0.5,xslant=-1},yslant=0.5,xslant=-1]
		\fill[\MphysicsA,fill opacity=.9] (0,2) rectangle (3,5);
        \draw[black,thick] (0,2) rectangle (3,5);
        \draw (1.5,3.5) node[scale=1.0]{\tiny$\AOO^{(1)}$};
    \end{scope}

    \begin{scope}[xshift=26,yshift=-30,scale=0.7,
    	    every node/.append style={
    	    yslant=0.5,xslant=-1},yslant=0.5,xslant=-1]
		\fill[\MphysicsB,fill opacity=.9] (3,0) rectangle (5,2);
        \draw[black,thick] (3,0) rectangle (5,2);
        \draw (4,1) node[scale=0.5]{\tiny$\All^{(1)}$};
    \end{scope}

    \begin{scope}[xshift=0,yshift=0,scale=1,
    	    every node/.append style={
    	    yslant=0.5,xslant=-1},yslant=0.5,xslant=-1]
		\fill[\MphysicsA,fill opacity=.9] (0,2) rectangle (3,5);
		\fill[\MphysicsB,fill opacity=.9] (3,0) rectangle (5,2);	
		\fill[\MphysicsCouplingA,fill opacity=.9] (0,0) rectangle (3,2);
		\fill[\MphysicsCouplingA,fill opacity=.9] (3,2) rectangle (5,5);
        \draw[black,thick] (0,2) rectangle (3,5);
        \draw[black,thick] (3,0) rectangle (5,2);	
        \draw[black,thick] (0,0) rectangle (3,2);
        \draw[black,thick] (3,2) rectangle (5,5);
        \draw (1.5,3.5) node{$\AOO$};
        \draw (1.5,1) node{$\AOl$};
        \draw (4,3.5) node{$\AlO$};
        \draw (4,1) node{$\All$}; 
    \end{scope}     
  \draw[black,thick,dotted] (-2,1) -- (-2,-3.7);   
  \draw[black,thick,dotted] (3,1.5) -- (3,0);   
  \end{scope}    
\end{tikzpicture}
\caption{Outer coupling iteration with nested multigrid methods. Example with four and two multigrid levels for $\AOO$ and $\All$.}
\label{fig:mgmphmodelsA}
\end{subfigure}
\begin{subfigure}[t]{0.45\textwidth}\centering
\tikzexternaldisable
\begin{tikzpicture}[scale=0.4,every node/.style={minimum size=1cm},on grid]
\begin{scope}[]

\draw[black,thick,dotted] (5,2.5) -- (2.5,-3);
\draw[black,thick,dotted] (-5,2.5) -- (-2.5,-3);
\draw[black,thick,dotted] (0,5) -- (0,-1.6);

    \begin{scope}[xshift=0,yshift=-120,scale=0.5,
    	    every node/.append style={
    	    yslant=0.5,xslant=-1},yslant=0.5,xslant=-1
    	             ]
		\fill[\MphysicsA,fill opacity=.9] (0,2) rectangle (3,5);
		\fill[\MphysicsB,fill opacity=.9] (3,0) rectangle (5,2);	
		\fill[\MphysicsCouplingA,fill opacity=.9] (0,0) rectangle (3,2);
		\fill[\MphysicsCouplingA,fill opacity=.9] (3,2) rectangle (5,5);
        \draw[black,thick] (0,2) rectangle (3,5);
        \draw[black,thick] (3,0) rectangle (5,2);	
        \draw[black,thick] (0,0) rectangle (3,2);
        \draw[black,thick] (3,2) rectangle (5,5);
        \draw (1.5,3.5) node[scale=0.5]{\tiny$\AOO^{(3)}$};
        \draw (1.5,1) node[scale=0.5]{\tiny$\AlO^{(3)}$};
        \draw (4,3.5) node[scale=0.5]{\tiny$\AOl^{(3)}$};
        \draw (4,1) node[scale=0.5]{\tiny$\All^{(3)}$}; 
    \end{scope}  
    
    \begin{scope}[xshift=0,yshift=-80,scale=0.666,
    	    every node/.append style={
    	    yslant=0.5,xslant=-1},yslant=0.5,xslant=-1
    	             ]
		\fill[\MphysicsA,fill opacity=.9] (0,2) rectangle (3,5);
		\fill[\MphysicsB,fill opacity=.9] (3,0) rectangle (5,2);	
		\fill[\MphysicsCouplingA,fill opacity=.9] (0,0) rectangle (3,2);
		\fill[\MphysicsCouplingA,fill opacity=.9] (3,2) rectangle (5,5);
        \draw[black,thick] (0,2) rectangle (3,5);
        \draw[black,thick] (3,0) rectangle (5,2);	
        \draw[black,thick] (0,0) rectangle (3,2);
        \draw[black,thick] (3,2) rectangle (5,5);
        \draw (1.5,3.5) node[scale=0.75]{\tiny$\AOO^{(2)}$};
        \draw (1.5,1) node[scale=0.75]{\tiny$\AlO^{(2)}$};
        \draw (4,3.5) node[scale=0.75]{\tiny$\AOl^{(2)}$};
        \draw (4,1) node[scale=0.75]{\tiny$\All^{(2)}$}; 
    \end{scope}  
    
    \begin{scope}[xshift=0,yshift=-40,scale=0.833,
    	    every node/.append style={
    	    yslant=0.5,xslant=-1},yslant=0.5,xslant=-1
    	             ]
		\fill[\MphysicsA,fill opacity=.9] (0,2) rectangle (3,5);
		\fill[\MphysicsB,fill opacity=.9] (3,0) rectangle (5,2);	
		\fill[\MphysicsCouplingA,fill opacity=.9] (0,0) rectangle (3,2);
		\fill[\MphysicsCouplingA,fill opacity=.9] (3,2) rectangle (5,5);
        \draw[black,thick] (0,2) rectangle (3,5);
        \draw[black,thick] (3,0) rectangle (5,2);	
        \draw[black,thick] (0,0) rectangle (3,2);
        \draw[black,thick] (3,2) rectangle (5,5);
        \draw (1.5,3.5) node{\tiny$\AOO^{(1)}$};
        \draw (1.5,1) node{\tiny$\AlO^{(1)}$};
        \draw (4,3.5) node{\tiny$\AOl^{(1)}$};
        \draw (4,1) node{\tiny$\All^{(1)}$}; 
    \end{scope}

    \begin{scope}[xshift=0,yshift=0,scale=1,
    	    every node/.append style={
    	    yslant=0.5,xslant=-1},yslant=0.5,xslant=-1
    	             ]
		\fill[\MphysicsA,fill opacity=.9] (0,2) rectangle (3,5);
		\fill[\MphysicsB,fill opacity=.9] (3,0) rectangle (5,2);	
		\fill[\MphysicsCouplingA,fill opacity=.9] (0,0) rectangle (3,2);
		\fill[\MphysicsCouplingA,fill opacity=.9] (3,2) rectangle (5,5);
        \draw[black,thick] (0,2) rectangle (3,5);
        \draw[black,thick] (3,0) rectangle (5,2);	
        \draw[black,thick] (0,0) rectangle (3,2);
        \draw[black,thick] (3,2) rectangle (5,5);
        \draw (1.5,3.5) node{$\AOO$};
        \draw (1.5,1) node{$\AlO$};
        \draw (4,3.5) node{$\AOl$};
        \draw (4,1) node{$\All$}; 
    \end{scope}     
  \draw[black,very thick,dashed] (0,0) -- (0,-4.2);    
  \end{scope}    
\end{tikzpicture}
\caption{Multiphysics multigrid approach with nested coupling iteration on all multigrid levels. }
\label{fig:mgmphmodelsB}
\end{subfigure}
\caption{Multigrid for block matrices.}
\label{fig:mgmphmodels}
\end{figure}

From a multigrid perspective, the most important question is
where to consider the coupling between the different equations within the overall solver layout.
In general, there are only two possible strategies to apply multigrid ideas to coupled block systems:
\begin{description}
\item[\textnormal{\textbf{Nested multigrid approach:}}] 
Multigrid methods can serve as local single field smoothers or solvers within well-known block preconditioners
such as the SIMPLE method ({\cf}~\cite{patankar1972}) and variants for Schur complement based preconditioners
or the block Gauss--Seidel (BGS) method. 
The coupling of the different fields or variables is only considered on the finest level in the outer (SIMPLE or BGS) iteration.
This approach is well known in literature, {\eg} for the Navier--Stokes equations~\cite{griebel1998,trottenberg2001},
fluid-structure interaction~\cite{gee2011,Tezduyar2006a} or general $n$-field problems~\cite{Verdugo2016a}.
The implementation is very easy and allows to use existing multigrid components in a standalone fashion within the solver.
A graphic representation of this approach is shown in Figure~\ref{fig:mgmphmodelsA}.
\item[\textnormal{\textbf{Fully coupled multigrid approach:}}]
Truly monolithic algebraic multigrid methods aim at coarsening the fully coupled fine level problem
such that the block structure of the fine level matrix is preserved and the coupling information is present on all coarser levels, {\cf}~\figref{fig:mgmphmodelsB}.
This is often achieved by using segregated transfer operators to preserve the characteristics of the sparsity pattern across all levels.
Then, each level utilizes block smoothers to address the coupling.
In~\cite{wabro2004}, a coupled AMG method is developed and analyzed
for a stabilized mixed finite element discretization of the Oseen equations.
Fully coupled multigrid methods for multiphysics problems have been described in~\cite{gee2011,langer2014numerical,Verdugo2016a}.
\end{description}

While the nested multigrid approach is easier to implement, it also allows for a high degree of modularity,
since the multigrid hierarchies used to approximate the block inverses of the block smoother on the fine level
can easily by swapped by any other method, either another type of multigrid algorithm,
or another type of multi-level scheme (e.g. based on domain decomposition),
or even any single-level approach if the block is of moderate size and scalability is not deteriorated.
This flexibility is particularly useful if the coupled blocks differ a lot in size
or if the user wants to apply a highly optimized solver for an individual block.
The fully coupled multigrid method does not offer such a  degree of flexibility,
yet it propagates the coupling conditions throughout the entire preconditioner.
Thus, one expects a stronger and more robust preconditioning effect,
since the coarse level corrections are aware of the coupling conditions.
This expectation will later be confirmed in the numerical experiments,
where the number of iterations for the fully coupled scheme is lower and more independent of the active contact nodes
than for the nested multigrid approach.

\subsection{Designing algebraic multigrid methods for contact problems}
\label{sec:MultigridForMultiphysics}

In the present context, we can interpret the mortar contact problem in saddle point formulation
as the coupling of two types of equations:
the structural equations and the contact equations which serve as constraint equations.
Since the contact constraint equations are only defined along the contact interface,
we can further classify the mortar contact problem as an interface-coupled problem (in contrast to volume-coupled problems).
This information is important for the choice of coarsening strategy.
The contact constraint equations are also responsible for the characteristic saddle point structure,
which needs special attention when choosing an appropriate coupling algorithm
between the structural equations and the contact constraints. 
Considering the class of fully coupled AMG schemes, 
the generalized saddle point problem~\eqref{eq:ch6_finalsaddlepointsystem_comprehensive}
has to be preserved on all multigrid levels
such that the contact constraints are considered on all levels.
Due to the constraints, this will require Schur complement based level smoothers on all levels.

Alltogether, the key ingredients for designing an algebraic multigrid method for contact problems in saddle point formulation are 
the coarsening strategy as proposed in \secref{sec:CoarseningOfContactConstraints}
and the level smoother and the coupling iteration as detailed in \secref{sec:BlockSmoothers}.


\providecommand{\precit}{\relaxationiterationidx} 
\providecommand{\primvar}[1][]{\dispi^{{#1}}}
\providecommand{\secondvar}[1][]{\lagveci^{{#1}}}
\providecommand{\primvarincinc}[1][]{\disp^{{#1}}}
\providecommand{\secondvarincinc}[1][]{\lagvec^{{#1}}}

\providecommand{\primrhs}[1][]{\res_{\disp}^{{#1}}}
\providecommand{\secondrhs}[1][]{\res_{\lagvec}^{{#1}}}

\providecommand{\Ablock}{{\stiffmat}} 
\providecommand{\Boneblock}{{\mortarCmat}} 
\providecommand{\Btwoblock}{{\mortarConstrmat}} 
\providecommand{\Cblock}{\mortarZonstrmat} %

\providecommand{\SchurBlock}{\schur}
\providecommand{\SchurApprox}{\apprx{\SchurBlock}}

\providecommand{\blockdampingparameter}{\alpha}

\providecommand{\deltaincr}[1][]{\delta{#1}}

\providecommand{\errormat}{\mathsf{E}}

\section{A coarsening strategy for mortar contact constraints}
\label{sec:CoarseningOfContactConstraints}

\subsection{Segregated transfer operators}
To keep the characteristic saddle point block structure \eqref{eq:ch6_finalsaddlepointsystem_comprehensive} on all multigrid levels, the common approach is to use \emph{segregated} transfer operators
\begin{equation}
\prolongator_{\level+1}=
\begin{pmatrix}
\prolongatordisp & \zeromat \\
\zeromat & \ptentlagv
\end{pmatrix}_{\level+1}
\quad
\textnormal{ and }
\quad
\restrictor_{\level+1}=
\begin{pmatrix}
\restrictordisp & \zeromat \\
\zeromat & \rtentlagv
\end{pmatrix}_{\level+1},
\label{eq:ch6_segregatedtransfers}
\end{equation}
as, {\eg} introduced in~\cite{adams2004,braess1997}.
The segregated block transfer operators \eqref{eq:ch6_segregatedtransfers} are put together from the transfer operator blocks for the different physical and mathematical fields. 
Here, $\prolongatordisp$ and~$\restrictordisp$ describe the transfer operator blocks
corresponding to the stiffness matrix block~$\stiffmat$ in~\eqref{eq:ch6_finalsaddlepointsystem_comprehensive}.
The transfer operators~$\ptentlagv$ and~$\rtentlagv$ define the level transfer for the Lagrange multipliers. 

The block diagonal structure in~\eqref{eq:ch6_segregatedtransfers} guarantees
that the primary displacement variables and the secondary Lagrange multipliers are not ``mixed up" on the coarser levels.
That is, the coarse level matrix still has the same block structure
with a clear distinction of momentum and constraint equations as for the fine level problem
since the columns and rows of the transfer operators~$\prolongator_{\level+1}$ and~$\restrictor_{\level+1}$
can be interpreted as some kind of basis functions.

For many volume-coupled problems, for example in thermo-structure-interaction problems~\cite{Danowski2013a}, it is straightforward to generate~$\ptentlagv$ and~$\rtentlagv$ to be consistent with~$\prolongatordisp$ and~$\restrictordisp$. That is, in context of smoothed aggregation algebraic multigrid we just use the same aggregates for building~$\prolongatordisp$ and~$\ptentlagv$ (and the same for the restrictors, respectively).

For interface-coupled problems with interface constraints it is more difficult.
Due to the saddle point structure of~\eqref{eq:ch6_finalsaddlepointsystem_comprehensive}, 
the nonzero pattern of the~$\mortarZonstrmat$ block is insufficient to generate valid aggregates for the Lagrange multipliers.
Consequently, we need a special routine for finding aggregates for the Lagrange multipliers~$\lagvec$ to be able to build the (non-smoothed) transfer operators~$\ptentlagv$ and~$\rtentlagv$. 
Nevertheless it seems natural to reflect the aggregation information of the structural equations along the contact interface in the choice of the aggregates for the corresponding Lagrange multipliers~$\lagvec$.


\subsection{Aggregation strategy for displacement variables}

In order to preserve the physics of the fine level contact problem, 
it is important to keep the two solid bodies separated in the matrix representation on all coarse levels. 
Therefore, we apply the standard aggregation strategy to a modified~$\Ablock$ block from~\eqref{eq:ch6_finalsaddlepointsystem_comprehensive}, 
where all off-diagonal entries representing connections between the two solid bodies are dropped 
--- in paricular the matrix blocks~$\stiffmat_{\inact \master}, \stiffmat_{\act \master}, \stiffmat_{\master \inact}$ and~$\stiffmat_{\master \act}$ --- 
in order to make sure that the resulting displacement aggregates $\aggs[\level]^{\disp}$ do not cross the contact interface (see Figure \ref{fig:ch6_aggregationinsaddlepointformulation}). 
Neglecting these blocks during aggregation guarantees that the two solid bodies are not melted together in the coarse matrix representation.
We stress that the modified $\Ablock$ is never formed explicitly, but rather the off-diagonal entries are dropped on the fly during the aggregation process.

\subsection{Aggregation strategy for Lagrange multipliers}
\label{sec:ch6_aggregationforlagrangemult}

In contrast to geometric multigrid methods, there is not so much literature on \aggbased AMG methods for contact problems in saddle point formulation.
The only publication, the authors are aware of covering all aspects of smoothed aggregation methods for structural contact problems in saddle point formulation, is~\cite{adams2004}, which also discusses a special aggregation strategy for the Lagrange multipliers.
To find aggregates~$\aggs[\level]^{\lagvec}$ for the Lagrange multipliers,
Adams~\cite{adams2004} proposes to apply the standard aggregation algorithm to the graph of a suitable matrix representing the Lagrange multipliers. 
However, this approach has some drawbacks:
First, the graph used for the aggregation of the Lagrange multipliers~$\lagvec$ has to be built explicitly to serve as input for the standard aggregation algorithm.
Secondly, one has to run the aggregation algorithm sequentially both for the displacement degrees of freedom and for the Lagrange multipliers.
For the second run of the aggregation method, one might have to use a different set of aggregation parameters to obtain optimal results,
which further increases the complexity for the user.
Algorithmically, the resulting aggregates~$\aggs[\level]^{\lagvec}$ for the Lagrange multipliers are built independently from the displacement aggregates~$\aggs[\level]^{\disp}$.

\begin{figure}
\begin{center}
\includegraphics[width=0.8\textwidth]{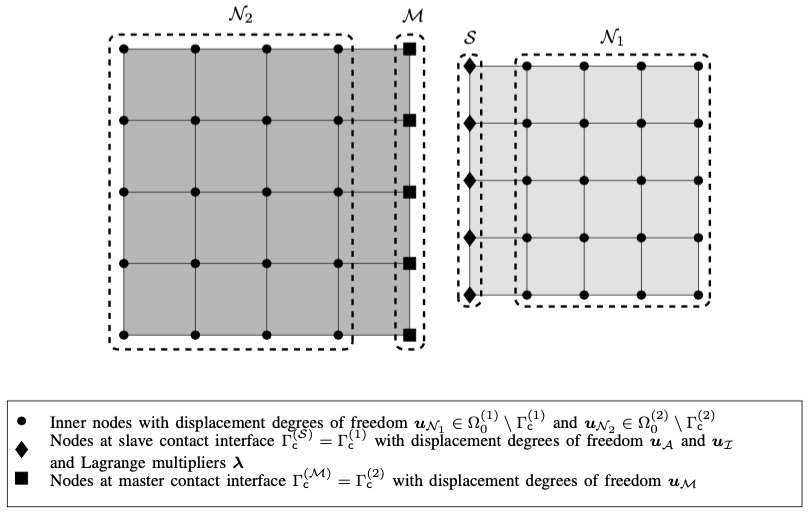}
\caption{Aggregation for contact example in saddle point formulation.}
\label{fig:ch6_aggregationinsaddlepointformulation}
\end{center}
\end{figure}

In this work, we propose a different approach to build aggregates~$\aggs[\level]^{\lagvec}$ for the Lagrange multipliers,
which does not suffer from above drawbacks.
Instead of explicitly building some helper matrix for the aggregation routine,
interface aggregates~$\aggs[\level]^{\lagvec}$ for the Lagrange multipliers are directly generated
using the aggregation information of the displacement variables (see Figure~\ref{fig:ch6_aggregationinsaddlepointformulation}).
The resulting interface aggregates for the Lagrange multipliers are by construction aligned with the corresponding displacement aggregates.
\begin{algorithm}
\SetKwProg{myproc}{Procedure}{}{}
\SetKwFunction{proc}{LagMultAggregation}
\myproc{\proc{$\aggs[\level]^{\disp}$,$\mortarDmat$}}{
\BlankLine
\emph{Initialize empty set and counter for aggregates $\aggs[\level]^{\lagvec}$}\newline
$\aggs[\level]^{\lagvec}\leftarrow \emptyset, ~ l\leftarrow 0$
\BlankLine
\emph{Initialize empty mapping of displacement aggregates to Lagrange multiplier aggregates}\newline
$\dispaggTOlagagg(k)\leftarrow\emptyset\quad \forall k=1,\ldots,\numnodes[{{\aggs[\level]}^{\disp}}]$
\BlankLine
\emph{Loop over slave displacement DOFs (rows of $\mortarDmat$)}\newline
\For{$i \in \slavedofs$}{
\BlankLine
\emph{Find displacement node $n^{\disp}$ id corresponding to displacement DOF $i$}\newline
$n^{\disp}\leftarrow \doftonode[i]$
\BlankLine
\emph{Find aggregate index $k$ that contains displacement node $n^{\disp}$}\newline
Find $k$ with $\aggs[\level]^{(k)}\in \aggs[\level]^{\disp}$ where $n^{\disp}\in\aggs[\level]^{(k)}$
\BlankLine
\emph{Loop over all Lagrange multipliers $j$}\newline
\For{$j \in \dofs[\lagvec]$}{
\BlankLine
\emph{Check whether Lagrange multiplier $j$ is coupled with row $i$}\newline
\If{$\mortarDmat_{i,j}\neq 0$}{
\BlankLine
\emph{Find pseudo node $n^{\lagmult}$ for Lagrange multiplier $j$}\newline
$n^{\lagmult} \leftarrow \doftonode[j]$
\BlankLine
\emph{Check whether to build a new Lagrange multiplier aggregate}\newline
\uIf{$\dispaggTOlagagg(k)=\emptyset$}{
\BlankLine
\emph{Increment internal aggregation counter}\newline
$l \leftarrow l+1$
\BlankLine
\emph{Build a new aggregate and add Lagrange multiplier node $n^{\lagmult}$}\newline
$\aggs[\level]^{(l)} \leftarrow \{n^{\lagmult}\}$
\BlankLine
\emph{Associate displacement aggregate $k$ with Lagrange multiplier aggregate $l$}\newline
$\dispaggTOlagagg(k) \leftarrow \{l\}$
\BlankLine
\emph{Add new aggregate to set of Lagrange multiplier aggregates $\aggs[\level]^{\lagvec}$}\newline
$\aggs[\level]^{\lagvec}\leftarrow \aggs[\level]^{\lagvec} \cup \aggs[\level]^{(l)}$
} 
\Else {
\BlankLine
\emph{Extend aggregate $0\leq\dispaggTOlagagg(k)\leq l$ with pseudo node}\newline
$\aggs[\level]^{(\dispaggTOlagagg(k))} \leftarrow \aggs[\level]^{(\dispaggTOlagagg(k))}\cup\{n^{\lagmult}\}$
} 
} 
} 
} 

\BlankLine
\emph{Return aggregates for Lagrange multipliers}\newline
\Return $\aggs[\level]^{\lagvec}$
} 
\caption{Aggregation algorithm for Lagrange multipliers.}
\label{alg:ch6_lagmultaggregation}
\end{algorithm}
The exact aggregation procedure is described in Algorithm~\ref{alg:ch6_lagmultaggregation}.
Assuming that the standard aggregates~$\aggs[\level]^{\disp}$ for the displacement degrees of freedom are available,
new aggregates~$\aggs[\level]^{\lagvec}$ are built by collecting the corresponding Lagrange multiplier degrees of freedom.
Beside the displacement aggregates~$\aggs[\level]^{\disp}$,
only the mortar matrix~$\mortarDmat$ is needed to algebraically reconstruct the contact interface and find the associated Lagrange multipliers.
The new aggregates~$\aggs[\level]^{\lagvec}$ for the Lagrange multipliers can be interpreted
as the natural extension of the displacement aggregates~$\aggs[\level]^{\disp}$ at the interface.
This facilitates to keep the ratio of coarse level nodes at the slave contact interface and the coarse Lagrange multipliers constant,
which also balances the ratio of contact constraints and inner structural displacement degrees of freedom over all multigrid levels.

The coarsening strategy outlined in Algorithm~\ref{alg:ch6_lagmultaggregation}
as well as all other AMG components of the presented saddle-point preconditioner for contact problems
have been implemented in {\muelu} \cite{BergerVergiat2019a}, the next-generation multigrid package within the {\trilinos} project \cite{TrilinosURL}.
For further details on the implementation,
we refer to the {\muelu} User's Guide \cite{BergerVergiat2019a} and the {\muelu} website \cite{MueLuURL}.

\section{Block smoothing methods for mortar contact problems}
\label{sec:BlockSmoothers}

Using saddle point preserving aggregation and segregated transfer operators as outlined in Section~\ref{sec:CoarseningOfContactConstraints}
to generate a fully coupled AMG hierarchy (see Section~\ref{sec:MultigridForMultiphysics}),
the coupling of structural equilibrium equations and contact constraints on all levels of a fully coupled AMG hierarchy
is now addressed by block smoothing methods on each level.
Schur complement based coupling iterations present themselves as ideal candidates to deal with the saddle point structure
resulting from the constraint-like contact equations.


Nevertheless, not all classical Schur complement based block smoothers for saddle point systems behave the same way
when applied to contact problems.
Specifically, it is important that the block smoothers account for the contact constraints in ~\eqref{eq:ch6_finalsaddlepointsystem_comprehensive}.
While each iteration's intermediate solution itself is not of great interest,
its satisfaction of the contact constraints impacts the overall number of required iterations to reach convergence,
since only an intermediate solution that satisfies the contact constraints
can then also be accepted as the final solution of the iterative process.
While an exact satisfaction of the contact constraints by the block smoothers is desirable,
practical computations require a compromise between the accuracy of the block smoothers and their computational effort
in order to obtain computationally competitive preconditioners.

In \secref{sec:ch6_blocksmoothers}, we first revisit some classical smoothing methods for saddle point systems.
Therefore, we assume exact inverses for the predictor and corrector step of each smoother
to focus on the systematic error resulting from the specific block structure of the smoother.
The assumption of exact inverses allows to assess the behavior of each block smoother via an error matrix,
such that the impact of the block smoother on the contact constraints can be characterized.
Afterwards, we introduce computationally cheaper variants with inexact block inverses in the predictor and corrector step
as a compromise between accuracy and performance in \secref{sec:ch6_cheapblocksmoothers},
however preventing the discourse on error matrices.
While the approximations due to the block structure of a specific smoother originate from the definition of the smoothing method,
the quality of the approximate block inverses can fully be controlled by the user.
Finally, \secref{sec:ComparisonOfContactSmoothers} discusses their application in the context of saddle point systems for contact formulations.

\subsection{Block smoothers for saddle point problems}
\label{sec:ch6_blocksmoothers}

Now, we study the systematic errors introduced by specific block smoothers and their impact on the contact constraints.
Therefore, we first assume exact mathematical block operations.
To carefully distinguish between systematic errors and errors stemming from practical Schur complement approximations,
we postpone the discussion of such approximations to \secref{sec:ch6_cheapblocksmoothers}.


The general block smoothing scheme can be written as
\begin{equation}
\begin{bmatrix}
\primvar[\precit+1] \\
\secondvar[\precit+1]
\end{bmatrix}
=
\begin{bmatrix}
\primvar[\precit] \\
\secondvar[\precit]
\end{bmatrix}
+
\relaxationinversemat^{-1}
\Biggl(
\begin{bmatrix}
\primrhs[\precit] \\
\secondrhs[\precit]
\end{bmatrix}
-
\begin{pmatrix}
\Ablock & \Boneblock \\
\Btwoblock & -\Cblock
\end{pmatrix}
\begin{bmatrix}
\primvar[\precit] \\
\secondvar[\precit]
\end{bmatrix}
\Biggr)
\label{eq:ch6_commonblocksmoothing}
\end{equation}
where~$\relaxationinversemat$ describes the $2\times 2$ block preconditioning matrix
approximating the $2\times 2$ block operator in~\eqref{eq:ch6_finalsaddlepointsystem_comprehensive}.
Typical block smoothers consist of an outer coupling iteration with nested subsolvers
to build the inverses of the diagonal blocks of $\relaxationinversemat$ in an algorithmic predictor-corrector scheme.
In the following, a few classical block smoothers from literature ({\eg}~\cite{notay2014}) are introduced, stating that this list is by far not complete.
All of them can be interpreted as block extensions of classical iterative smoothing methods
following the general block scheme \eqref{eq:ch6_commonblocksmoothing},
but depending on the definition of $\relaxationinversemat$ with a different effect on the contact constraints
by introducing certain systematic errors.
In general, the better~$\relaxationinversemat$ approximates the block operator from~\eqref{eq:ch6_finalsaddlepointsystem_comprehensive},
the lower the number of linear iterations will be when using the block smoother within a multigrid preconditioner.

\subsubsection{Uzawa smoother}
\label{sec:ch6_uzawa}
For the (inexact) Uzawa smoother, one chooses
\begin{equation}
\indUzawa{\relaxationinversemat} :=
\frac{1}{\blockdampingparameter}\begin{pmatrix}
\stiffmat & \zeromat\\ \Btwoblock & -\apprx{\schur}
\end{pmatrix}.
\label{eq:ch6_Uzawa}
\end{equation}
The parameter~$\blockdampingparameter>0$ is a damping parameter and~$\SchurApprox$ describes a cheap approximation of the Schur complement $\SchurBlock=\Cblock+\Btwoblock\Ablock^{-1}\Boneblock$.
For a theoretical review of Uzawa like smoothers, the reader is referred to \cite{bramble1997,elman1994,zulehner2002}.


With the off-diagonal coupling block~$\Btwoblock$ in~\eqref{eq:ch6_Uzawa}, the smoother performs a one-way coupling in the sense that the Lagrange multiplier increments now depend on the current increment of the displacement degrees of freedom. 
Algorithm \ref{alg:ch6_uzawa} represents the practical implementation as a predictor-corrector method. In each smoothing iteration, one calculates a prediction for the displacement increments~$\deltaincr{\primvarincinc[\precit+1]}$,
which are taken into account when solving for the corresponding Lagrange multiplier increments~$\deltaincr{\secondvarincinc[\precit+1]}$.

\begin{algorithm}
\SetKwProg{myproc}{Procedure}{}{}
\SetKwFunction{proc}{Uzawa}
\myproc{\proc{$\blockdampingparameter$, $\precit_{\max}$}}{
\BlankLine
\emph{Apply $\precit_{\max}$ smoothing sweeps with the Uzawa algorithm}\newline
\For{$\precit\leftarrow 0$~\KwTo$\precit_{\max}-1$}{
\BlankLine
\emph{Prediction step: solve for $\deltaincr{\primvarincinc[\precit+1]}$}\newline
$\Ablock~\deltaincr{\primvarincinc[\precit+1]}=\primrhs[\precit]-\Ablock\primvar[\precit]-\Boneblock\secondvar[\precit]$
\BlankLine
\emph{Correction step: solve for $\deltaincr{\secondvarincinc[\precit+1]}$}\newline
$-\SchurApprox~\deltaincr{\secondvarincinc[\precit+1]}=\secondrhs[\precit]-\Btwoblock\primvar[\precit]+\Cblock\secondvar[\precit]-\Btwoblock~
\deltaincr{\primvarincinc[\precit+1]}$
\BlankLine
\emph{Update step: update solution variables}\newline
$\primvar[\precit+1]\leftarrow \primvar[\precit]+\blockdampingparameter~\deltaincr{\primvarincinc[\precit+1]}$\newline
$\secondvar[\precit+1]\leftarrow \secondvar[\precit]+\blockdampingparameter~\deltaincr{\secondvarincinc[\precit+1]}$
} 
\BlankLine
\emph{Return smooth solution vector}\newline
\Return $\bigl(\primvar[\precit_{\max}],\secondvar[\precit_{\max}]\bigr)$
} 
\caption{Uzawa smoother.}
\label{alg:ch6_uzawa}
\end{algorithm}

Assuming the smoothing iteration has converged,
the error matrix for the Uzawa smoother is given as
\begin{equation}
\indUzawa{\errormat} : = A - \indUzawa{\relaxationinversemat} =
\begin{pmatrix}
\bigl(1-\frac{1}{\blockdampingparameter}\bigr) \Ablock & \Boneblock \\
\bigl(1-\frac{1}{\blockdampingparameter}\bigr) \Btwoblock & -\Cblock+\frac{1}{\blockdampingparameter}\apprx{\SchurBlock}
\end{pmatrix}.
\label{eq:ch6_errormatUzawa}
\end{equation}
With $\blockdampingparameter=1$,
$\SchurApprox=\Cblock+\Btwoblock\apprx{\Ablock}^{-1}\Boneblock$ being an approximation to the Schur complement,
and $\apprx{\Ablock}$ denoting an easy-to-invert approximation of~$\Ablock$,
{\eg} the diagonal of~$\Ablock$ as a cheap variant for the approximation~$\apprx{\Ablock}$, {\ie}~$\apprx{\Ablock}=\diag[(\Ablock)]$,
it is easy to verify that the error matrix of the Uzawa smoother reduces to
\begin{equation}
\indUzawa{\errormat} =
\begin{pmatrix}
\zeromat & \Boneblock\\
\zeromat & \Btwoblock\apprx{\Ablock}^{-1}\Boneblock
\end{pmatrix}.
\label{eq:ch6_errormatUzawa_special}
\end{equation}
That is, since the second block row in the error matices~\eqref{eq:ch6_errormatUzawa}
or~\eqref{eq:ch6_errormatUzawa_special} does not vanish, the Uzawa smoother by definition cannot exactly fulfill the contact constraints,
but adds a systematic error for the contact constraints.
This might have a negative impact on the overall performance of the iterative linear solver.

\subsubsection{Braess--Sarazin smoother}
\label{sec:ch6_braessarazin}

Originally introduced for the Stokes problem in \cite{braess1997},
the Braess--Sarazin smoother belongs to the class of block approximate smoothers and is based on the choice
\begin{equation}
\indBraessSarazin{\relaxationinversemat} :=
\begin{pmatrix}
\blockdampingparameter\apprx{\Ablock} & \Boneblock \\ \Btwoblock & -\Cblock
\end{pmatrix}
\label{eq:ch6_ABraessSarazin}
\end{equation}
for the block preconditioning matrix~$\relaxationinversemat$ in~\eqref{eq:ch6_commonblocksmoothing}.
Again, the parameter~$\blockdampingparameter>0$ denotes a scaling parameter and~$\apprx{\Ablock}$ refers to an easy-to-invert approximation of~$\Ablock$.

\begin{algorithm}
\SetKwProg{myproc}{Procedure}{}{}
\SetKwFunction{proc}{BraessSarazin}
\myproc{\proc{$\blockdampingparameter$, $\precit_{\max}$}}{
\BlankLine
\emph{Apply $\precit_{\max}$ smoothing sweeps with Braess--Sarazin algorithm}\newline
\For{$\precit\leftarrow 0$~\KwTo$\precit_{\max}-1$}{
\BlankLine
\emph{Prediction step: determine prediction $\primvar[\precit+\frac{1}{2}]$ by calculating}\newline
$\primvar[\precit+\frac{1}{2}]=\primvar[\precit]+\frac{1}{\blockdampingparameter}\apprx{\Ablock}^{-1}\bigl(\primrhs[\precit]-\Ablock\primvar[\precit]-\Boneblock\secondvar[\precit]\bigr)$
\BlankLine
\emph{Correction step: solve for $\deltaincr{\secondvarincinc[\precit+\frac{1}{2}]}$}\newline
$-\bigl(\Cblock+\frac{1}{\blockdampingparameter}\Btwoblock\apprx{\Ablock}^{-1}\Boneblock\bigr)~\deltaincr{\secondvarincinc[\precit+\frac{1}{2}]}=\secondrhs[\precit]+\Cblock\secondvar[\precit]-\Btwoblock\primvar[\precit+\frac{1}{2}]$
\BlankLine
\emph{Update step: update solution variables}\newline
$\secondvar[\precit+1]\leftarrow \secondvar[\precit]+\deltaincr{\secondvarincinc[\precit+\frac{1}{2}]}$\newline
$\primvar[\precit+1]\leftarrow \primvar[\precit+\frac{1}{2}]-\frac{1}{\blockdampingparameter}\apprx{\Ablock}^{-1}\Boneblock~\deltaincr{\secondvarincinc[\precit+\frac{1}{2}]}$
} 
\BlankLine
\emph{Return smooth solution vector}\newline
\Return $\bigl(\primvar[\precit_{\max}],\secondvar[\precit_{\max}]\bigr)$
} 
\caption{Braess--Sarazin smoother.}
\label{alg:ch6_braesssarazin}
\end{algorithm}

%

Assuming convergence of the smoother,
the error matrix for the Braess--Sarazin smoother is given as
\begin{equation}
\indBraessSarazin{\errormat} := A - \indBraessSarazin{\relaxationinversemat} =
\begin{pmatrix}
\Ablock-\blockdampingparameter\apprx{\Ablock} & \zeromat\\
\zeromat & \zeromat
\end{pmatrix}.
\label{eq:ch6_errormatBraessSarazin}
\end{equation}

With the second block row in the blocked operator~\eqref{eq:ch6_finalsaddlepointsystem_comprehensive}
being retained in~\eqref{eq:ch6_ABraessSarazin}, the Braess--Sarazin smoother seems to be a reasonable choice for dealing with contact constraints. The error matrix in~\eqref{eq:ch6_errormatBraessSarazin}~reveals that the quality of the block smoother only depends on the choice of $\apprx{\Ablock}$.

The implementation as a predictor-corrector method is based on the splitting of~\eqref{eq:ch6_ABraessSarazin} into
\begin{equation}
\begin{pmatrix}
\blockdampingparameter\apprx{\Ablock} & \Boneblock \\ \Btwoblock & -\Cblock
\end{pmatrix}
=
\begin{pmatrix}
\blockdampingparameter\apprx{\Ablock} & \zeromat \\ \Btwoblock & -\Cblock-\frac{1}{\blockdampingparameter}\Btwoblock\apprx{\Ablock}^{-1}\Boneblock
\end{pmatrix}
\begin{pmatrix}
\identitymatrix & \frac{1}{\blockdampingparameter}\apprx{\Ablock}^{-1}\Boneblock \\ \zeromat & \identitymatrix
\end{pmatrix}.
\end{equation}
As one can easily see from Algorithm~\ref{alg:ch6_braesssarazin}, the prediction step can be understood as one hard-coded sweep with a (damped) Jacobi iteration. In other words, the quality of the prediction for the displacement degrees of freedom must be considered rather poor. Exactly fulfilling contact constraints with respect to a rather poor prediction of the displacement variables might not be optimal for the overall performance of the preconditioner.


\subsubsection{SIMPLE variants}
\label{sec:simple}
Originally introduced in \cite{patankar1980,patankar1972}, the SIMPLE method is based on the approximate block factorization
\begin{equation}
\indSIMPLE{\relaxationinversemat} :=
\begin{pmatrix}
\Ablock & \zeromat \\
\Btwoblock & -\SchurApprox
\end{pmatrix}
\begin{pmatrix}
\identitymatrix & \apprx{\Ablock}^{-1} \Boneblock\\
\zeromat & \frac{1}{\blockdampingparameter}\identitymatrix
\end{pmatrix}
=
\begin{pmatrix}
\stiffmat & \stiffmat \apprx{\stiffmat}^{-1} \Boneblock \\
\Btwoblock & \Bigl(1-\frac{1}{\blockdampingparameter}\Bigr)\Btwoblock\apprx{\stiffmat}^{-1}\Boneblock-\frac{1}{\blockdampingparameter}\Cblock
\end{pmatrix}
\label{eq:ch6_ASIMPLE}
\end{equation}
for the iterative method in \eqref{eq:ch6_commonblocksmoothing}.
In \eqref{eq:ch6_ASIMPLE}, $\SchurApprox$ denotes an approximation of the Schur complement $\SchurBlock:=\Cblock+\Btwoblock\Ablock^{-1}\Boneblock$ with a cheap and easy-to-invert approximation $\apprx{\Ablock}$ of the block $\Ablock$.
Algorithm~\ref{alg:ch6_simple} shows the implementation of the SIMPLE method using the predictor-corrector scheme ({\cf} \cite{elman2008}).

\begin{algorithm}
\SetKwProg{myproc}{Procedure}{}{}
\SetKwFunction{proc}{SIMPLE}
\myproc{\proc{$\blockdampingparameter$, $\precit_{\max}$}}{
\BlankLine
\emph{Apply $\precit_{\max}$ smoothing sweeps with SIMPLE algorithm}\newline
\For{$\precit\leftarrow 0$~\KwTo$\precit_{\max}-1$}{
\BlankLine
\emph{Prediction step: solve for $\primvar[\precit+\frac{1}{2}]$}\newline
$\Ablock\primvar[\precit+\frac{1}{2}]=\primrhs[\precit]-\Boneblock\secondvar[\precit]$
\BlankLine
\emph{Correction step: solve for $\deltaincr{\secondvarincinc[\precit+\frac{1}{2}]}$}\newline
$-\SchurApprox~\deltaincr{\secondvarincinc[\precit+\frac{1}{2}]}=\secondrhs[\precit]+\Cblock\secondvar[\precit]-\Btwoblock\primvar[\precit+\frac{1}{2}]$
\BlankLine
\emph{Update step: update solution variables}\newline
$\secondvar[\precit+1]\leftarrow \secondvar[\precit]+\blockdampingparameter~\deltaincr{\secondvarincinc[\precit+\frac{1}{2}]}$\newline
$\primvar[\precit+1]\leftarrow \primvar[\precit+\frac{1}{2}]-\blockdampingparameter\apprx{\Ablock}^{-1}\Boneblock~\deltaincr{\secondvarincinc[\precit+\frac{1}{2}]}$
} 
\BlankLine
\emph{Return smooth solution vector}\newline
\Return $\bigl(\primvar[\precit_{\max}],\secondvar[\precit_{\max}]\bigr)$

} 
\caption{SIMPLE smoother.}
\label{alg:ch6_simple}
\end{algorithm}

For our applications, we found the diagonal matrix containing the row sums of~$\bigl|\Ablock\bigr|=\bigl(|a_{ij}|)_{i,j=1,\ldots,\numdofs[\Ablock]}$ to be a good approximation for block~$\Ablock$. This corresponds to the SIMPLEC method as introduced in \cite{vandoormal1984}.
That is, $\apprx{\Ablock}$ is defined as the diagonal lumping of $\bigl|\Ablock\bigr|$ with 
\begin{equation}
\apprx{\Ablock}=\diag[ \Bigl(\sum_{j=1}^{\numdofs[\Ablock]}|a_{ij}|\Bigr)],\quad i=1,\ldots,\numdofs[\Ablock].
\label{eq:ch6_simplecapprox}
\end{equation}
The default choice for~$\SchurApprox$ is consequently~$\SchurApprox=\blockdampingparameter\Cblock+\blockdampingparameter\Btwoblock \apprx{\Ablock}^{-1} \Boneblock$ with $\apprx{\Ablock}$
as defined in~\eqref{eq:ch6_simplecapprox}.
A more theoretical discussion on the mathematical consequences of approximations for the Schur complement~$\SchurBlock$ can be found in \cite{zulehner2000}.

For the SIMPLE preconditioner, the error matrix is calculated by
\begin{equation}
\indSIMPLE{\errormat} := A - \indSIMPLE{\relaxationinversemat} =
\begin{pmatrix}
\zeromat & \Boneblock-\Ablock\apprx{\Ablock}^{-1}\Boneblock\\
\zeromat & -\Cblock-\Btwoblock\apprx{\Ablock}^{-1}\Boneblock+\frac{1}{\blockdampingparameter}\SchurApprox
\end{pmatrix}.
\label{eq:ch6_errormatsimple}
\end{equation}
As one can see from~\eqref{eq:ch6_errormatsimple},
SIMPLE perturbs the Lagrange multipliers, but it does not affect the terms that operate on the primary displacement variables since the first block column in~\eqref{eq:ch6_errormatsimple} is zero.
Choosing~$\SchurApprox=\blockdampingparameter\Cblock+\blockdampingparameter\Btwoblock\apprx{\Ablock}^{-1}\Boneblock$,
the error matrix reduces to
\begin{equation}
\indSIMPLE{\errormat} =
\begin{pmatrix}
\zeromat & \Boneblock-\Ablock\apprx{\Ablock}^{-1}\Boneblock\\
\zeromat & \zeromat
\end{pmatrix}.
\label{eq:ch6_errormatsimple_special}
\end{equation}
That is, an appropriate approximation~$\SchurApprox$ of the Schur complement~$\schur$ allows to exactly satisfy the contact constraints within one smoothing sweep.
Depending on the choice for the approximation~$\apprx{\Ablock}$,
the SIMPLE method admits an error in the coupling between displacements and contact constraints.
However, compared to the Braess-Sarazin method from Section \ref{sec:ch6_braessarazin},
we put more focus on a good prediction for the displacements with a consistent update for fulfilling the contact constraints.


\subsection{Cheap variants of block smoothers}
\label{sec:ch6_cheapblocksmoothers}


As one can easily see from the Algorithms \ref{alg:ch6_uzawa} to \ref{alg:ch6_simple},
all block smoothing methods internally require inverses of the matrix blocks on the matrix diagonal and of the Schur complement operator $\SchurApprox$.
Specifically, there is one linear system to be solved in the prediction step and one during the correction step.
To keep the computational costs low in practical computations,
one does not solve for the block inverses exactly, but apply a cheap approximation,
{\eg} by using a fixed number of smoothing sweeps with a relaxation-based smoothing method such as symmetric Gauss--Seidel or an ILU sweep.
This allows for more flexibility for finding a good compromise between quality and performance,
as the user can decide how much effort should be put on finding a good prediction and fulfilling the Schur complement equation in the corrector step.
While the systematic errors introduced by the choice of the block smoother from Section~\ref{sec:ch6_blocksmoothers} are fixed,
the practitioner has full control over the quality of the block inverses.

The numerical examples in Section~\ref{sec:NumEx} show
that such an approximation leads to efficient and computationally reasonable block smoothing methods.
As a naming convention, the prefix ``Cheap" is added to the name of the block smoothing method
to indicate the usage of a cheap approximation for finding the inverse of the diagonal blocks
in addition to the systematic approximations of building the Schur complement operator as discussed in Section~\ref{sec:ch6_blocksmoothers}.


The block smoothing methods and their cheap variants are available in the {\xpetra} package of the {\trilinos} project \cite{TrilinosURL}.
For further details on the implementation,
we refer to the {\muelu} User's Guide \cite{BergerVergiat2019a} and the {\muelu} and {\xpetra} websites \cite{MueLuURL,XpetraURL}, respectively.

\subsection{Comparison of saddle point smoothing methods for contact problems}
\label{sec:ComparisonOfContactSmoothers}

Considering the block scheme of~\eqref{eq:ch6_finalsaddlepointsystem_comprehensive},
there are two main challenges for contact problems.
First, we have the two distinct sets of equations as described in Section~\ref{sec:finalalgebraicsystem}:
the structural equations formulated in cartesian coordinates
and the set of contact constraints formulated in normal-tangential coordinates relative to the contact surface.
Second, the coupling of those two distinct sets of structural equations and contact constraint equations.

Algorithmically, the coupling of structural equations at the contact interface via the off-diagonal blocks in~\eqref{eq:ch6_finalsaddlepointsystem_comprehensive} is only considered within the block smoothers from Section~\ref{sec:ch6_blocksmoothers}. Therefore, constraint smoothers ({\cf}~\cite{keller2000}) are the natural choice for contact problems, since the contact problem is implicitly governed by the contact constraint equations. Only solutions that are in alignment with the contact constraints are of interest.

Under certain preconditions as discussed in Section~\ref{sec:ch6_blocksmoothers}
and assuming sufficient smoother iterations to reach convergence of the smoother,
both the Braess--Sarazin method and the SIMPLE-based methods would exactly fulfill the contact constraints as one can see from equations~\eqref{eq:ch6_errormatBraessSarazin} or~\eqref{eq:ch6_errormatsimple_special}. Therefore, with a systematic error in the lower-right block of~\eqref{eq:ch6_errormatUzawa_special}, the Uzawa method seems to be less promising for contact problems.
In the Braess--Sarazin method, the approximation~$\apprx{\Ablock}=\diag[(\Ablock)]$ is hard-coded
with some scaling parameter~$\blockdampingparameter>0$ and consistently used within the approximate Schur complement operator,
which is defined by~$\SchurApprox=\Cblock+\frac{1}{\blockdampingparameter}\Btwoblock\apprx{\Ablock}^{-1}\Boneblock$.

In contrary to the Braess--Sarazin method,
the SIMPLE based methods keep the full~$\Ablock$ block whenever possible in the block factorization
and use~$\apprx{\Ablock}$ only where its inverse is required.
Consequently, in ``cheap" variants of the SIMPLE method,
more elaborate smoothing strategies can be used for the~$\Ablock$ block instead of a hard-coded Jacobi sweep. 
Therefore, one can think of the SIMPLE methods to allow for a more balanced quality of approximations for the displacement degrees of freedom and Lagrange multipliers for the contact constraints,
whereas the Braess--Sarazin method exhibits an imbalance in the sense
that the computational effort spent for approximating the constraints is much higher than for dealing with the displacement variables. 


\section{Numerical examples}
\label{sec:NumEx}

For the numerical examples, we use our in-house code BACI~\cite{BaciWebsite}
that internally uses various capabilities from the {\trilinos} project \cite{TrilinosURL}.
The implementation of the multigrid algorithms is based on Trilinos' MueLu package~\cite{BergerVergiat2019a,Wiesner2014a}.
In particular, all block smoothers from \secref{sec:BlockSmoothers} as well as the contact specific aggregation strategy for the Lagrange multiplier unknowns as described in \secref{sec:CoarseningOfContactConstraints} are readily available in MueLu.


\subsection{Two solid bodies example}
\label{sec:ch5_twosolidbodiesexample}

\providecommand{\angleZtext}{\alpha_z}
\providecommand{\angleYtext}{\alpha_y}



With the first example, we want to study the effect of the block smoothers from \secref{sec:BlockSmoothers} for contact problems.
Here, we not only compare different block smoothers,
but also highlight the effect of varying the quality of the sub-smoothing steps within the block smoother
versus increasing the number of outer coupling iterations.

Motivated by findings in our previous work~\cite{wiesner2017},
this example briefly revisits a detail that has been problematic in the context of contact problems in condensed contact formulations.
While the discrete global unknowns $(\disp,\lagvec)$ are -- as usual -- formulated with respect to the global Cartesian frame, the discrete Lagrange multiplier weights $\muvec$ and therefore the contact constraint equations
in~\eqref{eq:ch4_discretenormalcontact} and~\eqref{eq:ch4_discretefrictionless} are formulated
with respect to a local convective coordinate system.
This local system is defined at each slave node $j$ by a surface normal vector and two tangent vectors,
{\ie} by a triad of orthonormal basis vectors $(\boldsymbol{n})_j$, $(\boldsymbol{\tau}_\xi)_j$ and $(\boldsymbol{\tau}_\eta)_j$.
Although it represents a quite intuitive and natural choice in contact mechanics,
this local constraint formulation may lead to non-diagonally dominant system matrices
and therefore poses a serious challenge to the development of iterative linear solvers as has been elaborated in~\cite{wiesner2017}.
In the following, we will also investigate the susceptibility of the proposed saddle point preconditioners
to this phenomenon.

\subsubsection{Geometrical setup}

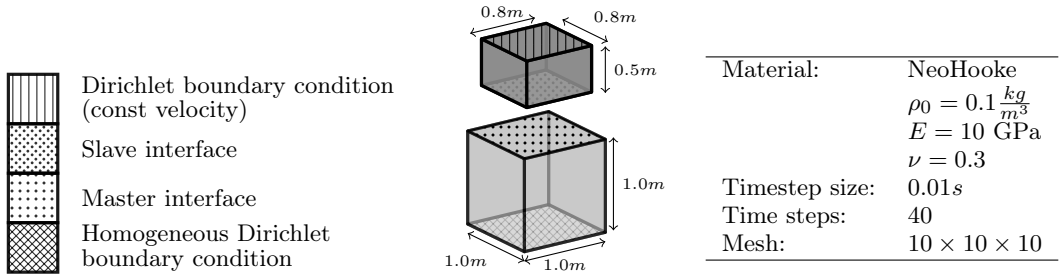
\begin{figure}
\begin{center}
\begin{tabular}{llr}
\tikzset{external/figure name={twoblocks}}
\tikzset{external/figure name/.add={}{_configuration}}

\newlength{\offsetO}
\setlength{\offsetO}{8pt}
\tikzexternaldisable
\begin{tikzpicture}[scale=0.65,
		curve/.style={red,densely dotted,thick},
		cube/.style={opacity=.6,very thick,fill=lightgray},
		cubepurple/.style={opacity=.9,very thick,fill=gray!110!black},
		cubeslave/.style={opacity=.9,very thick,fill=gray!110!black, pattern=crosshatch dots, pattern color=black},
		cubemaster/.style={opacity=.9,very thick,fill=gray!110!black, pattern=dots, pattern color=black},
		cubefixed/.style={opacity=.9,very thick,fill=gray!110!black, pattern=crosshatch, pattern color=black},		
		cubedbc/.style={opacity=.9,very thick,fill=gray!110!black, pattern=vertical lines, pattern color=black}]

		\draw[cubefixed] (0,-1) -- (0,0) -- (1,0) -- (1,-1) -- cycle;
		\node (master) [anchor=west] at (1.3,-0.25) {Homogeneous Dirichlet};
		\node (master) [anchor=west] at (1.3,-0.75) {boundary condition};
		\draw[cubemaster] (0,0) -- (0,1) -- (1,1) -- (1,0) -- cycle;
		\node (master) [anchor=west] at (1.3,0.5) {Master interface};
		\draw[cubeslave] (0,1) -- (0,2) -- (1,2) -- (1,1) -- cycle;
		\node (slave) [anchor=west] at (1.3,1.5) {Slave interface};
		\draw[cubedbc] (0,2) -- (0,3) -- (1,3) -- (1,2) -- cycle;
		\node (master) [anchor=west] at (1.3,2.75) {Dirichlet boundary condition};
		\node (master) [anchor=west] at (1.3,2.25) {(const velocity)};	
\end{tikzpicture} %
\tdplotsetmaincoords{110}{35}
\begin{tikzpicture}[scale=0.65,
		tdplot_main_coords,
		curve/.style={red,densely dotted,thick},
		cube/.style={opacity=.6,very thick,fill=lightgray},
		cubepurple/.style={opacity=.9,very thick,fill=gray!110!black},
		cubeslave/.style={very thick,fill=gray!110!black, pattern=crosshatch dots, pattern color=black},
		cubemaster/.style={pattern=dots, pattern color=black},
		cubefixed/.style={opacity=.6,very thick,fill=gray!110!black, pattern=crosshatch, pattern color=black},		
		cubedbc/.style={pattern=vertical lines, pattern color=black},
		graytransparent/.style={opacity=.5,fill=gray,thick},
		curve/.style={gray,thick},
		curveback/.style={gray,densely dotted,thick},
		axis/.style={->,black,thick},
		grid/.style={very thin,gray},
		line join=round]
		

		\coordinate (newOrigin) at (0,0,\offsetO);
		\tdplotsetrotatedcoordsorigin{(newOrigin)}


        \draw[cubefixed,tdplot_rotated_coords] (-1,-1,-2) -- (-1,1,-2) -- (1,1,-2) -- (1,-1,-2) -- cycle;
        \draw[cube,tdplot_rotated_coords] (1,-1,-2) -- (1,1,-2) -- (1,1,0) -- (1,-1,0) -- cycle;
        \draw[cube,tdplot_rotated_coords] (-1,-1,-2) -- (1,-1,-2) -- (1,-1,0) -- (-1,-1,0) -- cycle;
        \draw[cube,tdplot_rotated_coords] (-1,-1,-2) -- (-1,1,-2) -- (-1,1,0) -- (-1,-1,0) -- cycle;
        \draw[cube,tdplot_rotated_coords] (-1,1,-2) -- (1,1,-2) -- (1,1,0) -- (-1,1,0) -- cycle;
        \draw[cube,tdplot_rotated_coords] (-1,-1,0) -- (-1,1,0) -- (1,1,0) -- (1,-1,0) -- cycle;        
        \draw[cubemaster,tdplot_rotated_coords] (-1,-1,0) -- (-1,1,0) -- (1,1,0) -- (1,-1,0) -- cycle;
        \draw[<->,tdplot_rotated_coords] (-1,1,-2.2) --  node [midway,below] {\tiny $1.0 m$} (1,1,-2.2);
        \draw[<->,tdplot_rotated_coords] (-1,-1,-2.2) --  node [midway,below left] {\tiny $1.0 m$} (-1,1,-2.2);
        \draw[<->,tdplot_rotated_coords] (1.2,1,-2.0) --  node [midway,right] {\tiny $1.0 m$} (1.2,1,0.0);

        \draw[cubeslave,tdplot_rotated_coords] (-0.8,-0.8,1) -- (-0.8,0.8,1) -- (0.8,0.8,1) -- (0.8,-0.8,1) -- cycle;
        \draw[cubepurple,tdplot_rotated_coords] (0.8,-0.8,1) -- (0.8,0.8,1) -- (0.8,0.8,2) -- (0.8,-0.8,2) -- cycle;
        \draw[cubepurple,tdplot_rotated_coords] (-0.8,-0.8,1) -- (0.8,-0.8,1) -- (0.8,-0.8,2) -- (-0.8,-0.8,2) -- cycle;
        \draw[cubepurple,tdplot_rotated_coords] (-0.8,-0.8,1) -- (-0.8,0.8,1) -- (-0.8,0.8,2) -- (-0.8,-0.8,2) -- cycle;
        \draw[cubepurple,tdplot_rotated_coords] (-0.8,0.8,1) -- (0.8,0.8,1) -- (0.8,0.8,2) -- (-0.8,0.8,2) -- cycle;
        \draw[cubepurple,tdplot_rotated_coords] (-0.8,-0.8,2) -- (-0.8,0.8,2) -- (0.8,0.8,2) -- (0.8,-0.8,2) -- cycle;
        \draw[cubedbc,tdplot_rotated_coords] (-0.8,-0.8,2) -- (-0.8,0.8,2) -- (0.8,0.8,2) -- (0.8,-0.8,2) -- cycle;	
        \draw[<->,tdplot_rotated_coords] (-0.8,-1.2,2.0) --  node [midway,above] {\tiny $0.8 m$} (0.8,-1.2,2.0);
        \draw[<->,tdplot_rotated_coords] (1.2,-0.8,2) --  node [midway,above right] {\tiny $0.8 m$} (1.2,0.8,2);
        \draw[<->,tdplot_rotated_coords] (1.0,1,1.0) --  node [midway,right] {\tiny $0.5 m$} (1.0,1,2.0);
\end{tikzpicture}
\tikzexternalenable
&
\begin{minipage}{0.35\textwidth}
\vspace{-3cm}
\begin{tabular}{ll}\hline
Material: & NeoHooke \\%
& $\refdensity=0.1\frac{kg}{m^3}$ \\%
& $\YoungModulus=10$ GPa \\%
&$\nu=0.3$ \\%
Timestep size: & $0.01s$ \\%
Time steps: & $40$ \\%
Mesh: & $10\times 10\times 10$ \\ \hline
\end{tabular}
\end{minipage}

\end{tabular}

\caption{Two solid bodies example --- Geometric configuration and parameters.}
\label{fig:ch5_twosolidbodiesexamplesetupsurfaces}
\end{center}
\end{figure}

Since we are interested in the solver behavior, by intention we choose a simple 3D contact example as shown in Figure \ref{fig:ch5_twosolidbodiesexamplesetupsurfaces}.
There are two solid bodies with the same material parameters using a Neo-Hookean material
(density~$\refdensity=0.1\frac{kg}{m^3}$, Young's modulus~$\YoungModulus=10$ GPa, Poisson's ratio~$\nu=0.3$).
The initial gap between the two solid bodies is $0.02$ meters. The upper solid body (size: $0.8 m\times 0.8 m\times 0.5 m$) is moving down with constant velocity along the normal to the contact interface towards the lower fixed solid body (size: $1.0 m \times 1.0 m \times 1.0 m$).

\subsubsection{Experimental setup}
\begin{figure}
\begin{center}

\tdplotsetmaincoords{110}{35}
\tikzset{external/figure name={twoblocks}}
\tikzset{external/figure name/.add={}{_configuration}}

\setlength{\offsetO}{8pt}

\tikzexternaldisable

\begin{tikzpicture}[scale=0.65,
		tdplot_main_coords,
		curve/.style={red,densely dotted,thick},
		cube/.style={opacity=.6,very thick,fill=lightgray},
		cubepurple/.style={opacity=.9,very thick,fill=gray!110!black},
		graytransparent/.style={opacity=.5,fill=gray,thick},
		curve/.style={gray,thick},
		curveback/.style={gray,densely dotted,thick},
		axis/.style={->,black,thick},
		grid/.style={very thin,gray},
		line join=round]


		\pgfmathsetmacro{\sphereR}{4}
		\coordinate (newOrigin) at (0,0,0);
		\tdplotsetrotatedcoordsorigin{(newOrigin)}
		\foreach \angle in {0,22.5,45,67.5,90}
		{
		    \tdplotsinandcos{\sintheta}{\costheta}{\angle}%

		    \coordinate (P) at (0,0,\sphereR*\sintheta);

		    \tdplotdrawarc[curve]{(P)}{\sphereR*\costheta}{90}{180}{}{}
		}                
		\foreach \angle in {90,112.5,135,157.5,180}
		{
		    \tdplotsetthetaplanecoords{\angle}
                
		    \tdplotdrawarc[curve,tdplot_rotated_coords]{(0,0,0)}{\sphereR}{0}{90}{}{}
		}

        \draw[curveback,tdplot_rotated_coords] (0,0,0) -- (-\sphereR,0,0);
        \draw[curveback,tdplot_rotated_coords] (0,0,0) -- (0,\sphereR,0);
        \draw[curveback,tdplot_rotated_coords] (0,0,0) -- (0,0,\sphereR);
        \draw[curve,tdplot_rotated_coords] (-\sphereR,0,0) -- (-\offsetO+2pt,0,0);
        \draw[curve,tdplot_rotated_coords] (0,\sphereR,0) -- (0,\offsetO-2pt,0);
        \draw[curve,tdplot_rotated_coords] (0,0,\sphereR) -- (0,0,\offsetO-2pt);

        \draw[axis,tdplot_rotated_coords] (0,0,0) -- (1,0,0) node[anchor=west]{$x$};
        \draw[axis,tdplot_rotated_coords] (0,0,0) -- (0,1,0) node[anchor=north west]{$z$};
        \draw[axis,tdplot_rotated_coords] (0,0,0) -- (0,0,1) node[anchor=west]{$y$};

		\pgfmathsetmacro{\angleZ}{45.0}
		\pgfmathsetmacro{\angleY}{22.5}
		\tdplotsinandcos{\sinZ}{\cosZ}{\angleZ}
		\tdplotsinandcos{\sinY}{\cosY}{\angleY}
		
		{
		\pgfmathsetmacro{\myOffset}{\offsetO-2pt}
		\tdplotdrawarc[->]{(newOrigin)}{1.45}{180}{180-\angleY}{anchor=north}{$\angleYtext$}
		\tdplotsetthetaplanecoords{180-\angleY}
		\tdplotdrawarc[tdplot_rotated_coords,->]{(newOrigin)}{1.4}{0}{\angleZ}{anchor=south}{$\angleZtext$}
		
		\draw[curveback,tdplot_main_coords] (0,0,0) -- (-\sinZ*\cosY*\sphereR,\sinZ*\sinY*\sphereR,\cosZ*\sphereR);
		\draw[curve,tdplot_main_coords] (-\sinZ*\cosY*\sphereR,\sinZ*\sinY*\sphereR,\cosZ*\sphereR) -- (-\sinZ*\cosY*\myOffset,\sinZ*\sinY*\myOffset,\cosZ*\myOffset);
		\draw[curve,tdplot_main_coords] (0,0,0) -- (-\sinZ*\cosY*2,\sinZ*\sinY*2,0);
		\draw[graytransparent,tdplot_main_coords] (-\sinZ*\cosY*2,\sinZ*\sinY*2,\cosZ*2) -- (-\sinZ*\cosY*2,\sinZ*\sinY*2,0) -- (0,0,0) -- cycle;
		}
		

		\coordinate (newOrigin) at (0,0,\offsetO);
		\tdplotsetrotatedcoordsorigin{(newOrigin)}


        \draw[cube,tdplot_rotated_coords] (-1,-1,-2) -- (-1,1,-2) -- (1,1,-2) -- (1,-1,-2) -- cycle;
        \draw[cube,tdplot_rotated_coords] (1,-1,-2) -- (1,1,-2) -- (1,1,0) -- (1,-1,0) -- cycle;
        \draw[cube,tdplot_rotated_coords] (-1,-1,-2) -- (1,-1,-2) -- (1,-1,0) -- (-1,-1,0) -- cycle;
        \draw[cube,tdplot_rotated_coords] (-1,-1,-2) -- (-1,1,-2) -- (-1,1,0) -- (-1,-1,0) -- cycle;
        \draw[cube,tdplot_rotated_coords] (-1,1,-2) -- (1,1,-2) -- (1,1,0) -- (-1,1,0) -- cycle;
        \draw[cube,tdplot_rotated_coords] (-1,-1,0) -- (-1,1,0) -- (1,1,0) -- (1,-1,0) -- cycle;

        \draw[cubepurple,tdplot_rotated_coords] (-0.8,-0.8,0) -- (-0.8,0.8,0) -- (0.8,0.8,0) -- (0.8,-0.8,0) -- cycle;
        \draw[cubepurple,tdplot_rotated_coords] (0.8,-0.8,0) -- (0.8,0.8,0) -- (0.8,0.8,1) -- (0.8,-0.8,1) -- cycle;
        \draw[cubepurple,tdplot_rotated_coords] (-0.8,-0.8,0) -- (0.8,-0.8,0) -- (0.8,-0.8,1) -- (-0.8,-0.8,1) -- cycle;
        \draw[cubepurple,tdplot_rotated_coords] (-0.8,-0.8,0) -- (-0.8,0.8,0) -- (-0.8,0.8,1) -- (-0.8,-0.8,1) -- cycle;
        \draw[cubepurple,tdplot_rotated_coords] (-0.8,0.8,0) -- (0.8,0.8,0) -- (0.8,0.8,1) -- (-0.8,0.8,1) -- cycle;
        \draw[cubepurple,tdplot_rotated_coords] (-0.8,-0.8,1) -- (-0.8,0.8,1) -- (0.8,0.8,1) -- (0.8,-0.8,1) -- cycle;

		\coordinate (newOrigin) at (-\offsetO,0,0);
		\tdplotsetrotatedcoordsorigin{(newOrigin)}
		
		\tdplotsetrotatedcoords{0}{-90}{0}

        \draw[cube,tdplot_rotated_coords] (-1,-1,-2) -- (-1,1,-2) -- (1,1,-2) -- (1,-1,-2) -- cycle;     
        \draw[cube,tdplot_rotated_coords] (-1,-1,-2) -- (-1,1,-2) -- (-1,1,0) -- (-1,-1,0) -- cycle;
        \draw[cube,tdplot_rotated_coords] (-1,-1,-2) -- (1,-1,-2) -- (1,-1,0) -- (-1,-1,0) -- cycle;
        \draw[cube,tdplot_rotated_coords] (1,-1,-2) -- (1,1,-2) -- (1,1,0) -- (1,-1,0) -- cycle;
        \draw[cube,tdplot_rotated_coords] (-1,1,-2) -- (1,1,-2) -- (1,1,0) -- (-1,1,0) -- cycle;
        \draw[cube,tdplot_rotated_coords] (-1,-1,0) -- (-1,1,0) -- (1,1,0) -- (1,-1,0) -- cycle;

        \draw[cubepurple,tdplot_rotated_coords] (-0.8,-0.8,0) -- (-0.8,0.8,0) -- (0.8,0.8,0) -- (0.8,-0.8,0) -- cycle;        
        \draw[cubepurple,tdplot_rotated_coords] (-0.8,-0.8,0) -- (-0.8,0.8,0) -- (-0.8,0.8,1) -- (-0.8,-0.8,1) -- cycle;
        \draw[cubepurple,tdplot_rotated_coords] (-0.8,-0.8,0) -- (0.8,-0.8,0) -- (0.8,-0.8,1) -- (-0.8,-0.8,1) -- cycle;
        \draw[cubepurple,tdplot_rotated_coords] (0.8,-0.8,0) -- (0.8,0.8,0) -- (0.8,0.8,1) -- (0.8,-0.8,1) -- cycle;
        \draw[cubepurple,tdplot_rotated_coords] (-0.8,0.8,0) -- (0.8,0.8,0) -- (0.8,0.8,1) -- (-0.8,0.8,1) -- cycle;
        \draw[cubepurple,tdplot_rotated_coords] (-0.8,-0.8,1) -- (-0.8,0.8,1) -- (0.8,0.8,1) -- (0.8,-0.8,1) -- cycle;

		\coordinate (newOrigin) at (0,+\offsetO,0);
		\tdplotsetrotatedcoordsorigin{(newOrigin)}
		
		\tdplotsetrotatedcoords{-90}{-90}{0}

        \draw[cube,tdplot_rotated_coords] (-1,-1,-2) -- (-1,1,-2) -- (1,1,-2) -- (1,-1,-2) -- cycle;     
        \draw[cube,tdplot_rotated_coords] (-1,1,-2) -- (1,1,-2) -- (1,1,0) -- (-1,1,0) -- cycle;
        \draw[cube,tdplot_rotated_coords] (-1,-1,-2) -- (-1,1,-2) -- (-1,1,0) -- (-1,-1,0) -- cycle;
        \draw[cube,tdplot_rotated_coords] (-1,-1,-2) -- (1,-1,-2) -- (1,-1,0) -- (-1,-1,0) -- cycle;
        \draw[cube,tdplot_rotated_coords] (1,-1,-2) -- (1,1,-2) -- (1,1,0) -- (1,-1,0) -- cycle;
        \draw[cube,tdplot_rotated_coords] (-1,-1,0) -- (-1,1,0) -- (1,1,0) -- (1,-1,0) -- cycle;

        \draw[cubepurple,tdplot_rotated_coords] (-0.8,-0.8,0) -- (-0.8,0.8,0) -- (0.8,0.8,0) -- (0.8,-0.8,0) -- cycle;        
        \draw[cubepurple,tdplot_rotated_coords] (-0.8,0.8,0) -- (0.8,0.8,0) -- (0.8,0.8,1) -- (-0.8,0.8,1) -- cycle;
        \draw[cubepurple,tdplot_rotated_coords] (-0.8,-0.8,0) -- (-0.8,0.8,0) -- (-0.8,0.8,1) -- (-0.8,-0.8,1) -- cycle;
        \draw[cubepurple,tdplot_rotated_coords] (-0.8,-0.8,0) -- (0.8,-0.8,0) -- (0.8,-0.8,1) -- (-0.8,-0.8,1) -- cycle;
        \draw[cubepurple,tdplot_rotated_coords] (0.8,-0.8,0) -- (0.8,0.8,0) -- (0.8,0.8,1) -- (0.8,-0.8,1) -- cycle;
        \draw[cubepurple,tdplot_rotated_coords] (-0.8,-0.8,1) -- (-0.8,0.8,1) -- (0.8,0.8,1) -- (0.8,-0.8,1) -- cycle;

		\coordinate (newOrigin) at (-\sinZ*\cosY*\offsetO,\sinZ*\sinY*\offsetO,\cosZ*\offsetO);
		\tdplotsetrotatedcoordsorigin{(newOrigin)}
		
		\tdplotsetrotatedcoords{-\angleY}{-\angleZ}{0}

        \draw[cube,tdplot_rotated_coords] (-1,-1,-2) -- (-1,1,-2) -- (1,1,-2) -- (1,-1,-2) -- cycle;     
        \draw[cube,tdplot_rotated_coords] (1,-1,-2) -- (1,1,-2) -- (1,1,0) -- (1,-1,0) -- cycle;
        \draw[cube,tdplot_rotated_coords] (-1,-1,-2) -- (1,-1,-2) -- (1,-1,0) -- (-1,-1,0) -- cycle;
        \draw[cube,tdplot_rotated_coords] (-1,-1,-2) -- (-1,1,-2) -- (-1,1,0) -- (-1,-1,0) -- cycle;
        \draw[cube,tdplot_rotated_coords] (-1,1,-2) -- (1,1,-2) -- (1,1,0) -- (-1,1,0) -- cycle;
        \draw[cube,tdplot_rotated_coords] (-1,-1,0) -- (-1,1,0) -- (1,1,0) -- (1,-1,0) -- cycle;

        \draw[cubepurple,tdplot_rotated_coords] (-0.8,-0.8,0) -- (-0.8,0.8,0) -- (0.8,0.8,0) -- (0.8,-0.8,0) -- cycle;        
        \draw[cubepurple,tdplot_rotated_coords] (0.8,-0.8,0) -- (0.8,0.8,0) -- (0.8,0.8,1) -- (0.8,-0.8,1) -- cycle;
        \draw[cubepurple,tdplot_rotated_coords] (-0.8,-0.8,0) -- (0.8,-0.8,0) -- (0.8,-0.8,1) -- (-0.8,-0.8,1) -- cycle;
        \draw[cubepurple,tdplot_rotated_coords] (-0.8,-0.8,0) -- (-0.8,0.8,0) -- (-0.8,0.8,1) -- (-0.8,-0.8,1) -- cycle;
        \draw[cubepurple,tdplot_rotated_coords] (-0.8,0.8,0) -- (0.8,0.8,0) -- (0.8,0.8,1) -- (-0.8,0.8,1) -- cycle;
        \draw[cubepurple,tdplot_rotated_coords] (-0.8,-0.8,1) -- (-0.8,0.8,1) -- (0.8,0.8,1) -- (0.8,-0.8,1) -- cycle;

\end{tikzpicture}
\tikzexternalenable

\caption{Two solid bodies example --- Experimental setup to study independence of spatial orientation.}
\label{fig:ch5_twosolidbodiesexample}
\end{center}
\end{figure}
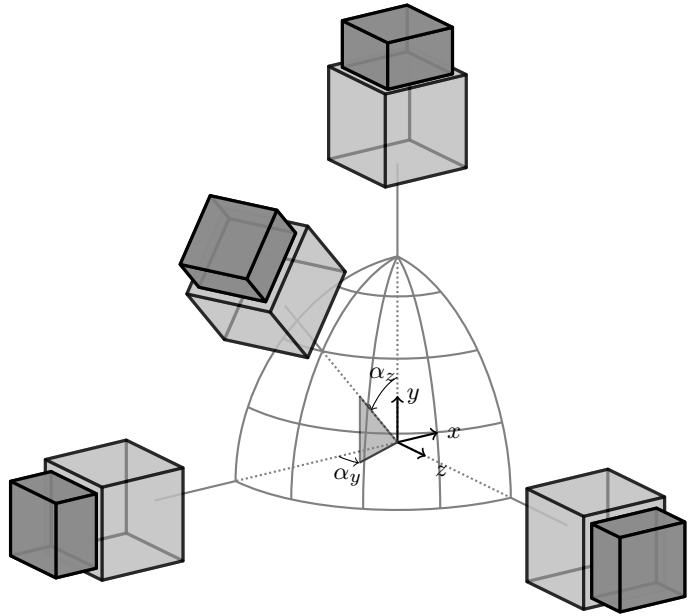

To investigate a potential impact of the contact formulation in different coordinate systems (cartesian coordinates for the structural degrees of freedom and normal-tangential coordinates for the Lagrange multipliers) on the linear solver, we perform a similar experiment as introduced in \cite{wiesner2017} and rotate the example setup around $\angleYtext$ and $\angleZtext$ as shown in Figure \ref{fig:ch5_twosolidbodiesexample}.
We expect the number of linear iterations to be independent of the rotation angles $\angleYtext$ and $\angleZtext$, since the underlying physics do not change. Any dependency of the linear solver on $\angleYtext$ and $\angleZtext$ would be a result of purely numerical effects and would turn out highly problematic for the iterative solution of large and complex contact problems.
For reasons of symmetry, it is sufficient to vary $\angleYtext$ and $\angleZtext$ within $0\leq \angleYtext$, $\angleZtext \leq \frac{\pi}{2}$.

\subsubsection{Discretization}
The spatial discretization is based on a $10\times 10\times 10$ mesh for each solid block
with altogether $6000$ displacement degrees of freedom and $300$ Lagrange multipliers
modeling the contact coupling constraints for the $10\times 10$ slave nodes at the contact interface (see Figure \ref{fig:ch5_twosolidbodiesexamplesetupsurfaces}). The simulation runs for $40$ time steps with a time step size of $0.01s$ on $4$ processors. After $6$ time steps ($\ttime=0.06s$) both bodies come into contact and are deformed. We assume frictionless contact here. With this example we reduce the contact-specific effects (such as the contact search based on an active set strategy) to a minimum, such that one can focus on the linear solvers. That is, the contact zone is not changing once the two solid bodies are in contact.

\subsubsection{Stopping criteria}
The nonlinear iteration inside each time step stops if either $\norm[e]{\dispi}<10^{-8}$ holds for the Newton increment of the displacement degrees of freedom,
or alternatively, if the conditions
\begin{equation}
\norm[e]{\res^{\disp}_{\nonlinstep}}<10^{-6}\wedge \norm[e]{\res^{\lagvec}_{\nonlinstep}}<10^{-4}
\end{equation} 
hold for the nonlinear residuals $\res^{\disp}_\nonlinstep$ and $\res^{\lagvec}_{\nonlinstep}$ in \eqref{eq:ch6_finalsaddlepointsystem} after applying $\nonlinstep$ Newton iterations.
Thereby, $\norm[e]{\bullet}$ denotes the Euclidian vector norm.
Those stopping criteria for the nonlinear solver are chosen to result in the same number of nonlinear iterations in each time step for an easier comparison of the linear solver behavior.

Within each Newton iteration, the saddle point system \eqref{eq:ch6_finalsaddlepointsystem} is solved iteratively using a preconditioned GMRES method
with a 3-level AMG preconditioner as described in Section \ref{sec:AMGForContact}.
The iterative process for the linear system is considered to be converged, if it is
\begin{equation}
\norm[e]{\frac{\res^\linstep}{\res^0}} < 10^{-8}
\label{eq:ch7_linearstoppingcriterion}
\end{equation}
for the full residual vector $\res^\linstep=\begin{bmatrix}\res^{\disp}\\ \res^{\lagvec}\end{bmatrix}$ in the linear iteration step $\linstep$. Here, the subscript $\nonlinstep$ for the nonlinear Newton iteration is dropped.

In this work, we focus on the behavior of the linear solver. 
Therefore, a fixed stopping criterion for all tested variants is chosen in \eqref{eq:ch7_linearstoppingcriterion}. 
This allows the comparison of different preconditioning techniques including their effect on the linear solution strategy. 
For real world problems, and especially for coupled multiphysics problems, the task of choosing appropriate stopping criteria for both the nonlinear and linear solver turns out to be quite challenging. Usually, one would choose a combination of different (length-scaled) norms for the partial vectors $\res^{\disp}$ and $\res^{\lagvec}$. In order to reduce the solver time in the inner linear solver, it is recommended to adapt the linear (relative) solver tolerance according to the residual norms of the outer nonlinear solver.

\begin{table}
\caption{Two solid bodies example ---
Average (maximum) number of linear GMRES iterations per nonlinear iteration (over all $40$ time steps)
for different combinations of rotation angles $\angleYtext$ and $\angleZtext$.
As preconditioner, a 3 level AMG method (\paamg~+~\paamg, minimum aggregate size: $6$ nodes) is used with different level smoothers.
Within the block smoothers on all multigrid levels, symmetric Gauss--Seidel is used for the structural degrees of freedoms and ILU for the Lagrange multipliers.}
\label{tab:april2013_6_BS}
\begin{subtable}[b]{\textwidth}
\caption{\textbf{Level smoother:} 3 CheapUzawa($0.7$)~with $1$ SGS ($0.7$) + ILU(0)}
\label{tab:april2013_uzawa}
\begin{center}
\scalebox{0.95}{%
\renewcommand{\arraystretch}{1.5}
\begin{tabular}{rr|ccccc|} \cline{3-7}
& & \multicolumn{5}{c|}{$\alpha_y$} \\ 
 & & $0$ & $\frac{1}{8}\pi$ & $\frac{1}{4}\pi$ & $\frac{3}{8}\pi$ & $\frac{1}{2}\pi$ \\ \hline 
\multicolumn{1}{|c}{\multirow{5}{*}{\rotatebox{90}{$\alpha_z$}}} & $0$ & 26.2 (30) & 26.0 (30) & 24.6 (29) & 24.6 (30) & 26.0 (29)\\ 
\multicolumn{1}{|c}{}& $\frac{1}{8}\pi$ & 26.8 (30) & 25.8 (37) & 25.3 (30) & 24.7 (30) & 27.3 (31)\\ 
\multicolumn{1}{|c}{}& $\frac{1}{4}\pi$ & 25.9 (31) & 25.1 (32) & 25.3 (30) & 28.7 (40) & 26.0 (31)\\ 
\multicolumn{1}{|c}{}& $\frac{3}{8}\pi$ & 25.9 (31) & 25.0 (30) & 24.9 (30) & 25.5 (33) & 25.3 (29)\\ 
\multicolumn{1}{|c}{}& $\frac{1}{2}\pi$ & 26.1 (30) & 25.4 (32) & 25.5 (30) & 26.8 (31) & 26.0 (30)\\ 
\hline 
 \end{tabular}%
  \renewcommand{\arraystretch}{1.0}
 } 
 \end{center}
\end{subtable}\\[0.03\textheight] %
\begin{subtable}[b]{\textwidth}
\caption{\textbf{Level smoother:} 3 CheapBraessSarazin($1.9$) with ILU($0$))}
\label{tab:april2013_bs}
\begin{center}
\scalebox{0.95}{%
\renewcommand{\arraystretch}{1.5}
\begin{tabular}{rr|ccccc|} \cline{3-7}
& & \multicolumn{5}{c|}{$\alpha_y$} \\ 
 & & $0$ & $\frac{1}{8}\pi$ & $\frac{1}{4}\pi$ & $\frac{3}{8}\pi$ & $\frac{1}{2}\pi$ \\ \hline 
\multicolumn{1}{|c}{\multirow{5}{*}{\rotatebox{90}{$\alpha_z$}}} & $0$ & 29.9 (37) & 29.6 (36) & 27.9 (33) & 30.7 (38) & 29.7 (37)\\ 
\multicolumn{1}{|c}{}& $\frac{1}{8}\pi$ & 41.2 (59) & 43.1 (68) & 42.0 (58) & 43.6 (63) & 41.4 (59)\\ 
\multicolumn{1}{|c}{}& $\frac{1}{4}\pi$ & 56.8 (82) & 64.8 (86) & 68.6 (95) & 64.3 (86) & 54.9 (75)\\ 
\multicolumn{1}{|c}{}& $\frac{3}{8}\pi$ & 40.9 (60) & 55.5 (74) & 73.1 (102) & 111.1 (141) & 41.3 (61)\\ 
\multicolumn{1}{|c}{}& $\frac{1}{2}\pi$ & 29.9 (37) & 41.0 (55) & 56.8 (79) & 41.3 (61) & 29.7 (36)\\ 
\hline 
 \end{tabular}%
   \renewcommand{\arraystretch}{1.0}
 } 
 \end{center}
\end{subtable}\\[0.03\textheight] %
\begin{subtable}[b]{\textwidth}
\caption{\textbf{Level smoother:} 3 \cheapsimplec($0.7$)~with $1$ SGS ($0.7$) + ILU(0)}
\label{tab:april2013_simplec}
\begin{center}
\scalebox{0.95}{%
\renewcommand{\arraystretch}{1.5}
\begin{tabular}{rr|ccccc|} \cline{3-7}
& & \multicolumn{5}{c|}{$\alpha_y$} \\ 
 & & $0$ & $\frac{1}{8}\pi$ & $\frac{1}{4}\pi$ & $\frac{3}{8}\pi$ & $\frac{1}{2}\pi$ \\ \hline 
\multicolumn{1}{|c}{\multirow{5}{*}{\rotatebox{90}{$\alpha_z$}}} & $0$ & 20.5 (26) & 19.3 (21) & 18.8 (23) & 19.3 (26) & 19.3 (20)\\ 
\multicolumn{1}{|c}{}& $\frac{1}{8}\pi$ & 20.1 (22) & 19.7 (27) & 20.0 (24) & 19.8 (21) & 21.2 (25)\\ 
\multicolumn{1}{|c}{}& $\frac{1}{4}\pi$ & 20.1 (25) & 19.9 (23) & 20.1 (23) & 22.0 (26) & 20.8 (30)\\ 
\multicolumn{1}{|c}{}& $\frac{3}{8}\pi$ & 19.7 (22) & 19.8 (22) & 19.8 (23) & 20.0 (26) & 19.7 (23)\\ 
\multicolumn{1}{|c}{}& $\frac{1}{2}\pi$ & 19.3 (20) & 19.6 (22) & 20.4 (27) & 20.7 (25) & 20.5 (28)\\ 
\hline 
 \end{tabular}%
    \renewcommand{\arraystretch}{1.0}
 } 
 \end{center}
\end{subtable}
\end{table}


\begin{table}
\caption{Two solid bodies example ---
Average (maximum) number of linear GMRES iterations per nonlinear iteration (over all $40$ time steps) for different rotation angles $\angleYtext$ and $\angleZtext$.
As preconditioner, a 3 level AMG method (minimum aggregate size: $6$ nodes) is used with different variants of \cheapsimplec.
Within the \cheapsimplec~method~on all multigrid levels, symmetric Gauss-Seidel sweeps are used for the structural degrees of freedoms and ILU for the Lagrange multipliers.}
\label{tab:ch6_twoblockspaamgvssaamg}

\begin{subtable}[b]{\textwidth}
\caption{\textbf{Level smoother:} 1 \cheapsimplec($0.7$)~with $1$ SGS ($0.7$) + ILU(0)}
\label{tab:CheapSIMPLEComparisonOne}
\begin{center}
\scalebox{0.95}{%
\renewcommand{\arraystretch}{1.5}
\begin{tabular}{rr|ccccc|} \cline {3-7}
& & \multicolumn{5}{c|}{$\alpha_y$} \\ 
 & & $0$ & $\frac{1}{8}\pi$ & $\frac{1}{4}\pi$ & $\frac{3}{8}\pi$ & $\frac{1}{2}\pi$ \\ \hline 
\multicolumn{1}{|c}{\multirow{5}{*}{\rotatebox{90}{$\alpha_z$}}} & $0$ & 37.5 (57) & 35.6 (43) & 35.2 (46) & 36.1 (44) & 35.7 (45)\\ 
\multicolumn{1}{|c}{} & $\frac{1}{8}\pi$ & 36.4 (45) & 37.6 (44) & 38.5 (47) & 36.3 (43) & 38.1 (53)\\ 
\multicolumn{1}{|c}{} & $\frac{1}{4}\pi$ & 39.1 (50) & 37.8 (47) & 37.0 (47) & 39.2 (48) & 37.0 (44)\\ 
\multicolumn{1}{|c}{} & $\frac{3}{8}\pi$ & 36.3 (43) & 37.0 (48) & 37.8 (46) & 37.2 (51) & 36.4 (43)\\ 
\multicolumn{1}{|c}{} & $\frac{1}{2}\pi$ & 36.2 (47) & 38.0 (51) & 39.8 (50) & 36.9 (50) & 36.6 (53)\\ 
\hline 
 \end{tabular}%
 \renewcommand{\arraystretch}{1.0}
 } 
 \end{center}
\end{subtable}\\[0.03\textheight]
\begin{subtable}[b]{\textwidth}
\caption{\textbf{Level smoother:} 1 \cheapsimplec($0.7$)~with $3$ SGS ($0.7$) + ILU(0)}
\label{tab:CheapSIMPLEComparisonTwo}
\begin{center}
\scalebox{0.95}{%
\renewcommand{\arraystretch}{1.5}
\begin{tabular}{rr|ccccc|} \cline {3-7}
& & \multicolumn{5}{c|}{$\alpha_y$} \\ 
 & & $0$ & $\frac{1}{8}\pi$ & $\frac{1}{4}\pi$ & $\frac{3}{8}\pi$ & $\frac{1}{2}\pi$ \\ \hline 
\multicolumn{1}{|c}{\multirow{5}{*}{\rotatebox{90}{$\alpha_z$}}} & $0$ & 29.5 (41) & 27.5 (34) & 28.4 (36) & 28.5 (37) & 27.1 (36)\\ 
\multicolumn{1}{|c}{} & $\frac{1}{8}\pi$ & 29.3 (37) & 30.1 (40) & 32.7 (40) & 28.8 (35) & 28.4 (38)\\ 
\multicolumn{1}{|c}{} & $\frac{1}{4}\pi$ & 30.0 (35) & 29.6 (37) & 28.5 (37) & 31.1 (38) & 29.2 (36)\\ 
\multicolumn{1}{|c}{} & $\frac{3}{8}\pi$ & 28.8 (36) & 27.8 (34) & 28.1 (34) & 29.1 (39) & 28.1 (34)\\ 
\multicolumn{1}{|c}{} & $\frac{1}{2}\pi$ & 28.5 (38) & 28.4 (34) & 29.7 (40) & 27.9 (35) & 27.1 (35)\\ 
\hline 
 \end{tabular}%
 \renewcommand{\arraystretch}{1.0}
 } 
 \end{center}
\end{subtable}\\[0.03\textheight]
\begin{subtable}[b]{\textwidth}
\caption{\textbf{Level smoother:} 3 \cheapsimplec($0.7$)~with $3$ SGS ($0.7$) + ILU(0)}
\label{tab:CheapSIMPLEComparisonThree}
\begin{center}
\scalebox{0.95}{%
\renewcommand{\arraystretch}{1.5}
\begin{tabular}{rr|ccccc|} \cline {3-7}
& & \multicolumn{5}{c|}{$\angleYtext$} \\ 
 & & $0$ & $\frac{1}{8}\pi$ & $\frac{1}{4}\pi$ & $\frac{3}{8}\pi$ & $\frac{1}{2}\pi$ \\ \hline 
\multicolumn{1}{|c}{\multirow{5}{*}{\rotatebox{90}{$\angleZtext$}}} & $0$ & 15.5 (16) & 17.0 (19) & 16.9 (20) & 15.3 (16) & 15.2 (16)\\ 
\multicolumn{1}{|c}{} & $\frac{1}{8}\pi$ & 16.2 (17) & 16.1 (17) & 16.3 (17) & 15.9 (17) & 16.1 (17)\\ 
\multicolumn{1}{|c}{} & $\frac{1}{4}\pi$ & 16.0 (17) & 16.0 (17) & 16.0 (17) & 15.6 (18) & 15.6 (16)\\ 
\multicolumn{1}{|c}{} & $\frac{3}{8}\pi$ & 15.7 (17) & 15.9 (17) & 15.9 (17) & 15.5 (19) & 15.6 (16)\\ 
\multicolumn{1}{|c}{} & $\frac{1}{2}\pi$ & 15.6 (16) & 15.8 (16) & 15.7 (16) & 15.4 (16) & 15.1 (16)\\ 
\hline 
 \end{tabular}%
 \renewcommand{\arraystretch}{1.0}
 } 
 \end{center}
\end{subtable}\hspace{0.02\textwidth}%
\end{table}

\subsubsection{Results}
\label{sec:example1results}
First, the effect of the different saddle point smoothers on the number of linear iterations is explored.
Table~\ref{tab:april2013_6_BS} summarizes the average number of linear iterations per time step for different combinations of the rotation angles $\angleYtext$ and $\angleZtext$.
The numbers in brackets denote the maximum number of linear iterations needed for solving one linear system during the full simulation,
roughly indicating the variation of the number of linear iterations within the simulation.
For the CheapUzawa smoother, the number of iterations does not show a dependence on the rotation angles $\angleYtext$ and $\angleZtext$.
Comparing the numbers from Table \ref{tab:april2013_uzawa} with the results for the CheapBraessSarazin smoother in Table \ref{tab:april2013_bs},
the CheapBraessSarazin smoother heavily suffers from the worse approximation of the displacement degrees of freedom
using one internal hard-coded Jacobi sweep (\cf~Section \ref{sec:ch6_braessarazin}).
The resulting iteration numbers show an obvious dependency on the rotation angles.
With a \cheapsimplec~block smoother, the number of iterations is lower than for the CheapUzawa smoother and independent from $\angleYtext$ and $\angleZtext$
when compared with the CheapBraessSarazin smoother (see Table \ref{tab:april2013_simplec}).
So, the linear solver has some benefit from the two-way coupling of displacements and Lagrange multipliers within the AMG preconditioner.
Compared to the Uzawa smoother, the additional computational costs for the \cheapsimplec~method are very low with only one additional matrix-vector product by $\apprx{\Ablock}^{-1}\Boneblock$ per iteration.
Therefore, \cheapsimplec~is the preferred level smoother for our further experiments
with some cheap approximations for the internal single fields using some sweeps with a (symmetric) Gauss--Seidel (SGS) method  for the structural block or incomplete LU factorization (ILU) for the Lagrange multipliers.

Table \ref{tab:ch6_twoblockspaamgvssaamg} illustrates how the number of \cheapsimplec~coupling iterations
and the quality of the single field smoothing methods within the \cheapsimplec~smoother
affect the number of linear iterations.
Improving the quality of the Schur complement approximations within \cheapsimplec~(see Tables~\ref{tab:CheapSIMPLEComparisonOne} vs. \ref{tab:CheapSIMPLEComparisonTwo})
as well as increasing the number of \cheapsimplec~coupling iterations (see Table~ \ref{tab:CheapSIMPLEComparisonThree})
unsurprisingly reduces the number of linear solver iterations.
Aside from the concrete parameter choices for the level smoother,
one can even further reduce the number of linear iterations with a reasonable transfer operator smoothing strategy
for the displacement block, {\eg} as indicated in~\eqref{eq:ProlongatorSmoothing}. 

By intention, we do not report solver timings, since this example is too small to perform reasonable time measurements,
especially when using~$4$ processors for altogether only $6300$ degrees of freedom. 

The intention of this example is to compare typical saddle point smoothers within a fully coupled AMG preconditioner. One can observe the expected behavior that increasing the number of smoothing sweeps reduces the number of linear GMRES iterations. However, in practice, the variant with a smaller number of GMRES iterations may not always be the fastest method.
This example shows that the proper choice of block level smoothing is essential for the overall performance of a saddle point multigrid method.
The particular choice of the block smoothing method gives the user full control over the quality of the coupling with field-specific parameters and allows for fine-grained adaptions and problem-specific optimizations.


With the experience from this example one can choose efficient level smoothers which provide results independent from the exact geometric configuration.


\subsection{Weak scaling behavior}
\label{sec:WeakScaling}


To assess the behavior of the proposed multilevel preconditioner applied to large-scale examples,
we now report a weak scaling study.
To exclude side effects (such as changes in the contact active set) and to fully focus on the behavior of the preconditioner and iterative linear solver,
we study a simplified and linear contact problem.

\subsubsection{Setup}
\label{sec:ScalingStudySetup}

We consider a small block (dimensions $0.8 m \times 0.8 m \times 0.4 m$) and a slightly bigger block (dimensions $1.0 m \times 1.0 m \times 0.5 m$),
where contact will occur between the large faces of both blocks.
To reduce the complexity of the contact problem and to exclude nonlinearities due to changes in the contact active set,
the faces opposite to the contact interface are fixed with Dirichlet boundary conditions,
while the blocks initially penetrate each other at the contact interface by $0.001$.
The smaller block acts as the slave side and its entire contact area is initialized as ``active''.
Application of the contact algorithms will then result in a slight compression of both blocks, such that the initial penetration vanishes.
This problem setup allows to distill the performance of the AMG preconditioner under uniform mesh refinement and weak scaling conditions.

Both blocks use a Neo--Hooke material with Young's modulus~$E=10$ MPa and Poisson's ratio $\nu=0.3$.
Denoting the mesh refinement factor with~$\meshfac$,
both blocks are discretized with $2\meshfac$ linear hexahedral elements along their longer edges and $\meshfac$ elements along the shorter edges.
The Lagrange multiplier field is discretized with standard shape functions, {\ie}, linear Lagrange polynomials.

\subsubsection{Solver and preconditioner settings}
\label{sec:ScalingStudySolverSetup}
As linear solver, a preconditioned GMRES method is used with a fully coupled multigrid approach as described in Section \ref{sec:MultigridForMultiphysicsBlock}.
The convergence criterion for the linear GMRES solver is set to
\begin{equation}
\norm[e]{\frac{\res^\linstep}{\res^0}} < 10^{-8}
\end{equation}
for the full residual vector $\res^\linstep=\begin{bmatrix}\res^{\disp}\\ \res^{\lagvec}\end{bmatrix}$ in the linear iteration step $\linstep$.
For the segregated block transfer operator as introduced in \eqref{eq:ch6_segregatedtransfers},
we combine smoothed aggregation (\saamg) with a prolongator smoothing factor~$\omega = 4/3$ for the displacement aggregates
and plain non-smoothed (\paamg) transfer operators for the Lagrange multiplier aggregates.
The restriction operators are built as the transposed of the prolongation operators.
Coarsening stops when the total number of rows in the saddle-point system drops below $5000$.

Following the guidance from Section \ref{sec:ComparisonOfContactSmoothers} and the findings from Section \ref{sec:example1results},
we apply $3$ sweeps of {\cheapsimple}($0.8$) as a level smoother, where both the predictor and corrector step are approximated by $1$ sweep of SGS each.
We use the same level smoother layout on all multigrid levels except of the coarsest level.
For the coarsest level, we compare two variants:
a direct solver (marked as ``LU'', requiring an expensive merging of the block matrix into a regular sparse matrix format and a subsequent LU factorization)
vs. the {\cheapsimple} level smoother as on all other multigrid levels (marked as ``{\cheapsimple}'').
We will study the impact of the coarse solver on scaling behavior, iteration counts, as well as AMG setup and V-cycle timings.

\subsubsection{Results}

For the weak scaling study, we target the load per rank to be $50k$ displacement DOFs.
A uniform mesh refinement as outlined in Table~\ref{tab:ScalingStudyMeshRefinement} is performed
with the finest mesh consisting of roughly $23.9$ million unknowns in the saddle-point system.
\begin{table}
\centering
\caption{Weak scaling study --- mesh refinement and hierarchy details}
\label{tab:ScalingStudyMeshRefinement}
\begin{tabular}{c|c|ccc|c|ccc}
$\nproc$ & $\meshfac$ & $\nDofPrimal$ & $\nDofDual$ & $\nDofTotal$ & $\nDofPrimal$/$\mathrm{proc}$ & $\numlevels$ & $\nDofTotalCoarsest$ & $\operatorComplexity$\\
\hline
$4$ & $20$ & $211806$ & $5043$ & $216849$ & $52951.5$ & $3$ & $948$ & $1.16$\\
$8$ & $25$ & $405756$ & $7803$ & $413559$ & $50719.5$ & $3$ & $1275$ & $1.18$\\
$16$ & $32$ & $836550$ & $12675$ & $849225$ & $52284.4$ & $3$ & $2661$ & $1.23$\\
$24$ & $36$ & $1183038$ & $15987$ & $1199025$ & $49293.2$ & $3$ & $3426$ & $1.24$\\
$48$ & $46$ & $2439018$ & $25947$ & $2464965$ & $50812.9$ & $4$ & $393$ & $1.26$\\
$72$ & $52$ & $3505950$ & $33075$ & $3539025$ & $48693.8$ & $4$ & $570$ & $1.27$\\
$96$ & $58$ & $4845906$ & $41067$ & $4886973$ & $50478.2$ & $4$ & $762$ & $1.28$\\
$144$ & $66$ & $7110978$ & $53067$ & $7164045$ & $49381.8$ & $4$ & $1071$ & $1.28$\\
$192$ & $73$ & $9594396$ & $64827$ & $9659223$ & $49970.8$ & $4$ & $1401$ & $1.29$\\
$240$ & $79$ & $12134880$ & $75843$ & $12210723$ & $50562.0$ & $4$ & $1713$ & $1.29$\\
$288$ & $84$ & $14566110$ & $85683$ & $14651793$ & $50576.8$ & $4$ & $2103$ & $1.30$\\
$336$ & $88$ & $16729686$ & $93987$ & $16823673$ & $49790.7$ & $4$ & $2421$ & $1.30$\\
$384$ & $92$ & $19097550$ & $102675$ & $19200225$ & $49733.2$ & $4$ & $2709$ & $1.30$\\
$432$ & $96$ & $21678918$ & $111747$ & $21790665$ & $50182.7$ & $4$ & $3090$ & $1.30$\\
$480$ & $99$ & $23760600$ & $118803$ & $23879403$ & $49501.2$ & $4$ & $3884$ & $1.30$\\
\end{tabular}
\end{table}
Therein, $\nproc$ denotes the number of MPI ranks,
$\meshfac$ the mesh refinement factor introduced in \secref{sec:ScalingStudySetup},
$\nDofPrimal$, $\nDofDual$, and $\nDofTotal$ the overall number of displacement, Lagrange multiplier, and total number of unknowns,
$\nDofPrimal$/$\mathrm{proc}$ the average load per MPI rank.
The mesh refinement factor~$\meshfac$ is chosen such that the targeted average number of displacement unknowns per MPI rank is met.
The number of levels $\numlevels$, the actual size of the coarse level system $\nDofTotalCoarsest$
as well as the operator complexity~$\operatorComplexity$
(defined as the ratio of non-zero entries of all multigrid level matrices~$A_{\level}$ with $\level = 0, \hdots, \numlevels - 1$
and the number of non-zeros on the finest level)
are reported in the last three columns of Table~\ref{tab:ScalingStudyMeshRefinement}.
The operator complexity~$\operatorComplexity$ slightly increases with larger problem sizes,
but does not grow beyond~$\operatorComplexity\approx 1.30$.

\begin{figure}
\includegraphics[width=1.0\textwidth]{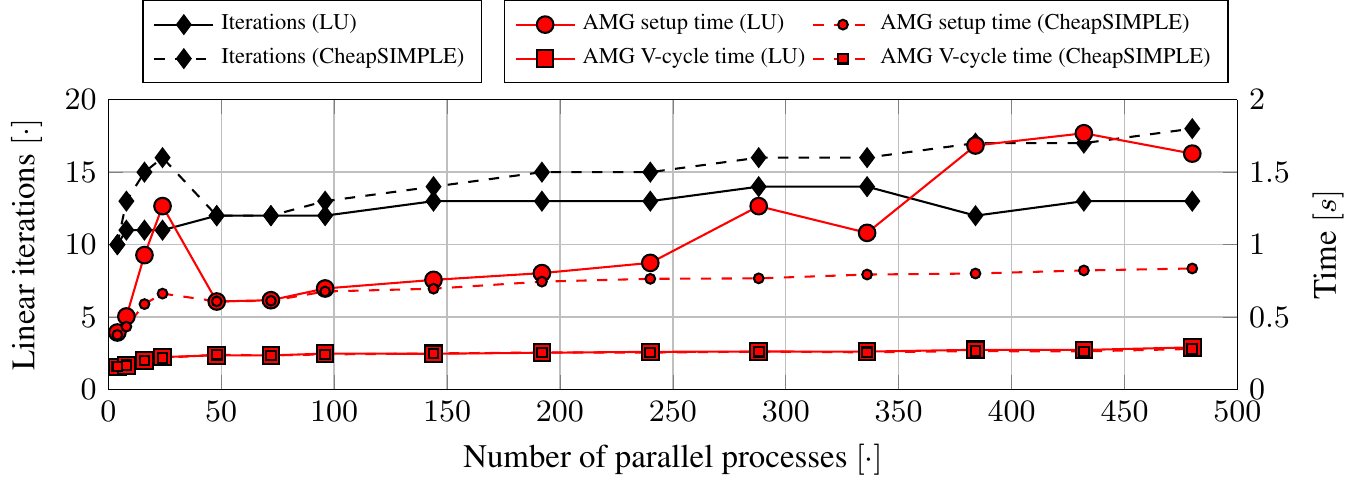}

\caption{Weak scaling study --- Linear solver iterations and AMG setup and V-cycle timings demonstrate proper weak scaling behavior.
The type of coarse solver is indicated in parentheses.}
\label{fig:ScalingDiagram}
\end{figure}

The scaling study is run on our in-house cluster (20 nodes with 2x Intel Xeon Gold 5118 (Skylake-SP) 12 core CPUs, 480 cores in total, Mellanox Infiniband Interconnect).
Solver performance in terms of the number of GMRES iterations as well as wall clock time spent in the AMG preconditioner construction and its application
({\ie} sweeping through the V-cycle once per GMRES iteration)
are summarized in Figure~\ref{fig:ScalingDiagram}.
\tikzset{external/figure name={scaling_study}}
\tikzset{external/figure name/.add={}{_diagram}}
With respect to iteration counts, both the direct solver as well as the block smoother
yield the GMRES iteration count to be fairly independent of the overall mesh size,
especially when taking the slight deviations of the load per core as detailed in Table~\ref{tab:ScalingStudyMeshRefinement} into account.

Using a direct solver on the coarsest multigrid level perfectly deals with all left-over error modes that are not properly handled by the block smoothers on the finer multigrid levels. Therefore, the ``LU'' variant results in a lower iteration count and is less sensitive to the actual size of the coarse level system.
In contrast, when using the {\cheapsimple} block smoother as coarse level solver, we observe a somewhat higher iteration count together with an obvious impact of the number of multigrid levels on the iteration count (\eg, when moving from a 3-level hierarchy on 24 MPI ranks to a 4-level hierarchy on 48 MPI ranks). Increasing the number of multigrid levels helps to compensate the lack of coupling of the {\cheapsimple} block smoother compared to a direct solve on the coarsest level.

We show the {\cheapsimple} variants for the coarse solve since this variant in practice might be superior with respect to the overall performance:
Regarding AMG setup time, the cost for the direct coarse solver becomes more expensive with a growing coarse system size,
since (i) more matrix entries need to be moved from a block sparse matrix layout to a regular sparse matrix layout
and (ii) a larger matrix needs to be factorized.
For the {\cheapsimple} block smoother as coarse level solver, no additional communication is required on the coarsest level.
Hence, the mild slope in the graph for preconditioner setup time for the {\cheapsimple} coarse level solver originates only from the Galerkin product during hierarchy construction.
In terms of preconditioner application, {\ie} time spent in the AMG V-cycle,
both methods are almost identical and scale very well.

While proper weak scalability up to $23.9$ million unknowns on 480 MPI ranks has been demonstrated,
the actual choice of the coarse solver in practical applications also depends on other factors such as the frequency of rebuilding the AMG hierarchy
or the balance of setup time, V-cycle time, and cost of the additional GMRES iterations, which all together impact the overall time to solution.
In the next example, we will put attention on effects for the linear solver caused by changes in the active set of contact nodes for larger problems.


\subsection{1000 deformable rings}
\label{sec:ch6_1000ringsexample}

Even though only a 2D example of moderate size,
the 1000 rings example comes with frequent changes in the contact active set and, thus, tests the robustness of the proposed multigrid methods.
Particularly, we are interested in the comparison of the fully coupled multigrid approach and the nested multigrid apparoch
as described in~\secref{sec:MultigridForMultiphysicsBlock}.

\subsubsection{Setup}
\begin{figure}
\begin{center}
\begin{subfigure}[b]{\textwidth}
\tikzexternaldisable
\begin{tikzpicture}
\node (agg15) at (0,0) {
\includegraphics[width=0.5\textwidth]{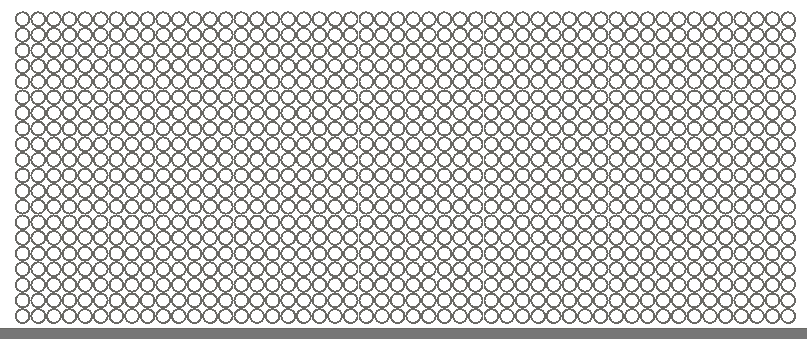}
};
\node [right=0.5cm of agg15] (closeagg15) {\framebox{
\includegraphics[width=0.45\textwidth]{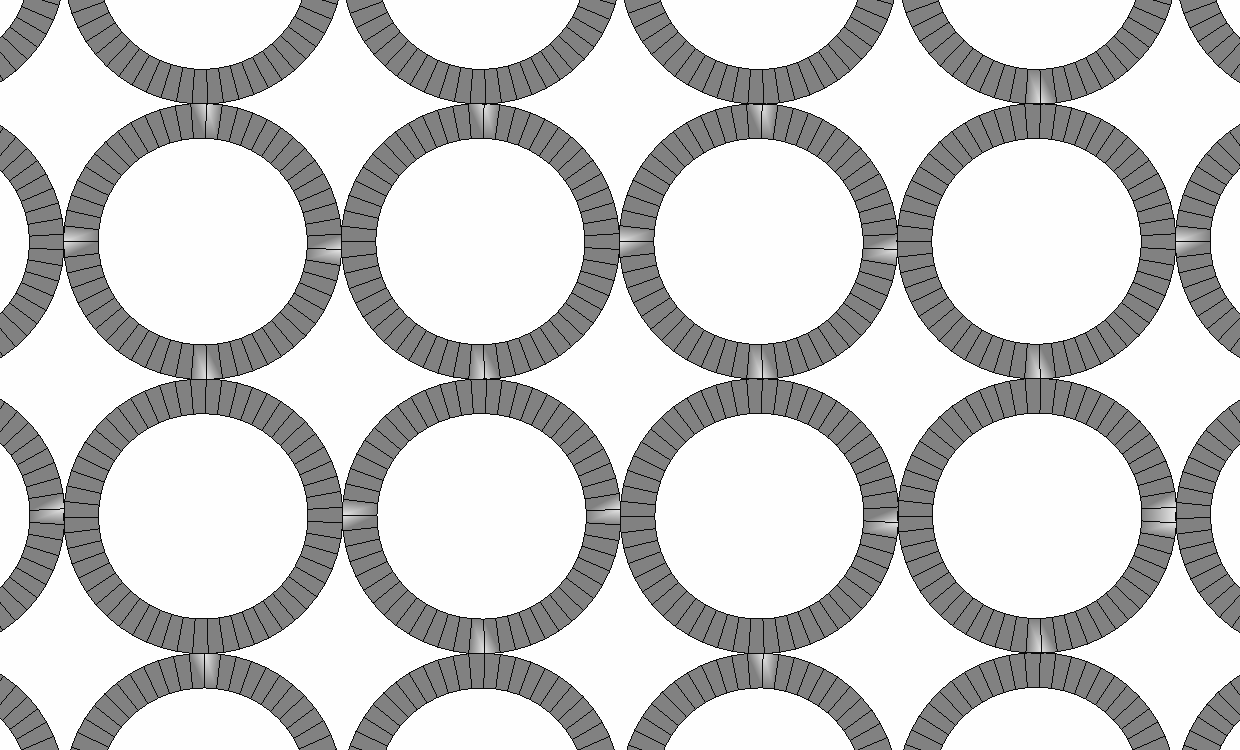} 
}
};
\draw (0.8,0.025) -- (closeagg15.west);
\draw (-0.03,-0.2) rectangle (0.7,0.25);
\end{tikzpicture}
\tikzexternalenable
\caption{Initial configuration with close-up view of the meshed deformable rings at $\ttime=0.0s$}
\label{fig:RingsInitialConfiguration}
\end{subfigure}
\scalebox{0.92}{
\begin{subfigure}[b]{0.49\textwidth}
\begin{tikzpicture}
\node (A) at (0.5\textwidth,3) {
\includegraphics[width=1.0\textwidth]{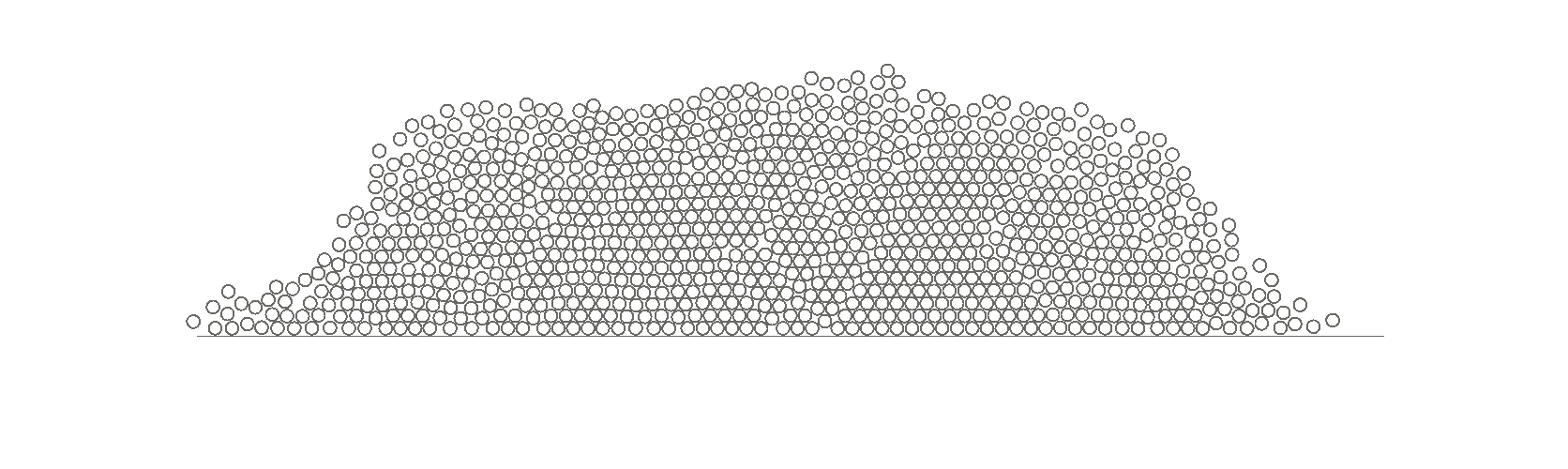}
};
\draw [fill=gray] (0.12\textwidth,2) rectangle (0.89\textwidth,2.2);
\end{tikzpicture}
\caption{Configuration at $\ttime=0.5s$}
\end{subfigure}
} 
\scalebox{0.92}{
\begin{subfigure}[b]{0.49\textwidth}
\begin{tikzpicture}
\node (A) at (0.5\textwidth,3) {
\includegraphics[width=1.0\textwidth]{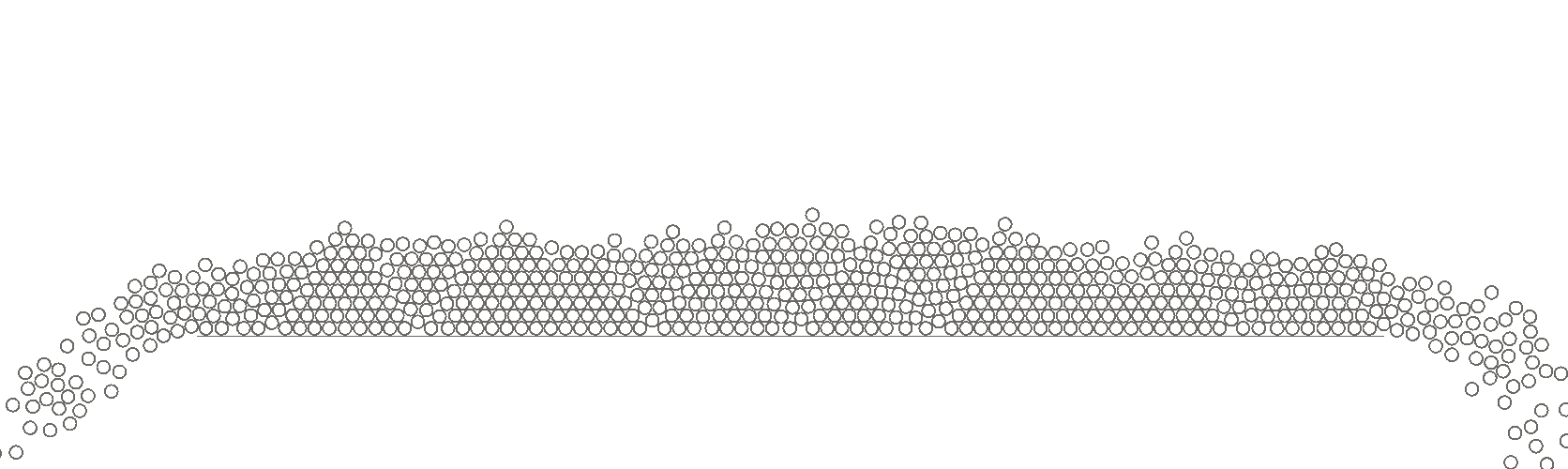}
};
\draw [fill=gray] (0.12\textwidth,2) rectangle (0.89\textwidth,2.2);
\end{tikzpicture}
\caption{Configuration at $\ttime=2.0s$}
\end{subfigure}
} 
\end{center}
\caption{1000 deformable rings --- Characteristic stages at different times.}
\label{fig:ch6_1000ringssetup}
\end{figure}
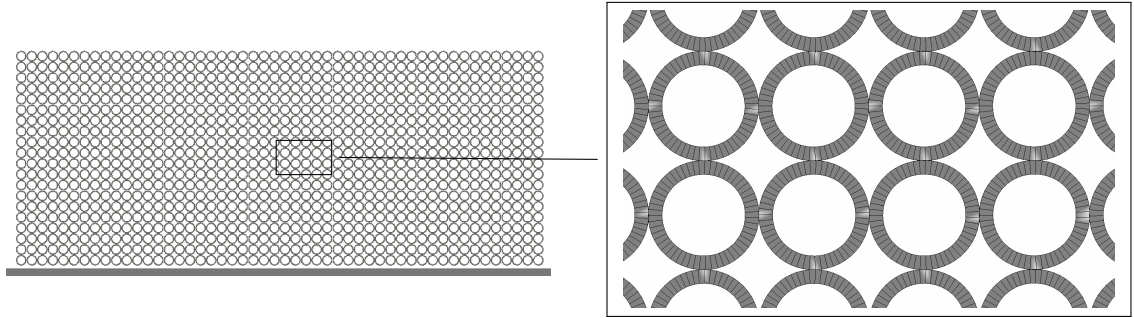
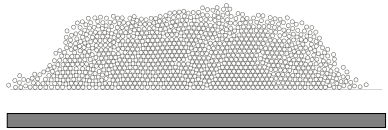
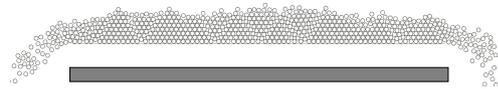

This example consists of $1000$ deformable rings (Neo-Hookean material with $\YoungModulus=210$ GPa, $\nu=0.3$ and $\refdensity=7.83\cdot 10^{-6}~\frac{kg}{m^3}$) arranged in a rectangle (see Figure~\ref{fig:ch6_1000ringssetup}).
A gravitational force is inducing an acceleration towards a rigid wall.
The simulated time extends to~$2.0s$ with a time step size of~$\Delta\ttime=0.0005s$, yielding $4000$ time steps in total.
The full mesh consists $110000$ nodes with $110$ nodes for each deformable ring.

\subsubsection{Stopping criteria}
\label{sec:ThousandRingsToppingCriteria}
In each time step, the nonlinear system is handled by a semi-smooth Newton method. 
As convergence criteria, we have chosen
\begin{equation}
\norm[e]{\dispi}<10^{-8}~\wedge~ \Bigl(\norm[e]{\res^{\disp}_{\nonlinstep}}<10^{-8} ~\wedge~ \norm[e]{\res^{\lagvec}_{\nonlinstep}}<10^{-6}\Bigr).
\label{eq:ch6_stoppingcriteriarings}
\end{equation}
Here, $\res^{\disp}_{\nonlinstep}$ and $\res^{\lagvec}_{\nonlinstep}$ denote the nonlinear residual for the displacement and Lagrange multiplier variables after $\nonlinstep$ Newton iterations, respectively.
Similarly, $\dispi$ denotes the solution increment for the displacement variables in the $i$-th Newton iteration.

For solving the linear systems arising during the simulation a GMRES solver is applied with different variants of AMG preconditioners listed in Table \ref{tab:ch6_amgvariants_1000rings}. 
The relative tolerance of convergence for the GMRES solver is set to $\norm[e]{\frac{\res^\linstep}{\res^0}} < 10^{-8}$ with $\res^\linstep=\begin{bmatrix}\res^{\disp}\\ \res^{\lagvec}\end{bmatrix}$ being the full residual vector in the linear iteration step $\linstep$. Here, the subscript $\nonlinstep$ for the nonlinear Newton step is dropped.

The stopping criteria~\eqref{eq:ch6_stoppingcriteriarings} for the nonlinear solver are carefully chosen in such a way that the simulations with all the tested preconditioner variants shown in Table \ref{tab:ch6_amgvariants_1000rings} always result in the same number of nonlinear iterations. This way we can directly compare the number of linear iterations of all tested solver variants which allow to draw some conclusions on the multigrid preconditioners. 


\subsubsection{Results}
Table \ref{tab:ch6_amgvariants_1000rings} provides an overview of the chosen preconditioner parameters for the level smoothers. For both the nested multigrid schemes and the fully coupled multigrid schemes we apply the level smoother on all multigrid levels including the coarsest level.
For each class of multigrid preconditioners,
only those variants are presented, that give the best overall timings and are able to accomplish the all $4000$ time steps of the full simulation.

The multigrid parameters are chosen to be the same for all preconditioner variants:
the minimum size of the aggregates is set to $6$ nodes for the two-dimensional problem
and the maximum coarse level size is set to $1000$ degrees of freedom, yielding a $3$ level multigrid method. 

For the fully coupled AMG variants, different transfer operator strategies are compared,
namely the non-smoothed (\paamg) transfer operators and the energy minimization approach with local damping parameters for transfer operator smoothing
denoted by \EMIN, {\cf}~\cite{sala2008}.
For the nested AMG variants, we compare transfer operators based on {\paamg} and {\saamg}.
All the simulations have been run on $16$ cores (spread over 2 Intel Xeon E5-2670 Octocore CPUs).



\begin{table}
\centering
\caption{1000 deformable rings --- Different AMG variants.}
\label{tab:ch6_amgvariants_1000rings}

\renewcommand{\arraystretch}{1.3}
\begin{tabular}{|l|l|} \hline
\multicolumn{2}{|c|}{Preconditioner type} \\ \hline
Full multigrid based methods & Nested multigrid based methods \\

\hline

\begin{minipage}{0.47\textwidth}
\vspace{1em}
\textbf{\paamg~(\cheapsimple)}\\
\begin{tabular}{ll}
Transfer operators: & \paamg \\
Level smoother: & 1 \cheapsimplec (0.8)\\
\enspace$-$~Pred. smoother: & 3 SGS ($0.8$) \\
\enspace$-$~Corr. smoother: & ILU ($0$)
\end{tabular}
\end{minipage}

&

\begin{minipage}{0.47\textwidth}
\vspace{1em}
\textbf{\cheapsimple~(\paamg)}\\
\begin{tabular}{ll}
Block prec.: & 1 \cheapsimplec (0.8)\\
\enspace$-$~Pred. smoother:  & AMG \\
$\qquad-$~Transfer op.:   & \paamg \\
$\qquad-$~Level sm.: & 1 SGS ($0.8$) \\
\enspace$-$~Corr. smoother:  & KLU
\end{tabular}
\end{minipage}

\\ \hline

\begin{minipage}{0.47\textwidth}
\vspace{1em}
\textbf{\EMIN~(\cheapsimple)}\\
\begin{tabular}{ll}
Transfer operators: & \EMIN\\
Level smoother: & 1 \cheapsimplec (0.8)\\
\enspace$-$~Pred. smoother: & 3 SGS ($0.8$) \\
\enspace$-$~Corr. smoother: & ILU ($0$)
\end{tabular}
\end{minipage}

&

\begin{minipage}{0.47\textwidth}
\vspace{1em}
\textbf{\cheapsimple~(\saamg)}\\
\begin{tabular}{ll}
Block prec.: & 1 \cheapsimplec (0.8)\\
\enspace$-$~Pred. smoother:  & AMG \\
$\qquad-$~Transfer op.:   & \saamg~($0.8$) \\
$\qquad-$~Level sm.: & 1 SGS ($0.8$) \\
\enspace$-$~Corr. smoother:  & KLU
\end{tabular}
\end{minipage}

\\ \hline

\end{tabular}
\renewcommand{\arraystretch}{1.0}

\end{table}

\begin{figure}
\begin{subfigure}[b]{1.0\textwidth}
\includegraphics[width=1.0\textwidth]{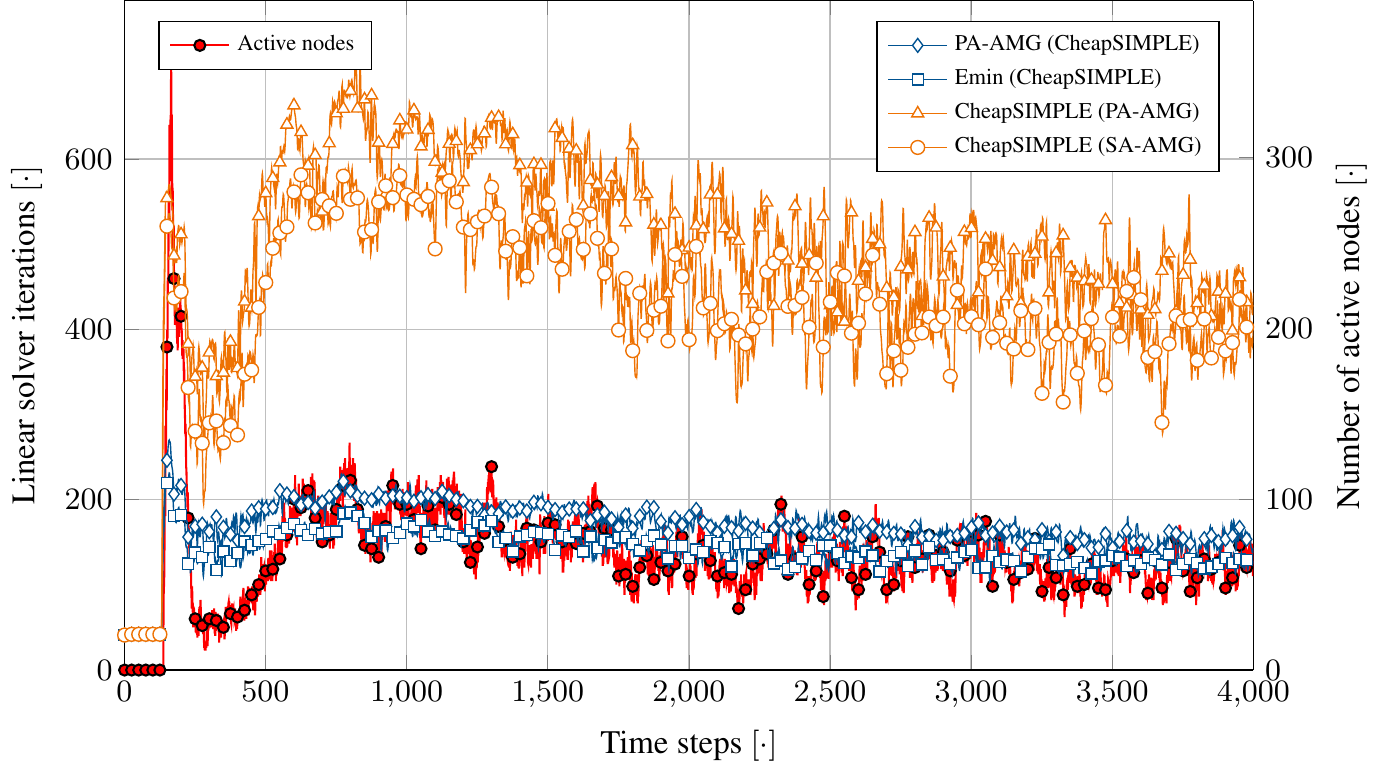}
\caption{Number of linear GMRES iterations for all nonlinear iterations over time steps for different preconditioner variants. The red curve shows the number of active nodes for each time step.}
\label{fig:ch6_ringsiterations}
\end{subfigure}\vspace{1cm}

\begin{subfigure}[b]{1.0\textwidth}
\includegraphics[width=1.0\textwidth]{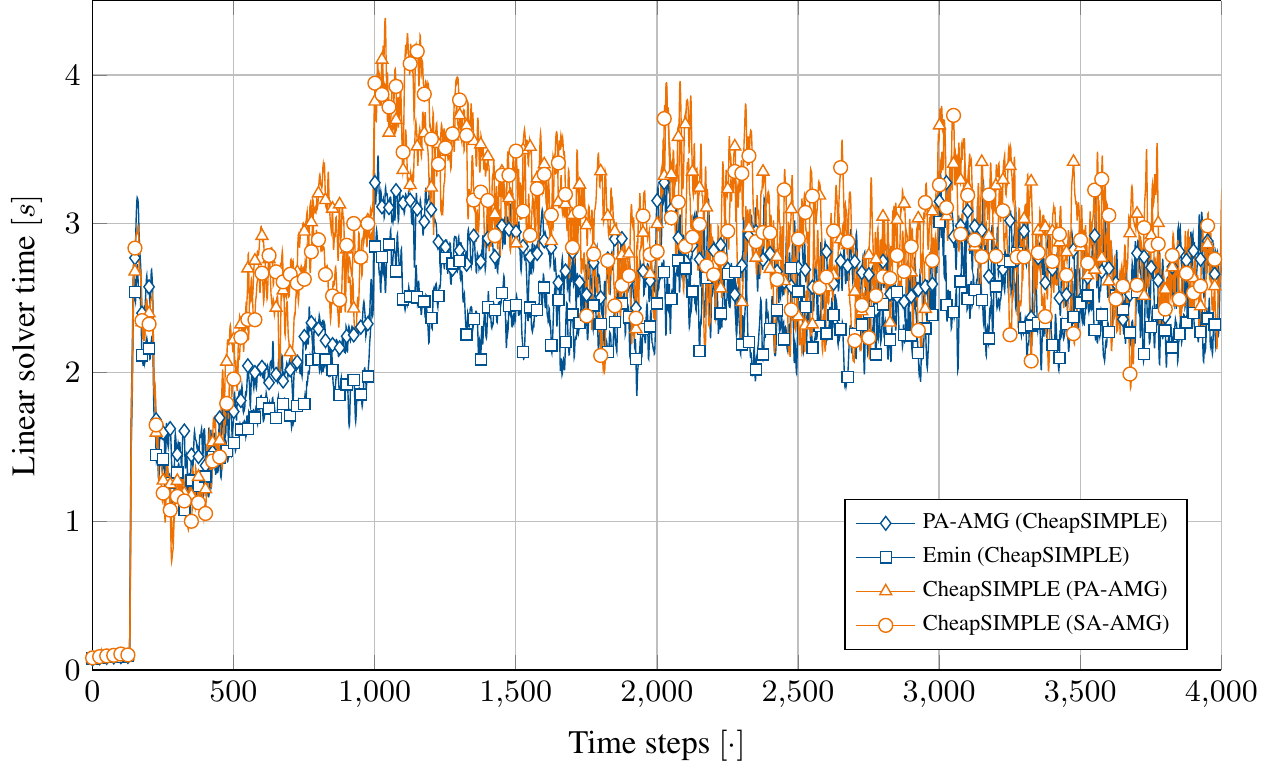}
\caption{Overall linear solver time over time steps.}
\label{fig:ch6_ringstimings}
\end{subfigure}

\caption{1000 collapsing rings example --- Results for different AMG preconditioner variants.}
\label{fig:ch6_results}
\end{figure}

Figure \ref{fig:ch6_ringsiterations} shows for each time step the overall number of linear iterations for all nonlinear iterations stacked onto each other.
Compared to the \cheapsimple~based methods, the fully coupled AMG variants need a significantly lower number of linear iterations to solve the problem in each time step.
This can be explained by the better approximation of the displacement degrees of freedom using $3$ instead of $1$ damped Gauss--Seidel sweeps. One can also see how the number of linear iterations correlates with the number of active contact nodes. Obviously, the fully coupled AMG variants are less sensitive to the changes in the number of active contact nodes than the \cheapsimple~based methods.
In Figure \ref{fig:ch6_ringstimings}, the overall linear solver time is given for each time step. Again, the fully coupled AMG variants perform better, although the difference is less pronounced
due to the higher computational effort of fully coupled AMG implementations.
The number of nonlinear iterations per time step is 2 for the initial phase without contact and then varies between 4 and 6, when contact is active.
It is the same for all preconditioner variants due to the particular choice of stopping criteria of the nonlinear solver (see Section \ref{sec:ThousandRingsToppingCriteria}).
Moreover, the number of nonlinear iterations is independent of the size of the active set.
 
\tikzset{external/figure name={8_rings1000_sp}}
\tikzset{external/figure name/.add={}{_history}}

Finally, overall timings for the linear solver including preconditioner setup are reported in Table \ref{tab:ch6_1000rings_overalltimings}.
The overall solver time only varies slightly for all given variants. 
The variants \EMIN~(\cheapsimple) and \cheapsimple~(\saamg) need more setup time due to the additional effort of prolongator smoothing.
\EMIN~(\cheapsimple) can fully amortize the additional setup cost during the solve phase due to coarse grid corrections, that respect the contact constraints,
while \cheapsimple~(\saamg) does not have this benefit and, thus, requires the largest overall solver time.
To show a fair comparison, the preconditioner has been rebuilt in every nonlinear iteration step.
Of course, re-use strategies might increase amortization of setup costs and reduce overall solver timings when solving actual problems.
\begin{table}
\centering
\caption{1000 deformable rings --- Exemplary timings in $\left[s\right]$ of the different preconditioning variants from Table \ref{tab:ch6_amgvariants_1000rings} for the full simulation ($4000$ time steps).}
\label{tab:ch6_1000rings_overalltimings}
\begin{tabular}{|l|rrr|} \hline
Method & Setup costs & Solver time  & Overall solver time \\ \hline
\paamg~(\cheapsimple) & 11870 & 10013 & 21883 \\ 
\EMIN~(\cheapsimple) & 12820 & 8679  & 21499 \\ 
\cheapsimple~(\paamg) & 11730 & 11103 & 22833 \\ 
\cheapsimple~(\saamg) & 12300 & 10763 & 23063 \\ \hline 
\end{tabular}
\end{table}


\subsection{Two tori impact example}
\label{sec:ch6_twotoriimpactexample}
Inspired by a similar analysis in \cite{yang2008}, with the two tori impact example we test the proposed multigrid variants from~\secref{sec:MultigridForMultiphysicsBlock} on a complex 3D contact example with more than 1 million unknowns.
Please refer to \cite{popp2012diss} for the detailed problem setup of the two tori impact example with geometry and load conditions.

\subsubsection{Setup}
\begin{figure}
\begin{center}
\framebox{
\begin{subfigure}[b]{0.22\textwidth}
\caption{$\ttime=0.0s$}
\label{fig:ch5_twotori_initial}
\includegraphics[width=1.0\textwidth]{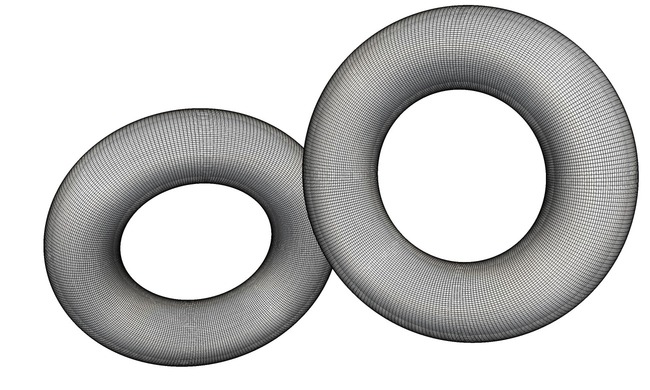}
\end{subfigure}%
}
\framebox{
\begin{subfigure}[b]{0.22\textwidth}
\caption{$\ttime=2.5s$}
\includegraphics[width=1.0\textwidth]{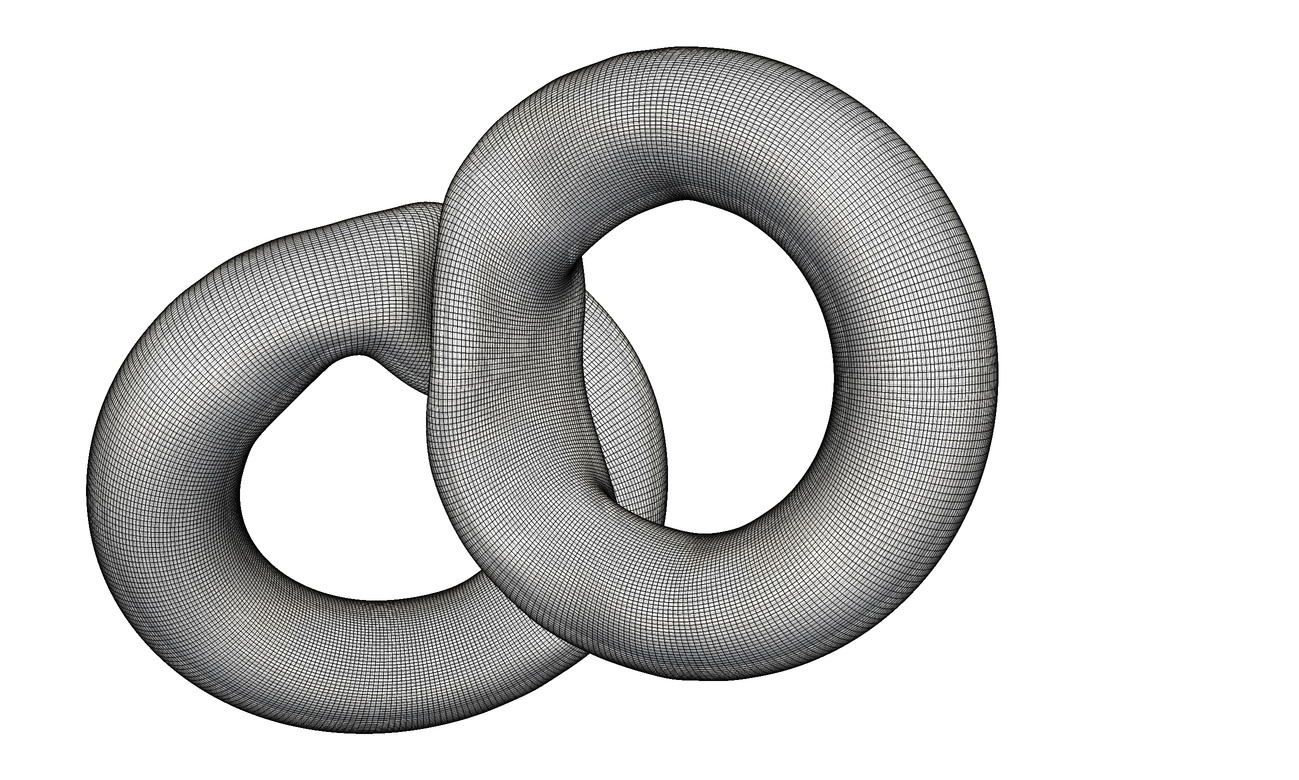}
\end{subfigure}%
}
\framebox{
\begin{subfigure}[b]{0.22\textwidth}
\caption{$\ttime=5.0s$}
\includegraphics[width=1.0\textwidth]{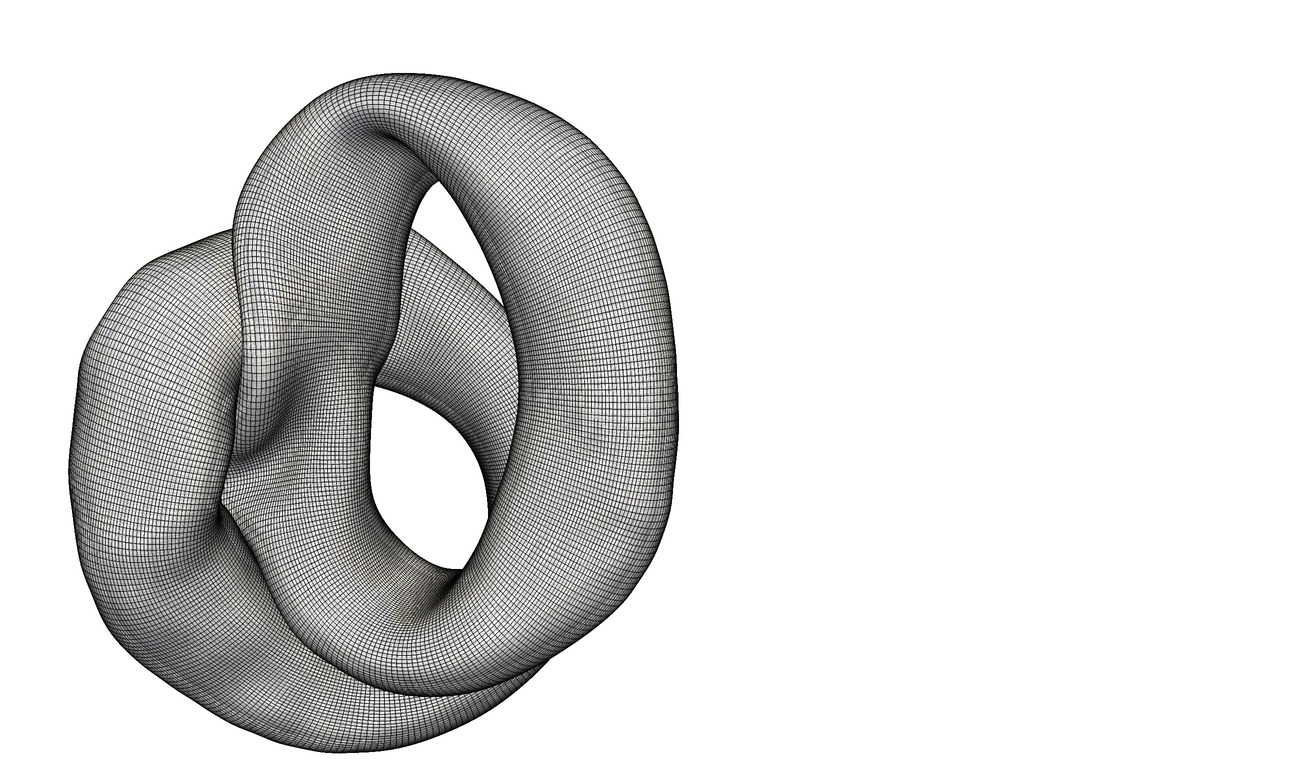}
\end{subfigure}%
}
\framebox{
\begin{subfigure}[b]{0.22\textwidth}
\caption{$\ttime=7.5s$}
\includegraphics[width=1.0\textwidth]{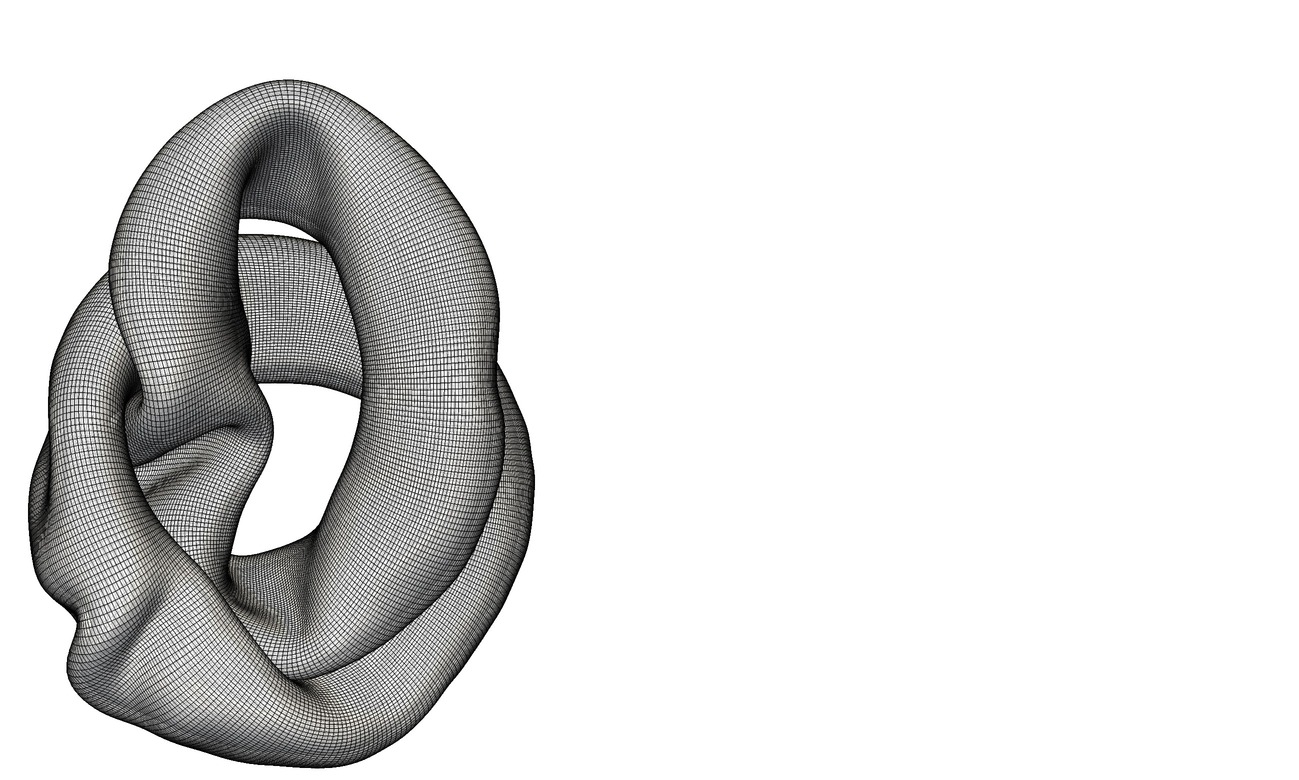}
\end{subfigure}%
}
\caption{Two tori impact example --- Characteristic stages of deformation.}
\label{fig:ch5_twotori}
\end{center}
\end{figure}

The example consists of two thin-walled tori with a Neo-Hookean material model ($\YoungModulus=2250$ GPa, $\nu=0.3$, $\refdensity=0.1~\frac{kg}{m^3}$) with a major and minor radius of $76m$ and $24m$ and a wall thickness of $4.5m$.
The left torus in Figure \ref{fig:ch5_twotori} lies in the $xy$-plane with resting initial conditions.
The right torus has been rotated around the $y$-axis by $45$ degrees and has an initial velocity of~$1.0~\frac{m}{s}$ directed towards the left torus.
The simulated time is $10s$ divided into $200$ time steps with a time step size of $0.05s$ using a generalized-$\alpha$ time integration scheme~\cite{Chung1993a}.
The mesh consists of $284672$ first-order hexahedral elements with $350208$ nodes.

With the rather complex geometry and contact configuration, that heavily and frequently changes over time,
this example can be considered as a representative test for the robustness and efficiency of the tested numerical methods.

\subsubsection{Stopping criteria}
The stopping criteria for the semi-smooth Newton method are chosen as
\begin{equation}
\norm[e]{\dispi}<10^{-7}~\wedge~ \Biggl(\norm[e]{\frac{\res^{\disp}_{\nonlinstep}}{\res^{\disp}_{0}}}<10^{-8} ~\wedge~ \norm[e]{\res^{\lagvec}_{\nonlinstep}}<10^{-4}\Biggr).
\label{eq:ch6_stoppingcriteriatwotori}
\end{equation}
Here, $\res^{\disp}_{\nonlinstep}$ and $\res^{\lagvec}_{\nonlinstep}$ denote the (nonlinear) residual for the displacement variables and Lagrange multipliers in the $i$-th Newton iteration, respectively.
The quantity $\dispi$ describes the solution increment for the displacement variables only.
Again, the stopping criteria in~\eqref{eq:ch6_stoppingcriteriatwotori} for the nonlinear solver are specifically chosen to produce the same number of nonlinear iterations for all tested solver variants. This allows for an easy comparison of the number of linear iterations during the simulation.

As linear solver a preconditioned GMRES method is used with the convergence criterion
\begin{equation}
\norm[e]{\frac{\res^\linstep}{\res^0}} < 10^{-8}
\end{equation}
for the full residual vector $\res^\linstep=\begin{bmatrix}\res^{\disp}\\ \res^{\lagvec}\end{bmatrix}$ in the linear iteration step $\linstep$.

\subsubsection{Results}
An overview of the different tested preconditioner variants is given in Table~\ref{tab:ch6_amgvariants_twotori}.
We study variants with the fully coupled multigrid approach, the nested multigrid approach, and a SIMPLE based variant without multigrid at all. 
For the fully coupled AMG variants, the transfer operators for the displacement blocks are varied. Particularly, non-smoothed transfer operators (\paamg) are compared to smoothed aggregation transfer operators (\saamg).
For the multigrid schemes there is no special treatment of the coarsest level. For the fully coupled multigrid schemes as well as for the nested multigrid method we apply the same level smoother on all multigrid levels including the coarsest level.
All the simulations have been run on $16$ cores (spread over 2 Intel Xeon E5-2670 Octocore CPUs).

\begin{table}
\begin{center}
\caption{Two tori impact example --- Different AMG variants.}
\label{tab:ch6_amgvariants_twotori}

\renewcommand{\arraystretch}{1.3}
\scalebox{0.8}{
\begin{tabular}{|l|l|} \hline
Full multigrid based methods & SIMPLE based methods \\

\hline

\begin{minipage}{0.47\textwidth}
\vspace{1em}
\textbf{\paamg~(\cheapsimple)}\\
\begin{tabular}{ll}
Transfer operators: & \paamg \\
Level smoother: & 1 \cheapsimplec (0.8)\\
\enspace$-$~Pred. smoother: & 1 SGS ($0.8$) \\
\enspace$-$~Corr. smoother: & ILU ($0$)
\end{tabular}
\end{minipage}

&

\begin{minipage}{0.47\textwidth}
\vspace{1em}
\textbf{\cheapsimple~(SGS)}\\
\begin{tabular}{ll}
Transfer operators:   & $-$\\
Block prec.: & 2 \cheapsimplec (0.8)\\
\enspace$-$~Pred. smoother: & 3 SGS ($0.8$) \\
\enspace$-$~Corr. smoother: & ILU ($0$)
\end{tabular}
\end{minipage}

\\ \hline

\begin{minipage}{0.47\textwidth}
\vspace{1em}
\textbf{\saamg~(\cheapsimple)}\\
\begin{tabular}{ll}
Transfer operators: & \saamg~($0.4$)\\
Level smoother: & 1 \cheapsimplec (0.8)\\
\enspace$-$~Pred. smoother: & 1 SGS ($0.8$) \\
\enspace$-$~Corr. smoother: & ILU ($0$)
\end{tabular}
\end{minipage}

&

\begin{minipage}{0.47\textwidth}
\vspace{1em}
\textbf{\cheapsimple~(\saamg)}\\
\begin{tabular}{ll}
Block prec.: & 2 \cheapsimplec (0.8)\\
\enspace$-$~Pred. smoother:  & AMG \\
$\qquad-$~Transfer op.:   & \saamg~($0.4$) \\
$\qquad-$~Level sm.: & 2 SGS ($0.8$) \\
\enspace$-$~Corr. smoother:  & ILU ($0$)
\end{tabular}
\end{minipage}
\\ \hline

\end{tabular}
} 
\renewcommand{\arraystretch}{1.0}

\end{center}
\end{table}

\begin{figure}
\begin{subfigure}{1.0\textwidth}
\includegraphics[width=\textwidth]{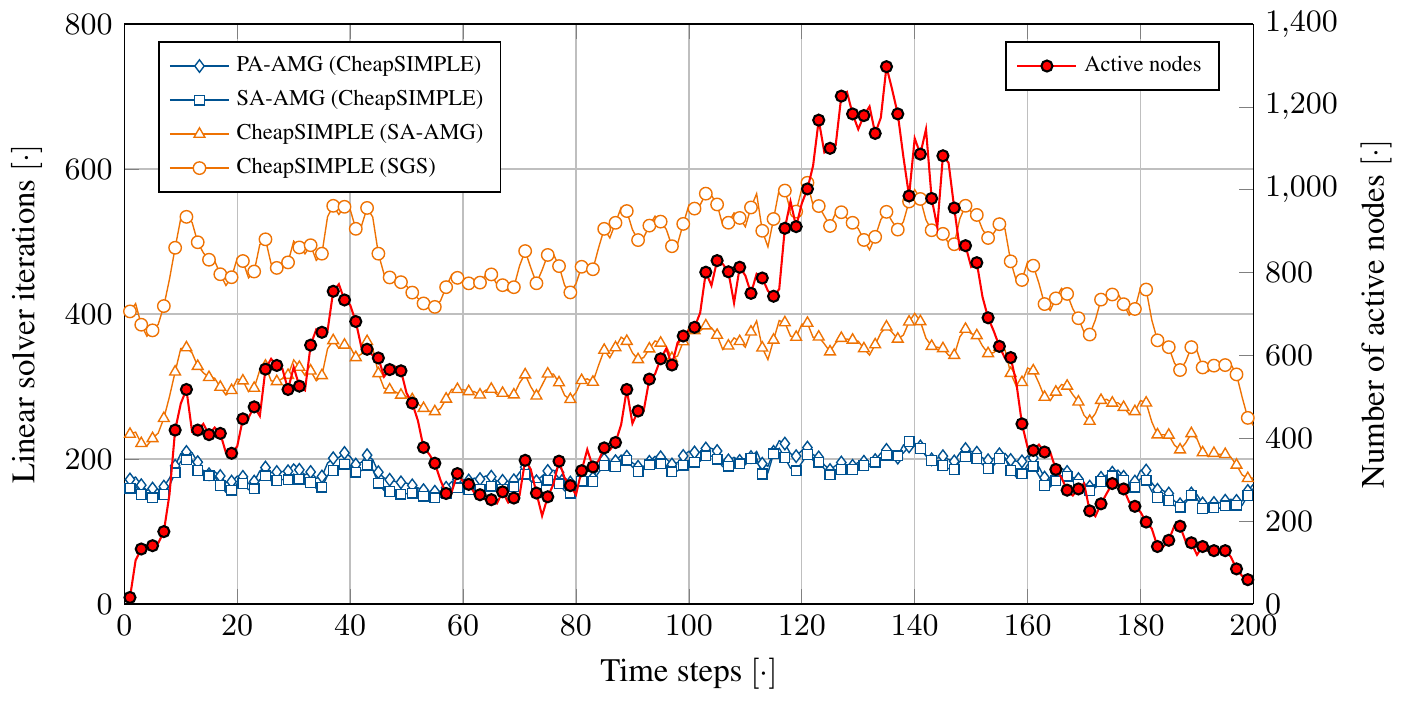}
\caption{Number of linear GMRES iterations for all nonlinear iterations over time steps for different preconditioner variants. The red curve shows the number of active nodes for each time step.}
\label{fig:ch6_twotoriiterations}
\end{subfigure}\vspace{3em}

\begin{subfigure}{1.0\textwidth}
\includegraphics[width=\textwidth]{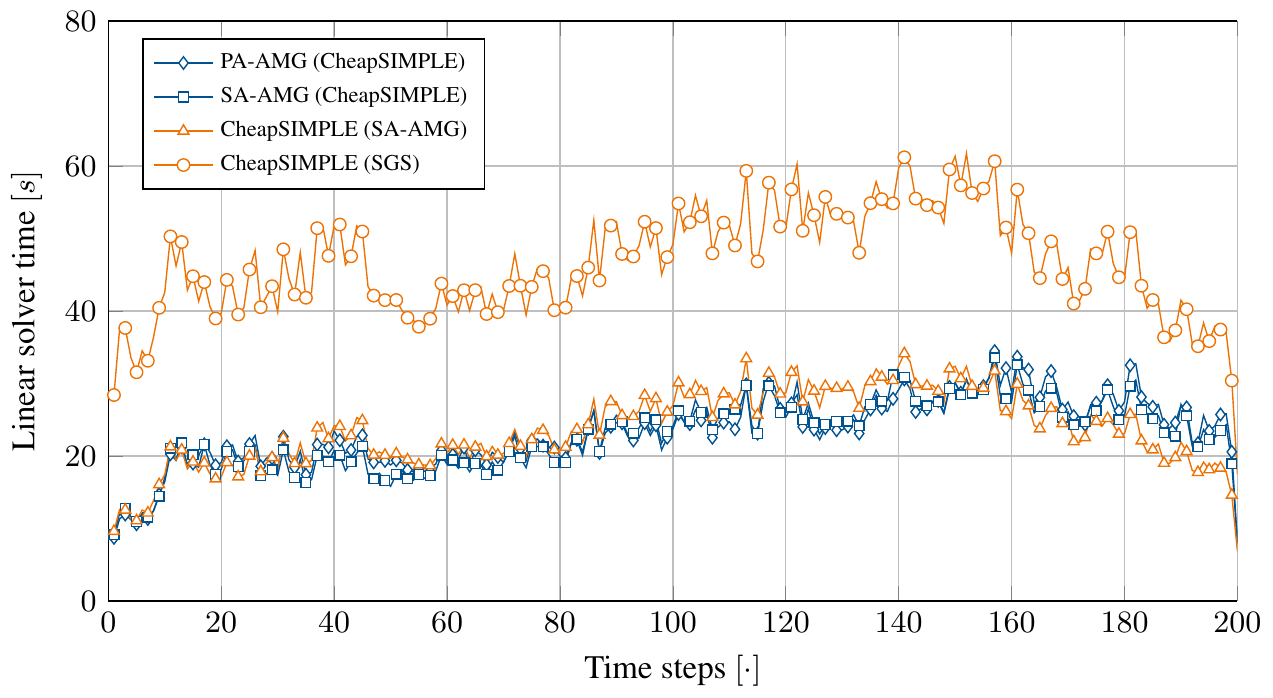}
\caption{Overall linear solver time over time steps.}
\label{fig:ch6_twotoritimings}
\end{subfigure}
\caption{Two tori impact example --- Results for different saddle point preconditioner variants.}
\label{fig:ch6_twotoriresults}
\end{figure}

Figure \ref{fig:ch6_twotoriiterations} shows the overall number of linear iterations performed to solve all nonlinear iterations for each time step.
All preconditioner variants require exactly the same number of nonlinear iterations per time step due to the particular choice of stopping criteria in \eqref{eq:ch6_stoppingcriteriatwotori},
ranging between 6 and 10 nonlinear iterations per time step, while also the number of nonlinear iterations is independent of the size of the active set.
Obviously, the SIMPLE based methods need more linear iterations than the AMG based methods.
In this example, there is nearly no difference between the non-smoothed transfer operator variant \paamg~(\cheapsimple) and the smoothed transfer operator variant \saamg~(\cheapsimple).
Furthermore, there is no clear and obvious correlation between the number of linear iterations and the number of active nodes.
This indicates that the fully coupled multigrid method is robust and efficient with regard to the increasing complexity of the contact configuration over time,
which is not the case for cheaper methods such as the SIMPLE based methods.
Evidently, one can see a significant drop in the linear iterations for the last $20$ time steps of the simulation,
which may correspond to the small number of nodes in contact.

When looking at the corresponding solver timings over the time steps in Figure \ref{fig:ch6_twotoritimings}, one finds the \cheapsimple~(\saamg) method to be very close to the AMG based methods \paamg~(\cheapsimple) and \saamg~(\cheapsimple). 
For the AMG based methods, one sweep with a \cheapsimplec~method is applied on each level,
which internally uses one sweep with a symmetric Gauss--Seidel iteration for the primary variable and one ILU sweep for the constraint equation.
That is, quite a lot of time is invested in the coupling on all levels with the comparably expensive ILU method.
In contrast to the AMG based method,
the \cheapsimple~(\saamg)~method uses $2$ sweeps with a \cheapsimple~preconditioner for the coupling (on the finest level only).
Internally, a $3$ level AMG multigrid is used with $2$ symmetric Gauss--Seidel sweeps for the level smoother and an ILU sweep for the constraint correction equation.
These parameters have been found to result in a reasonably low number of linear iterations.
For this example the experiment shows that the \cheapsimple~(\saamg) method needs twice as many iterations as the \saamg~(\cheapsimple) method,
but the costs per iteration are only half of the costs of the \saamg~(\cheapsimple).
Nevertheless, the AMG based methods seem to have a small advantage, when the number of nodes in contact increases.

\tikzset{external/figure name={twotori_sp}}
\tikzset{external/figure name/.add={}{_history}}

Last but not least, the overall timings for the linear solver are reported in Table \ref{tab:ch6_twotori_overalltimings}.
Except for the \cheapsimple~(SGS) variant, which is far away from the others, there is no clear winner.
The setup costs are quite close, since the same transfer operators have to be built for all methods with only a small difference for smoothed versus non-smoothed transfer operators.
To show a fair comparison, the preconditioner has been rebuilt in every nonlinear iteration step.
Of course, re-use strategies might increase amortization of setup costs and reduce overall solver timings when solving actual problems.

\begin{table}
\caption{Two tori impact example --- Exemplary timings in $\left[s\right]$ of the different preconditioning variants from Table \ref{tab:ch6_amgvariants_twotori} for the full simulation over $200$ time steps.}
\label{tab:ch6_twotori_overalltimings}
\centering
\begin{tabular}{|l|rrr|} \hline
Method & Setup costs & Solver time  & Overall solver time \\ \hline
\paamg~(\cheapsimple) & 11630 & 4658 & 16288 \\ 
\saamg~(\cheapsimple) & 12250 & 4564 & 16814 \\ 
\cheapsimple~(\saamg) & 12130 & 4731 & 16861 \\ 
\cheapsimple~(SGS)    & 10270 & 9320 & 19590 \\ \hline 
\end{tabular}
\end{table}

\section{Conclusion}

We have presented algebraic multigrid schemes designed for saddle point problems arising from contact problems using mortar finite element methods.
The new fully coupled multigrid scheme has the advantage that the contact constraints are considered on all multigrid levels,
which significantly reduces the number of iterations.
It gives the user full control over the coupling process by appropriately  choosing the solver parameters.
Additionally, we have proposed a novel aggregation method for the Lagrange multipliers, which reuses existing aggregation information at the contact interface.
We have demonstrated the robustness and efficiency of the overall multigrid method for large examples with increasingly complex contact configurations over time
as well as its weak scalability up to $23.9$ million unknowns on 480 MPI ranks.

\footnotesize
\bibliographystyle{myplainnat}
\setlength{\bibsep}{0pt}
\bibliography{mortar,amg,refs,nonsym,scaling,ap,mayrmt}

\end{document}